\documentclass[twocolumn]{aastex631}

\usepackage{graphicx}
\usepackage{dcolumn}
\usepackage{bm}
\usepackage[nointegrals]{wasysym}
\usepackage{verbatim}
\usepackage{color}
\usepackage{amsmath}
\usepackage[utf8]{inputenc}
\usepackage[shortlabels]{enumitem}
\usepackage{wrapfig}
\usepackage{scalerel}
\usepackage[caption=false]{subfig}
\usepackage{multirow} 

\newcommand{\dzreion}{\Delta z_\mathrm{rei}}
\begin{document}
\captionsetup[subfigure]{labelformat=empty}

\preprint{APS/123-QED}

\title{The Atacama Cosmology Telescope: DR6 Power Spectrum Foreground Model and Validation}

\author[0000-0001-9571-6148]{Benjamin~Beringue} \thanks{\url{beringue@apc.in2p3.fr}} \affiliation{Universite Paris Cite, CNRS, Astroparticule et Cosmologie, F-75013 Paris, France}
\affiliation{School of Physics and Astronomy, Cardiff University, The Parade, Cardiff, Wales, UK CF24 3AA}

\author[0000-0002-7611-6179]{Kristen M.~Surrao} \thanks{\url{k.surrao@columbia.edu}}
\affiliation{Department of Physics, Columbia University, New York, NY 10027, USA} 

\author[0000-0002-9539-0835]{J.~Colin Hill}
\affiliation{Department of Physics, Columbia University, New York, NY 10027, USA} \affiliation{Flatiron Institute, 162 5th Avenue, New York, NY 10010 USA}

\author[0000-0002-2287-1603]{Zachary~Atkins} \affiliation{Joseph Henry Laboratories of Physics, Jadwin Hall, Princeton University, Princeton, NJ, USA 08544}

\author[0000-0001-5846-0411]{Nicholas Battaglia} \affiliation{Department of Astronomy, Cornell University, Ithaca, NY 14853, USA}
\affiliation{Universite Paris Cite, CNRS, Astroparticule et Cosmologie, F-75013 Paris, France}

\author[0000-0003-4922-7401]{Boris~Bolliet} \affiliation{Department of Physics, Madingley Road, Cambridge CB3 0HA, UK} \affiliation{Kavli Institute for Cosmology Cambridge, Madingley Road, Cambridge CB3 0HA, UK}

\author[0000-0003-0837-0068]{Erminia~Calabrese}
\affiliation{School of Physics and Astronomy, Cardiff University, The Parade, Cardiff, Wales, UK CF24 3AA}

\author[0000-0002-9113-7058]{Steve~K.~Choi}
\affiliation{Department of Physics and Astronomy, University of California, Riverside, CA 92521, USA}

\author[0000-0002-7633-3376]{Susan E. Clark}
\affiliation{Department of Physics, Stanford University, Stanford, CA 94305, USA}
\affiliation{Kavli Institute for Particle Astrophysics \& Cosmology, P.O. Box 2450, Stanford University, Stanford, CA 94305, USA}

\author[0000-0003-2856-2382]{Adriaan~J.~Duivenvoorden} \affiliation{Max-Planck-Institut fur Astrophysik, Karl-Schwarzschild-Str. 1, 85748 Garching, Germany}

\author[0000-0002-7450-2586]{Jo~Dunkley} \affiliation{Joseph Henry Laboratories of Physics, Jadwin Hall, Princeton University, Princeton, NJ, USA 08544} \affiliation{Department of Astrophysical Sciences, Peyton Hall, Princeton University, Princeton, NJ USA 08544}

\author[0000-0002-8340-3715]{Serena~Giardiello} \affiliation{School of Physics and Astronomy, Cardiff University, The Parade, Cardiff, Wales, UK CF24 3AA}

\author[0000-0003-3155-245X]{Samuel~Goldstein} \affiliation{Department of Physics, Columbia University, New York, NY 10027, USA}

\author[0000-0001-7449-4638]{Brandon S.~Hensley}
\affiliation{Jet Propulsion Laboratory, California Institute of Technology, 4800 Oak Grove Drive, Pasadena, CA 91109, USA}

\author[0000-0002-0965-7864]{Ren\'ee~Hlo\v{z}ek} \affiliation{Dunlap Institute for Astronomy and Astrophysics, University of Toronto, 50 St. George St., Toronto, ON M5S 3H4, Canada} \affiliation{David A. Dunlap Dept of Astronomy and Astrophysics, University of Toronto, 50 St George Street, Toronto ON, M5S 3H4, Canada}

\author[0000-0002-9429-0015]{Hidde T. Jense}
\affiliation{School of Physics and Astronomy, Cardiff University, The Parade, Cardiff, Wales, UK CF24 3AA}

\author[0000-0003-0238-8806]{Darby~Kramer} \affiliation{School of Earth and Space Exploration, Arizona State University, Tempe, AZ, USA 85287}

\author[0000-0002-2613-2445]{Adrien La Posta}
\affiliation{Department of Physics, University of Oxford, Denys Wilkinson Building, Keble Road, Oxford OX1 3RH, United Kingdom}

\author[0000-0002-6849-4217]{Thibaut~Louis} \affiliation{Universit\'e Paris-Saclay, CNRS/IN2P3, IJCLab, 91405 Orsay, France}

\author[0009-0006-5846-6016]{Yogesh Mehta}
\affiliation{School of Earth and Space Exploration, Arizona State University, Tempe, AZ, USA 85287}

\author[0000-0001-6606-7142]{Kavilan~Moodley} \affiliation{Astrophysics Research Centre, School of Mathematics, Statistics and Computer Science, University of KwaZulu-Natal, Durban 4001, South Africa}

\author[0000-0002-4478-7111]{Sigurd~Naess} \affiliation{Institute of Theoretical Astrophysics, University of Oslo, Norway}

\author[0000-0001-6541-9265]{Bruce~Partridge}
\affiliation{Department of Physics and Astronomy, Haverford College, Haverford PA, USA}

\author[0000-0001-7805-1068]{Frank~J.~Qu} \affiliation{Department of Physics, Stanford University, Stanford, CA} \affiliation{Kavli Institute for Particle Astrophysics and Cosmology, 382 Via Pueblo Mall Stanford, CA  94305-4060, USA} \affiliation{Kavli Institute for Cosmology Cambridge, Madingley Road, Cambridge CB3 0HA, UK}

\author[0000-0002-0418-6258]{Bernardita~Ried~Guachalla} 
\affiliation{Department of Physics, Stanford University, Stanford, CA 94305-4085, USA} 
\affiliation{Kavli Institute for Particle Astrophysics and Cosmology, 382 Via Pueblo Mall Stanford, CA  94305-4060, USA}
\affiliation{SLAC National Accelerator Laboratory 2575 Sand Hill Road Menlo Park, California 94025, USA}

\author[0000-0002-9674-4527]{Neelima Sehgal} \affiliation{Physics and Astronomy Department, Stony Brook University, Stony Brook, NY 11794, USA}

\author[0000-0002-8149-1352]{Crist\'obal Sif\'on}
\affiliation{Instituto de F\'isica, Pontificia Universidad Cat\'olica de Valpara\'iso, Casilla 4059, Valpara\'iso, Chile}

\author[0000-0002-7020-7301]{Suzanne T. Staggs}
\affiliation{Joseph Henry Laboratories of Physics, Jadwin Hall, Princeton University, Princeton, NJ, 08544,  USA}

\author[0000-0001-6778-3861]{Hy~Trac} \affiliation{McWilliams Center for Cosmology, Carnegie Mellon University, Department of Physics, 5000 Forbes Ave., Pittsburgh PA, USA, 15213}

\author[0000-0002-3495-158X]{Alexander~Van~Engelen} \affiliation{School of Earth and Space Exploration, Arizona State University, Tempe, AZ, USA 85287}

\author[0000-0002-7567-4451]
{Edward~J.~Wollack} \affiliation{NASA/Goddard Space Flight Center, Greenbelt, MD, USA 20771}

\date{\today}
\begin{abstract}

We discuss the model of astrophysical emission at millimeter wavelengths used to characterize foregrounds in the multi-frequency power spectra of the Atacama Cosmology Telescope (ACT) Data Release 6 (DR6), expanding on~\citet{dr6_lcdm}. We detail several tests to validate the capability of the DR6 parametric foreground model to describe current observations and complex simulations, and show that cosmological parameter constraints are robust against model extensions and variations. We demonstrate consistency of the model with pre-DR6 ACT data and observations from \emph{Planck} and the South Pole Telescope. We evaluate the implications of using different foreground templates and extending the model with new components and/or free parameters. In all scenarios, the DR6 $\Lambda$CDM and $\Lambda$CDM+$N_{\rm eff}$ cosmological parameters shift by less than $0.5\sigma$ relative to the baseline constraints. Some foreground parameters shift more; we estimate their systematic uncertainties associated with modeling choices. From our constraint on the kinematic Sunyaev-Zel'dovich power, we obtain a conservative limit on the duration of reionization of $\dzreion < 4.4$, assuming a reionization midpoint consistent with optical depth measurements and a minimal low-redshift contribution, with varying assumptions for this component leading to tighter limits. Finally, we analyze realistic non-Gaussian, correlated microwave sky simulations containing Galactic and extragalactic foreground fields, built independently of the DR6 parametric foreground model. Processing these simulations through the DR6 power spectrum and likelihood pipeline, we recover the input cosmological parameters of the underlying cosmic microwave background field, a new demonstration for small-scale CMB analysis. These tests validate the robustness of the ACT DR6 foreground model and cosmological parameter constraints. 

\end{abstract}

\tableofcontents

\section{Introduction}

Observations of the cosmic microwave background (CMB) temperature and polarization anisotropy are of paramount importance to establishing a robust description of early universie cosmology~\citep[e.g.,][]{wmap_spergel_2003,Planck:2018_pslkl,Balkenhol_SPT,dr6_lcdm}. Constraints are predominantly derived via fitting models to the multi-frequency auto- and cross-power spectra of polarized microwave sky maps.  Parameters describing these models are fit using Bayesian techniques, e.g., using Markov Chain Monte Carlo (MCMC) methods to draw samples from the parameter posterior distributions, usually under the assumption of a Gaussian likelihood for the data on intermediate and small scales.  These multi-frequency power spectra are composed of contributions from the blackbody CMB --- which contains the cosmological information of interest --- and various foreground fields.  Such analyses thus require models for the foreground components' power spectra, with associated free parameters that must be inferred from the data themselves, due to the lack of exact knowledge regarding the foregrounds' amplitude, spectral dependence, and angular scale dependence. 

These foregrounds include both Galactic and extragalactic fields, with Galactic foregrounds predominantly affecting large scales, and extragalactic foregrounds predominantly affecting small scales. Galactic foregrounds include thermal dust, synchrotron, free-free, and anomalous microwave emission. Extragalactic foregrounds include the thermal and kinematic Sunyaev--Zel'dovich effects, the cosmic infrared background (CIB), and radio point source emission.  Recent studies have also shown that extragalactic CO emission may also be non-negligible on small scales at relevant frequencies~\citep{Maniyar_2023, Kokron:2024, meta_CO_25}. Past works have developed analytic models for several of these foregrounds~\citep[e.g.,][]{Reichardt_2012,Dunkley_2013, Louis:2017, choi_atacama_2020, George_SPT:2014, Reichardt_SPT:2020, Planck:2013_pslkl, Planck:2015_pslkl, Planck:2018_pslkl, Rosenberg_npipe:2022}. In addition, simulations of these foregrounds have been developed and calibrated to data, such as the \verb|PySM| Galactic foreground simulation suite~\citep{Thorne_2017, Pan-ExperimentGalacticScienceGroup:2025vcd}; the \verb|AGORA|~\citep{Omori:2022uox} and \verb|WebSky|~\citep{Websky} CMB and extragalactic foreground simulation suites; and the Planck Sky Model~\citep{Delabrouille:2009, Delabrouille:2011}.

The Atacama Cosmology Telescope (ACT) Data Release 6 (DR6) provides some of the tightest constraints on cosmological parameters to date. \citet{dr6_maps} (hereafter N25), \citet{dr6_lcdm} (hereafter L25), and \citet{dr6_extended} (hereafter C25) describe the DR6 maps, power spectra and cosmological parameter constraints in the $\Lambda$CDM and extended models. Given the high sensitivity of the ACT DR6 data, particularly to the small-scale power spectra, it is essential to validate the robustness and flexibility of the DR6 foreground model using existing data, state-of-the-art simulations, and ultimately the DR6 data themselves, in order to ensure the robustness of the DR6 cosmological parameter constraints.

The primary goal of this paper is to fully describe and validate the foreground model for the ACT DR6 multi-frequency auto- and cross-power spectra presented in L25. This model is used in the multi-frequency likelihood \texttt{MFlike}\footnote{\href{https://github.com/ACTCollaboration/act_dr6_mflike}{\texttt{act\_dr6\_mflike}, version 1.0.0}} employed in the DR6 cosmological parameter analyses, as well as in the extraction of the ``CMB-only'' (foreground-marginalized) \texttt{ACT-lite}\footnote{\href{https://github.com/ACTCollaboration/DR6-ACT-lite}{\texttt{DR6-ACT-lite}, version 1.0.1}} likelihood (L25, C25).  

The paper is organized as follows. In \S \ref{sec.fg_modeling} the foreground power spectrum models used in the DR6 likelihood are described. In \S \ref{sec:existing_data} consistency of the DR6 foreground models with existing data, including pre-DR6 ACT, \emph{Planck}, and South Pole Telescope (SPT) observations, is shown.  In \S \ref{sec:fg_tests} various modifications to the baseline foreground model from L25 are discussed, showing that the DR6 $\Lambda$CDM cosmological parameter constraints are robust to these changes. Such tests are also repeated with the $\Lambda$CDM+$N_{\rm eff}$ extended cosmological model. In \S \ref{sec.ng_sims} we describe realistic microwave sky simulations that include all non-negligible sky components, with non-Gaussian structures and correlations amongst the components, and in \S \ref{sec.ng_sims_params} we show that the DR6 foreground models are flexible enough to recover the input cosmological parameters in these simulations with no significant biases at the current sensitivity of the data. We conclude in \S \ref{sec.conclusion}. 

In general, we quote parameter results with two significant figures on the error bar, though we later discuss uncertainty on the uncertainty. To derive parameter constraints, we run MCMC chains using \texttt{cobaya}~\citep{Torrado_2021}, computing theoretical predictions up to $\ell_{max} = 9000$. We ensure convergence by requiring the Gelman–Rubin statistic $R-1$ fall below $0.01$. The MCMC chains are analyzed and posterior distributions are plotted using the \texttt{GetDist} package~\citep{getdist_2019}, after discarding the first 30\% of samples as burn-in.

\section{Modeling of Foreground Power Spectra}
\label{sec.fg_modeling}

The ACT DR6 foreground model was originally introduced in L25. In this section, we provide a more detailed description of that model as well as model extensions initially explored in L25. For completeness, here we also briefly summarize the likelihood and modeling framework used in L25.

The cosmological likelihood (up to an additive constant) used for the ACT DR6 power spectrum analysis is a Gaussian likelihood describing the multi-frequency auto- and cross- power spectra:
\begin{equation}
    -2\ln L = \left(\mathbf{D^{th}} - \mathbf{D^{data}}\right)^T\mathbf{\Sigma^{-1}}\left(\mathbf{D^{th}} - \mathbf{D^{data}}\right).
\end{equation}
Several ingredients come into the computation of this likelihood, which is simply $L = \exp(-\chi^2/2)$ : $\mathbf{D^{th}}$ is the model vector, ordered to match the data vector $\mathbf{D^{data}}$. The difference between the two is weighted by the inverse covariance matrix $\mathbf{\Sigma^{-1}}$, described in detail in L25 and in~\citet{Atkins:2024jlo}. The assumptions that come into the construction of $\mathbf{D^{th}}$ are the focus of this work. 

Following~\citet{Dunkley_2013} and~\citet{choi_atacama_2020} --- and as summarized in L25 --- the total theory model for the cross-power spectrum between observable $X \subset \{T,E\}$ at frequency $\nu_i$ and observable $Y$ at frequency $\nu_j$  is given by: 
\begin{equation}
    D_\ell^{{\rm th}, X_iY_j} \equiv \frac{\ell(\ell+1)}{2\pi}C_\ell^{{\rm th},X_iY_j} = D_\ell^{{\rm CMB}, XY} + D_\ell^{{\rm FG}, X_iY_j},
\end{equation}
capturing the fact that observed anisotropies of the temperature and polarization fields receive, in addition to the primary CMB, contributions from Galactic and extragalactic foregrounds, where the latter also include secondary anisotropies of the CMB itself. The conventional conversion to $D_\ell$ yields the power per logarithmic interval in $\ell$. The DR6 baseline model for $D_\ell^{{\rm FG}, XY}$ follows~\citet{choi_atacama_2020} and~\citet{Dunkley_2013}, with some minor extensions. In this section, we describe in greater detail the different components of the model from L25, and the various extensions that we test (in L25 and in more detail here), some of which are ultimately included in the baseline DR6 foreground model in L25. 

The likelihood is implemented in the \texttt{MFLike}\footnote{\href{https://github.com/simonsobs/LAT_MFLike/tree/v1.0.0}{\texttt{LAT\_MFLike}, version 1.0.0}} software developed for the Simons Observatory, and the \texttt{fgspectra}\footnote{\href{https://github.com/simonsobs/fgspectra/tree/v1.3.0}{\texttt{fgspectra}, version 1.3.0}} software is used to produce the foreground model spectra. To allow for faster sampling of the cosmological parameters, this work uses high-accuracy emulators throughout for \verb|CAMB|~\citep{Lewis_1999} introduced in~\cite{Jense_2025} to compute the lensed CMB power spectra, $D_\ell^{{\rm CMB}, XY}$, which L25 did not.\footnote{The Einstein-Boltzmann codes used in L25 and C25 are described in those works; validation of the primary theory codes (\texttt{CAMB}, \texttt{CLASS}, and emulators thereof) can be found in Appendix A of C25.} Unless otherwise stated, we vary the cosmological (and instrumental systematics) parameters along with the ones describing the foregrounds, with the theoretical settings and model assumptions described in L25 and C25. We consider two cosmological models, $\Lambda$CDM and $\Lambda$CDM+$N_{\rm eff}$, to explore the baseline ACT DR6 cosmology and a model testing new physics operating on small scales, respectively. The cosmological parameter set to describe these models includes the physical baryon and dark matter densities, $\Omega_b h^2$ and  $\Omega_c h^2$, the amplitude and spectral index of primordial scalar perturbations, $A_s$ and $n_s$, both defined at a pivot scale $k_0=0.05$\,Mpc$^{-1}$, the Hubble constant, $H_0$, the optical depth to reionization, $\tau$, and the effective number of relativistic species $N_{\rm eff}$, which is fixed to 3.044 in $\Lambda$CDM and free to vary in $\Lambda$CDM+$N_{\rm eff}$. We assume three neutrino species, with two massless and one with mass 0.06 eV. For the optical depth to reionization, $\tau$, in the $\Lambda$CDM model we impose a prior $\tau=0.0566 \pm 0.0058$, based on the {\it Planck} \href{https://web.fe.infn.it/~pagano/low_ell_datasets/sroll2/}{\texttt{Sroll2}} low-E likelihood~\citep{pagano/etal:2020}. For analysis of the $\Lambda$CDM+$N_{\rm eff}$ model, we add the full {\texttt{Sroll2}} likelihood along with the DR6 likelihood to constrain $\tau$.

\subsection{Formalism}
\label{subsec:fg_model}
Each component cross-power spectrum that contributes to $D_\ell^{{\rm FG}, XY}$ is modeled in the following schematic form:

\begin{equation}\label{eq:spectra_model}
    D_\ell^{X_iY_j} = a \frac{f(\nu_i) f(\nu_j)}{f^2(\nu_0)} D_\ell^0(\nu_0) \ .
\end{equation}

The multipole dependence of the power spectrum, $D_\ell^0(\nu_0)$, is normalized to unity at a reference frequency $\nu_0$ and angular scale $\ell_0$. Unless otherwise stated, $\ell_0=3000$ and $\nu_0=150~{\rm GHz}$ are used. This fiducial power spectrum is then multiplied by an overall amplitude $a$ and scaled by a factor $f(\nu)$ for each component as shown in Equation~\ref{eq:spectra_model}. The frequency dependence of the specific intensity of each astrophysical source, $ \delta I_\nu$, is determined by its spectral energy distribution (SED).  Each function $f(\nu)$ is expressed in temperature units (${\rm K}_{\rm CMB}$) referenced to fluctuations in the intensity of the CMB radiation via  the derivative of the blackbody spectrum as follows:
\begin{align}
    \delta I_\nu &= g^{-1}(\nu) f(\nu)\: \mathrm{with} \\  
    g^{-1}(\nu) &\equiv \left.\frac{\partial B_\nu(T)}{\partial T}\right\rvert_{T = T_{\rm CMB}}, \nonumber\\
    &= \frac{2 k_B^3 T_{\rm CMB}^2}{c^2h^2}\frac{x^4e^x}{\left(e^x-1\right)^2},
\end{align}
where we have defined $x\equiv h\nu/k_BT_{\rm CMB}$. Here $h$ is Planck's constant, $k_B$ is the Boltzmann constant, and $T_{\rm CMB}$ is the CMB temperature of 2.726 K~\citep[][]{Fixsen2009}.

\subsection{Passband Integration and Beam Chromaticity}

The power spectra above describe the contribution at a single, specific frequency, whereas the DR6 array-bands are sensitive to broader frequency ranges, described by their bandpass transmission functions $\tau^\alpha (\nu)$ (here, $\alpha$ indexes the array-bands: f090, f150, f220, where the three digits indicate the approximate observing band central frequency in GHz; see N25). Each component then has its frequency response integrated over the passbands to properly account for the variations of $f(\nu)$ across $\tau^\alpha(\nu)$. Foregrounds are not the only signals that vary across the passbands; the beams also have a frequency dependence, referred to as beam chromaticity. Following~\cite{Giardiello_2024}, chromatic beam window functions $b_\ell^\alpha(\nu)$ are included to account for this effect. As shown in L25, the modeled power spectrum for a given foreground component $c$, accounting for beam chromaticity and integrated over the passbands, is computed via
\begin{equation}
    D_{\ell,c}^{X_\alpha Y_\beta} = \int d\nu_i d\nu_j \tilde{B}_\ell^\alpha(\nu_i)\tilde{B}_\ell^\beta(\nu_j) D_{\ell,c}^{X_i Y_j}
\end{equation}
where $\tilde{B}_\ell^\alpha(\nu)$ are the normalized passbands, defined by\footnote{The passbands $\tau$ are reported as the averaged detector response to a source with a Rayleigh-Jeans ($\nu^2$) spectrum. The $\nu^{-
2}$ factor is included to correct for the source spectrum.}:
\begin{equation}
    \tilde{B}_\ell^\alpha(\nu) = \frac{b_\ell^\alpha(\nu)g^{-1}(\nu)\tau^\alpha(\nu)\nu^{-2}}{\int d\nu'b_\ell^\alpha(\nu')g^{-1}(\nu')\tau^\alpha(\nu')\nu'^{-2}} \quad .
\end{equation} 

Beam errors are included directly in the covariance matrix, allowing us to marginalize the constraints over the uncertainty in the measurement of the beam~\citep{beams_inprep}.

Finally, uncertainties are allowed in the determination of the passbands $\tau^\alpha(\nu)$. As in L25, we assume these errors can be well approximated by an effective frequency shift of the whole passband by a fixed value $\Delta_\nu^\alpha$: $\tau^\alpha(\nu) \rightarrow \tau^\alpha(\nu + \Delta^\alpha_\nu)$. For simplicity, it is assumed that the bandpass shifts cause only higher-order corrections to the beams, which are ignored ($b_\ell^\alpha(\nu + \Delta_\nu^\alpha) \approx b_\ell^\alpha(\nu)$). Further details on this formalism and its implementation in \texttt{MFLike} can be found in~\cite{Giardiello_2024}.

\subsection{Sunyaev-Zel'dovich Effects}
\label{subsec:sz}

The Sunyaev-Zel'dovich (SZ) effect is a secondary anisotropy of the CMB caused by the Compton (or inverse-Compton) scattering of CMB photons off free electrons in the late-time universe~\citep{Sunyaev_1970}. This effect includes two main components (with other, smaller contributions that are neglected): (i) the thermal SZ (tSZ) effect refers to the scattering of CMB photons off hot electrons and is particularly abundant in galaxy groups and clusters; (ii) the kinematic SZ (kSZ) effect refers to the Doppler boosting of CMB photons as they scatter off free electrons moving with a non-zero velocity along the line-of-sight. The kSZ signal can be further decomposed into contributions arising from bulk flows of newly-ionized electrons during the reionization epoch (reionization kSZ) and contributions generated by scattering off electrons in the ionized gas around and in between galaxies in the low-redshift universe (late-time kSZ).

The ACT DR6 tSZ power spectrum is modeled as follows:
\begin{equation}
    D_{\ell,\mathrm{tSZ}}^{T_i T_j} = a_\mathrm{tSZ} D_{\ell, \ell_0}^{\mathrm{tSZ}}\left[\frac{\ell}{\ell_0}\right]^{\alpha_\mathrm{tSZ}} \frac{f_{\rm tSZ}(\nu_i)f_{\rm tSZ}(\nu_j)}{f_{\rm tSZ}^2(\nu_0)} \,,
    \label{eq:tSZ}
\end{equation}
where $D_{\ell, \ell_0}^{\mathrm{tSZ}}$ is a tSZ power spectrum template normalized to unity at $\ell_0$ and frequency $\nu_0$ from the hydrodynamical simulations of~\cite{Battaglia_2010}. The tSZ spectral function in CMB-referenced thermodynamic units ($x \equiv h\nu/(k_B T_{\rm CMB})$)
\begin{equation}
\label{eq.tSZ_SED}
f_{\rm tSZ}(\nu) = x\:\mathrm{coth}\left( \frac{x}{2} \right) - 4
\end{equation}
describes the frequency dependence of the signal (here the non-relativistic limit is assumed;~\citealp{Birkinshaw_1999}). The parameter $\alpha_{\rm tSZ}$ in Equation~\eqref{eq:tSZ} is a new parameter introduced in L25, not previously included in the analysis of ACT data (or \emph{Planck} and SPT analyses), which allows for a different scale dependence of the tSZ signal, as compared to that in the template. This effective parametrization is able to capture a large variety of $\ell$-dependent shapes, including those from the BAryons and HAloes of MAssive Systems (\verb|BAHAMAS|) simulation suite \citep{Mccarthy:2017yqf} for different active galactic nucleus (AGN) gas heating temperatures, as well as SPT templates~\citep{Reichardt_SPT:2020}. As found in L25, and further described here in \S \ref{subsec:alpha_tsz}, we find evidence in the DR6 power spectrum analysis for a nonzero value of this new parameter, which provides new information about the intracluster medium astrophysics governing the tSZ signal, as it scales with halo mass as $M^{5/3}$. 

The kSZ signal adds a blackbody component to the CMB temperature anisotropies. Its contribution to the TT power spectrum is modeled via a template $D_{\ell, \ell_0}^\mathrm{kSZ}$ describing the kSZ power spectrum from~\cite{Battaglia_2013_kszpatchy}, such that
\begin{equation}
    D_{\ell,\mathrm{kSZ}}^{T_i T_j} = a_\mathrm{kSZ} D_{\ell, \ell_0}^\mathrm{kSZ}.
\end{equation}
We note that previous ACT power spectrum analyses, including L25, only modeled the late-time kSZ contribution, without an explicit additional model for the reionization kSZ contribution. This choice was motivated by the strong degeneracy between the two signals, which exhibit similar $\ell$-dependence~\cite[e.g.,][]{Alvarez2016,Park2016}, making them difficult to disentangle with ACT data alone. As such, the amplitude inferred for the kSZ template captures the sum of both contributions. We verify this approach by including separate late-time and reionization templates in the non-Gaussian sky simulations described in \S\ref{sec.ng_sims}.

\begin{table*}[htb]
	\centering
	\begin{tabular}{c|l|c|c}
		\hline\hline
	       & \textbf{Description} & {\textbf{Default Prior}} & \textbf{Extension}\\
		\hline \hline
		\boldmath{$a_{\rm tSZ}$} & Thermal SZ amplitude at $\ell=3000$ at $150~{\rm GHz}$ & $ \ge 0$ & --- \\ \hline
        
		\boldmath{$\alpha_{\rm tSZ}$} & Thermal SZ template spectral index &$-5 \le \alpha_{\rm tSZ} \le 5$ & --- \\ \hline
        
		\boldmath{$a_{\rm kSZ}$} & Kinematic SZ amplitude at $\ell=3000$ &$\ge 0$ \\ \hline
        
		\boldmath{$a_c$} & Clustered CIB amplitude at $\ell=3000$ at $150~{\rm GHz}$ & $\ge 0$ & --- \\ \hline
        
		\boldmath{$a_p$} & Poisson CIB amplitude at $\ell=3000$ at $150~{\rm GHz}$ &$ \ge 0$  & ---  \\  \hline
		
        \boldmath{$\beta_c$} & Clustered CIB spectral index &$0 \le \beta_c \le 5$ & --- \\  \hline

    	\boldmath{$\xi_{yc}$} & tSZ--CIB correlation coefficient  & $0 \le \xi_{yc} \le 0.2$ & $0 \le \xi_{yc} \le 1$  \\
        &at $\ell=3000$ at $150~{\rm GHz}$ & & \\ \hline
        
		\boldmath{$a_s^\mathrm{TT}$} & Poisson radio source amplitude in TT  &$ \ge 0$ & --- \\ 
        &at $\ell=3000$ at $150~{\rm GHz}$ & & \\ \hline

		\boldmath{$\beta_s$} & Radio source spectral index & $-3.5 \le \beta_s \le -1.5$ \\ \hline
        
		\boldmath{$a_g^\mathrm{TT}$} & Galactic dust amplitude in TT  & $(7.95 \pm 0.32)~{\rm \mu K^2}$ & ---   \\ 
        &at $\ell=500$ at $150~{\rm GHz}$ & & \\ \hline 
        
		\boldmath{$a_s^\mathrm{TE}$} & Poisson radio source amplitude in TE  & $-1 \le a_s^\mathrm{TE} \le 1$ & ---  \\ 
        &at $\ell=3000$ at $150~{\rm GHz}$ & & \\ \hline

		\boldmath{$a_g^\mathrm{TE}$} & Galactic dust amplitude in TE at $\ell=500$ at $150~{\rm GHz}$ & $(0.42 \pm 0.03)~{\rm \mu K^2}$ & ---\\ \hline
        
		\boldmath{$a_s^\mathrm{EE}$} & Poisson radio source amplitude in EE  & $0 \le a_s^\mathrm{EE} \le 1$ & --- \\ 
        &at $\ell=3000$ at $150~{\rm GHz}$ & & \\ \hline
        
		\boldmath{$a_g^\mathrm{EE}$} & Galactic dust amplitude in EE at $\ell=500$ at $150~{\rm GHz}$ & $(0.168 \pm 0.017)~{\rm \mu K^2}$ & ---  \\ 
        
        \hline \hline

        \boldmath{$\alpha_{g}^\mathrm{TE/EE}$} & Galactic dust $C_\ell$ power-law index in TE/EE  & $\alpha_g^\mathrm{TE/EE} = -0.4$ & $\alpha_g^{\rm TE/EE} \in [-2, 1]$ \\ 
        &for $\ell>500$ & & \\ \hline
        
        \boldmath{$\alpha_c$} & Clustered CIB $C_\ell$ power-law index for $\ell>3000$ & $\alpha_c = 0.8$ & $\alpha_c = 0.6$, $\alpha_c = 1$, $\alpha_c \in [0.5, 1]$ \\ 

        \hline

        \boldmath{$\beta_p$} & Poisson CIB spectral index & $\beta_p = \beta_c $ & $0 \le \beta_p \le 5$   \\ \hline

        \boldmath{$\beta_s^E$} & Radio source spectral index in polarization & $\beta_s^E = \beta_s $ & $-3.5 \le \beta_s^E \le -1.5$  \\ \hline

        \boldmath{$\xi_{ys}$} &  tSZ--radio correlation coefficient & $\xi_{ys} = 0$ & $ 0 \le \xi_{ys} \le 0.2$  \\ 
        & at $\ell=3000$ at $150~{\rm GHz}$ & & \\ \hline
        
        \boldmath{$\xi_{cs}$} &  CIB--radio correlation coefficient  & $\xi_{cs} = 0$ & $ 0 \le \xi_{cs} \le 0.2$  \\
        & at $\ell=3000$ at $150~{\rm GHz}$ & & \\ \hline

        \boldmath{$r^{\rm CIB}_{\nu_i\times \nu_j}$} &  CIB decorrelation between $\nu_i$ and $\nu_j$ & $r^{\rm CIB}_{\nu_i\times \nu_j} = 1$ & $ 0.8 \le r^{\rm CIB}_{\nu_i\times \nu_j} \le 1.0$, (for $\nu_i\neq\nu_j$)  \\ \hline

        \boldmath{$r^{\rm radio}_{\nu_i\times \nu_j}$} &  Radio decorrelation between $\nu_i$ and $\nu_j$ & $r^{\rm radio}_{\nu_i\times \nu_j} = 1$ & $ 0.8 \le r^{\rm radio}_{\nu_i\times \nu_j} \le 1.0$, (for $\nu_i\neq\nu_j$) \\

		\hline\hline
	\end{tabular}
	\caption{The 14 parameters of the baseline DR6 foreground model (top section of the table, as in L25), as well as additional parameters used in the foreground tests (bottom section of the table). For each parameter, we provide the default prior used in the baseline model. We also provide priors used in extensions where the prior is changed from the default. We note that the central value and width of the Gaussian prior on the dust amplitudes are based on the unrounded results of the fit presented in Eq.~\ref{eq:dust_amps}.} 
	\label{tab:fg_par_summary}
\end{table*}

\subsection{Cosmic Infrared Background}
\label{subsec.cib}

The CIB refers to the unresolved background of thermal emission from dusty galaxies over cosmic time~\cite[e.g.,][]{2005ARA&A..43..727L}. It receives contributions from both clustered and Poisson-distributed galaxies. The spatial statistics of the latter are modeled with a flat angular power spectrum (in $C_\ell$) and with a frequency dependence following a modified blackbody (MBB) SED with spectral index $\beta_p$ and temperature fixed to $T_d = 9.6~\mathrm{K}$:
\begin{equation}
    D_{\ell, \mathrm{CIB \text{--} p}}^{T_i T_j} = a_p \left[ \frac{\ell(\ell+1)}{\ell_0(\ell_0 + 1)} \right] \frac{\mu(\nu_i; \beta_p, T_d)\mu(\nu_j; \beta_p, T_d)}{\mu^2(\nu_0; \beta_p, T_d)} \, ,
\label{eqn:cib_p}
\end{equation}
with $\mu(\nu_i; \beta_p, T_d)$ a modified blackbody SED converted to CMB-referenced thermodynamic units:
\begin{equation}
\label{eq.MBB_SED}
\mu(\nu; \beta_p, T) \equiv \nu^{\beta_p} B_{\nu}(T) g(\nu) \, .
\end{equation} 
Both the overall amplitude $a_p$ and the spectral index $\beta_p$ are varied in the baseline model from L25.

The clustered component of the CIB emission is described by
\begin{equation}
    D_{\ell, \mathrm{CIB \text{--} c}}^{T_i T_j} = a_c D_{\ell, \ell_0}^\mathrm{CIB \text{--} c} \frac{\mu(\nu_i; \beta_c, T_d)\mu(\nu_j; \beta_c, T_d)}{\mu^2(\nu_0; \beta_c, T_d)},
    \label{eqn:cib_c}
\end{equation}
where $\beta_c$ is an MBB spectral index and $D_{\ell, \ell_0}^\mathrm{CIB \text{---} c}$ is a template power spectrum normalized to unity at $\ell = \ell_0$ and $\nu = \nu_0$.  This template is constructed as in \cite{choi_atacama_2020} using the CIB power measured from \emph{Planck}~\citep{planck_collaboration_planck_2014} up to $\ell=3000$, and extended with a power-law (defined in $C_\ell$) $\propto \ell^{\alpha_c}$ at smaller scales, motivated by earlier ACT and SPT measurements~\citep{Reichardt_2012, Dunkley_2013}. The DR6 baseline model fixes $\alpha_c = -0.8$ and $\beta_c = \beta_p$.  We investigate the impact of relaxing these assumptions and changing the temperature to $T_d = 25~{\rm K}$, as in~\citet{Reichardt_SPT:2020}, in \S \ref{sec:fg_tests}. 

The baseline model rigidly rescales the template across frequencies, assuming perfect correlation between channels. However, we expect that different frequencies probe different redshifts for which the dusty galaxy populations might be different, resulting in decorrelation between channels. This is supported by both observations (see, e.g.,~\citealp{planck_collaboration_planck_2014, Lenz:2019ugy}) and simulations (see, e.g.,~\citealp{Websky, Omori:2022uox}). The decorrelation is modeled empirically by multiplying the cross-frequency spectra for $\nu_i \neq \nu_j$ by an extra factor. The baseline model assumes that the three extra factors ($r^\mathrm{CIB}_{90\times 150}$, $r^\mathrm{CIB}_{90\times 220}$, and $r^\mathrm{CIB}_{150\times 220}$) are unity.  Here, $r^\mathrm{CIB}_{90\times 150}$ is the additional factor multiplying the cross-frequency spectra for 90 and 150 GHz, and similarly for the other factors.  Allowing these parameters to vary between $0.8$ and $1.0$ is considered as a model extension.

\begin{figure}
    \centering
    \includegraphics[width=0.99\linewidth]{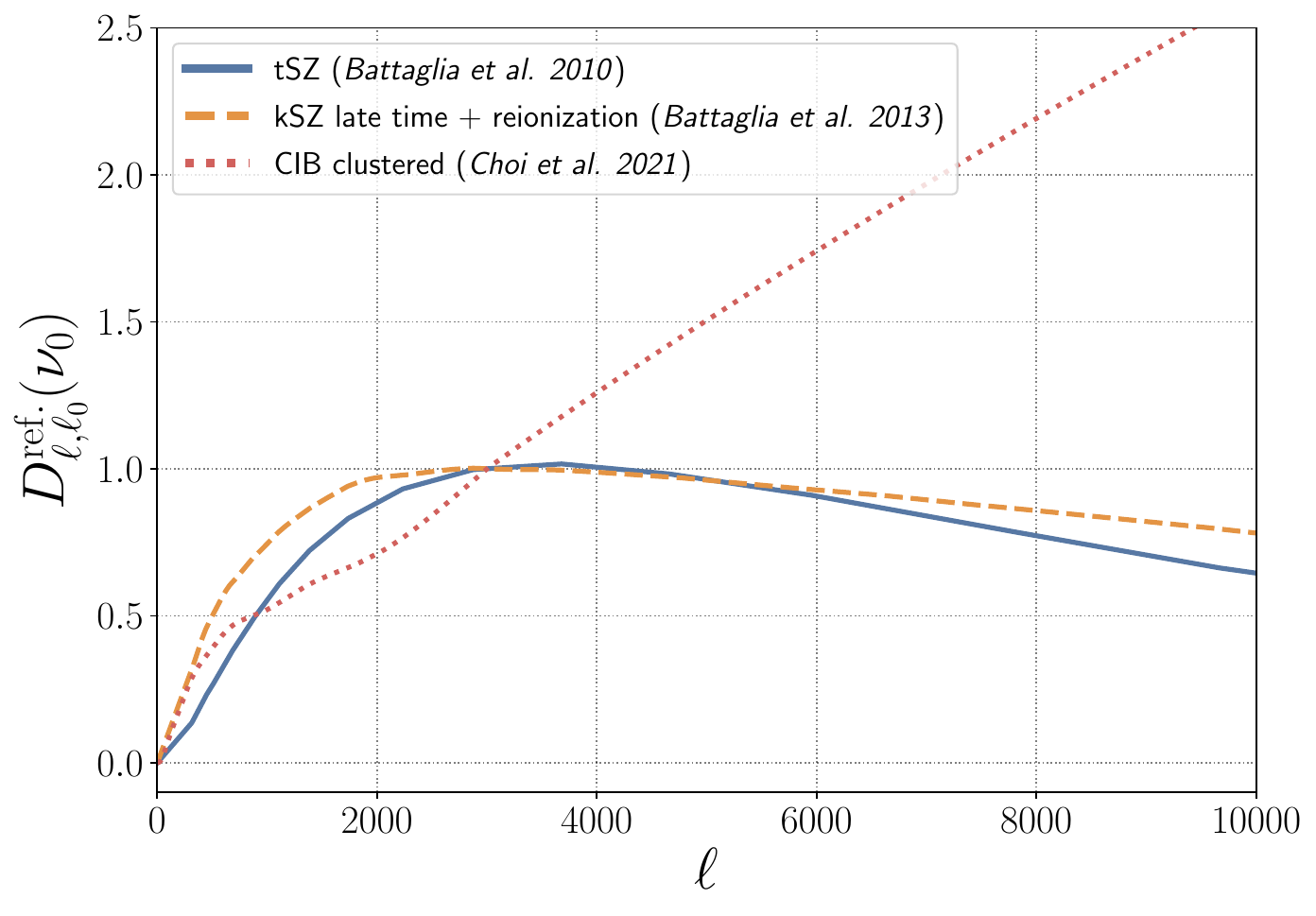}
    \caption{Templates for the $\ell$-dependence of the thermal and kinetic SZ and clustered CIB components of the baseline foreground model, as described in L25. These templates are normalized to unity at $\ell_0 = 3000$ and at a reference frequency $\nu_0 = 150~{\rm GHz}$ (for the tSZ and CIB), which is shown here.}
    \label{fig:templates_SZ_CIB}
\end{figure}

The tSZ, kSZ, and clustered CIB power spectrum templates are shown in Figure~\ref{fig:templates_SZ_CIB}. The tSZ and kSZ spectra exhibit broadly similar shapes, and are primarily distinguished by their distinct frequency dependences. As discussed in \S\ref{subsec:alpha_tsz} and in L25, we find tentative evidence that the shape of tSZ spectrum deviates slightly from the template in Figure~\ref{fig:templates_SZ_CIB}.

\subsection{Radio Galaxy Emission}

The redshifted emission of unresolved radio sources is expected to contribute to both the temperature and polarization anisotropy power spectra~\cite[e.g.,][]{2012AdAst2012E..52T}. Indeed, radio-source emission is dominated by synchrotron emission, which is intrinsically polarized. Even though the polarization fraction can be quite high, effects such as beam depolarization (averaging over multiple, differently oriented sky regions) or Faraday rotation can greatly reduce the polarization levels. Nevertheless, the TT, TE, and EE contributions are modeled as a flat angular power spectrum (in $C_\ell$), since radio sources are assumed to be mostly Poisson-distributed (see e.g., \citet{Sharp_2010, Hall_2010}). It is assumed that the radio source SED is a power-law in flux units~\citep{DeZotti_2009}, such that the power spectrum of this component is as follows:
\begin{equation}\label{eq:radio}
D_{\ell,\mathrm{radio}}^{X_i Y_j} = a_\mathrm{s}^{XY}\left[\frac{\ell(\ell + 1)}{\ell_0(\ell_0 + 1)}\right]\left[\frac{g(\nu_i)g(\nu_j)}{g^2(\nu_0)}\right] \left[ \frac{\nu_i\nu_j}{\nu_0^2} \right]^{\beta_s+2} \,,
\end{equation}
where $\beta_s$ is the spectral index of the adopted SED for the radio sources. In L25, it is assumed that the radio emission has the same frequency dependence in polarization as in temperature, but with separate amplitudes, $a_\mathrm{s}^\mathrm{TT}$, $a_\mathrm{s}^\mathrm{TE}$, and $a_\mathrm{s}^\mathrm{EE}$. The TE amplitude is allowed to span negative and positive values in order to capture possible anti-correlation between emission in T and E. As an extension of the baseline model, $\beta_s$ is allowed to have different values in temperature and polarization ($\beta_s^E$ is the new free parameter used to describe its value in polarization when allowing it to be different from the value in temperature). 

Similar to the CIB emission, radio galaxy emission is expected to show frequency decorrelation. As an extension to the baseline model, three extra factors are introduced ($r^\mathrm{radio}_{90\times 150}$, $r^\mathrm{radio}_{90\times 220}$, $r^\mathrm{radio}_{150\times 220}$), which are varied between $0.8$ and $1.0$.

\subsection{Cross-Correlations}

We expect several of the signals described above to be physically correlated, as they (at least partially) trace the same large-scale structure. These correlations can decrease the total power spectrum if the two signals are anti-correlated. For instance, since dark matter halos host both hot gas and galaxies, a non-zero correlation is expected between the tSZ signal and the clustered component of the CIB~\citep{Addison_2012}. In L25, the power spectrum of the tSZ--CIB cross-correlation is modeled as
\begin{align}
\label{eq.tSZCIB}
    &D_{\ell,\mathrm{tSZ}\times\mathrm{CIB}}^{T_i T_j} = - \xi_{yc} \sqrt{a_c a_\mathrm{tSZ}}D_{\ell, \ell_0}^{\mathrm{tSZ}\times\mathrm{CIB}} \nonumber\\
    &\times \left( \frac{f_{\rm tSZ}(\nu_i)\mu(\nu_j; \beta_c, T_d) + f_{\rm tSZ}(\nu_j)\mu(\nu_i; \beta_c, T_d)}{f_{\rm tSZ}(\nu_0)\mu(\nu_0; \beta_c, T_d)} \right) \,,
\end{align}
where $D_{\ell, \ell_0}^{\mathrm{tSZ}\times\mathrm{CIB}}$ is a template normalized to unity at $\ell_0$ and frequency $\nu_0$, $f_{\rm tSZ}(\nu)$ is the tSZ spectral function from Equation~\eqref{eq.tSZ_SED}, $\mu$ is the CIB spectral function from Equation~\eqref{eq.MBB_SED}, and $\xi_{yc}$ is the correlation coefficient between the tSZ and CIB components ($\xi_{yc}=1$ would mean that the two components are fully correlated). The minus sign accounts for the negative value of $f_{\rm tSZ}$ at $\nu_0 = 150{\; \rm GHz}$. We note that the DR6 modeling of the tSZ--CIB correlation only includes the clustered component of the CIB, since, as noted in~\cite{Addison_2012}, we do not expect the correlation between Poisson-distributed CIB sources and the tSZ signal (tracing massive clusters) to be statistically relevant. This differs from the SPT modeling in e.g.,~\citet{Reichardt_SPT:2020}, where both Poisson and clustered components are included in the tSZ--CIB correlation, leading to different interpretation and values for $\xi_{yc}$.

The template $D_{\ell, \ell_0}^{\mathrm{tSZ}\times\mathrm{CIB}}$ used in previous ACT analyses is derived from the analytic model of~\cite{Addison_2012}, which is adopted as a fiducial model in the DR6 foreground model. As an extension, using another template is considered, derived from the \verb|AGORA| simulations~\citep{Omori:2022uox}. Broadening the prior range for $\xi_{yc}$ is also considered. Based on previous measurements~\citep[e.g.,][]{Dunkley_2013}, the baseline prior range is restricted to $\left[0,0.2\right]$; as an extension, broadening it to $\left[ 0,1\right]$ is considered.

In addition, as an extension, a model for the tSZ--radio cross-correlation (in TT only, not TE) following that used for the tSZ--CIB cross-correlation above is constructed: 
\begin{align}
\label{eq.tSZradio}
    D_{\ell,\mathrm{tSZ \times rad}}^{T_i T_j} &= -\xi_{ys} \sqrt{a_{\rm tSZ} a_s} D_{\ell,\ell_0}^{\rm tSZ \times rad} \notag \nonumber 
    \\&\qquad \times \left( \frac{ f_{\rm tSZ}(\nu_i) \mu_2(\nu_j) + f_{\rm tSZ}(\nu_j) \mu_2(\nu_i)}{f_{\rm tSZ}(\nu_0) \mu_2(\nu_0)}  \right) \,,
\end{align}
where $\xi_{ys}$ is the correlation coefficient, $D_{\ell,\ell_0}^{\rm tSZ \times rad}$ is the template cross-power spectrum, and $\mu_2(\nu)$ is the radio source SED: $\mu_2(\nu) \equiv g(\nu) \nu^{\beta_s+2} $
. A template for $D_{0,\ell}^{\rm tSZ \times rad}$ from the \verb|AGORA| maps is constructed. A flux density cut of 15 mJy at $150~{\rm GHz}$ is applied to the source population in the maps to match that used in the ACT DR6 analysis (L25). The tSZ field is that corresponding to the \verb|BAHAMAS| 8.0 model in \verb|AGORA|. The predicted correlation coefficient from \verb|AGORA| is $\xi_{ys} \approx 0.03$; this prediction is quite uncertain, with the Websky simulation predicting $\approx 0.1$~\citep{Websky}.  A similar prior to the one for $\xi_{yc}$ is adopted, with $\xi_{ys}\in \left[0,0.2\right]$.

Similarly, as another extension, a model for the CIB--radio cross-correlation (in TT only, not TE) following that used for the tSZ--CIB  and tSZ--radio cross-correlations above is constructed:
\begin{align}
\label{eq.CIBradio}
    D_{\ell,\mathrm{CIB \times rad}}^{T_i T_j} &= \xi_{cs} \sqrt{a_c a_s} D_{\ell,\ell_0}^{\rm CIB \times rad} \notag \nonumber 
    \\&\qquad \times \left( \frac{\mu(\nu_i) \mu_2(\nu_j) + \mu(\nu_j) \mu_2(\nu_i)}{\mu(\nu_0) \mu_2(\nu_0)} \right) \,,
\end{align}
where $\xi_{cs}$ is the correlation coefficient (with a prior of $\xi_{cs} \in [0,0.2]$), $D_{\ell,\ell_0}^{\rm CIB \times rad}$ is the template cross-power spectrum, $\mu(\nu;\beta_c, T_d)$ is the CIB modified blackbody SED from Equation~\ref{eqn:cib_c}, and $\mu_2(\nu)$ is the radio source SED. Note the positive sign in Equation~\eqref{eq.CIBradio} compared to the negative sign in Equations~\eqref{eq.tSZCIB} and~\eqref{eq.tSZradio}, which is due to the tSZ SED at 150 GHz. As for the tSZ--radio cross-correlation, the theoretical template for the CIB--radio cross-correlation comes from a measurement of this cross-power spectrum in the \verb|AGORA| simulation.

As in the tSZ--CIB cross-correlation model, we assume that only the clustered CIB component correlates with the radio source population.  Investigating this assumption in detail is left to future work.

\begin{figure}[!t]
    \centering
    \includegraphics[width=0.99\linewidth]{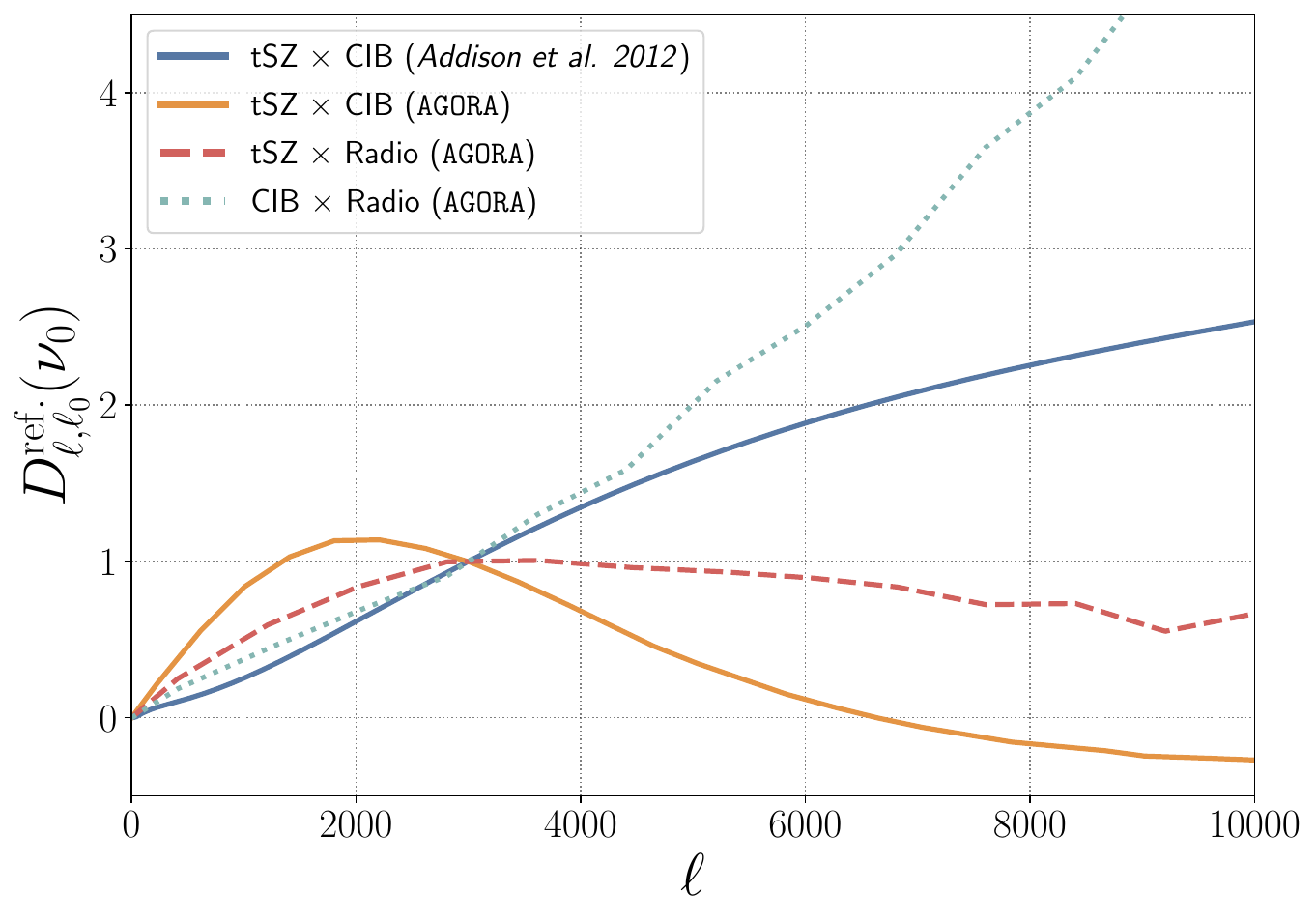}
    \caption{Templates for $\ell$-dependence of the cross-correlation terms considered in this work. The templates are normalized to unity at $\ell = 3000$ and at the reference frequency $\nu_0 = 150~{\rm GHz}$, which is shown here. The \texttt{AGORA} tSZ--CIB correlation is negative for $\ell \gtrsim 6500$. As discussed in~\citet{Omori:2022uox}, the shape of this template depends on the redshift limit in the tSZ and CIB maps used to measure this correlation. The CIB--radio cross-power spectrum is predicted to increase even more steeply on small scales than the fiducial tSZ--CIB cross-power spectrum.}
    \label{fig:correlation_templates}
\end{figure}

The normalized templates for the three cross-correlations are shown in Figure~\ref{fig:correlation_templates}. The CIB--radio cross-power spectrum is predicted to increase even more steeply on small scales than the tSZ--CIB cross-power spectrum.

The tSZ--radio or radio--CIB cross-correlations are not included in the baseline DR6 model, but including them as an extension, either in addition to or instead of the tSZ--CIB correlation, is considered.

\subsection{Galactic Dust}
\label{subsec:dust}

Galactic foregrounds, if not masked or modeled carefully, can bias cosmological inference. The most relevant sources of Galactic emission at the ACT frequencies are~\cite[e.g.,][]{2011piim.book.....D}: (i) Galactic dust emission, referring to the thermal emission of dust grains; (ii) Galactic synchrotron, the radiation emitted by charged particles spiralling in the Galactic magnetic field; (iii) free-free emission, the {\it Bremsstrahlung} radiation emitted by accelerating (or decelerating) charged particles in the interstellar medium; and (iv) anomalous microwave emission (AME), the diffuse astrophysical signal believed to originate from electric dipole radiation emitted by rapidly spinning dust grains in the interstellar medium.

The analysis mask used in L25 is the combination of a survey mask defining the DR6 footprint, a point-source mask, and a Galactic mask (G70) derived from the {\it Planck} $353~{\rm GHz}$ measurements and extended to mask additional bright dust clouds. The masked regions include areas close to the Galactic plane, where Galactic foreground emission is the brightest. Given the frequency bands observed in DR6 (f090, f150, f220), it is a reasonable assumption to neglect free-free emission, synchrotron emission, and AME. The steepness of the synchrotron SED~\citep{rybicki_1985} makes this signal subdominant to the dust emission at the DR6 frequencies. Besides, both AME and free-free emission have not been detected to be polarized~\citep[see, e.g.,][]{Thorne_2017} and their temperature signal is expected to be small compared to that of Galactic dust on the DR6 footprint.  We explicitly verify this assumption in \S\ref{sec.ng_sims}, in which we validate the DR6 foreground model on a full suite of non-Gaussian simulations that include Galactic dust, synchrotron emission, and AME.  In the simulation-based analysis, we see no appreciable levels of Galactic foregrounds aside from dust in the DR6 footprint, as shown in Figure~\ref{fig:ng_sim_comp_ps}.

The only Galactic foreground that is included in the DR6 model is therefore the thermal emission of Galactic dust, which is modeled in both intensity and polarization. Because dust grains are not perfect spheres, their short axis tends to align with the Galaxy’s magnetic field, leading to a polarization of this signal perpendicular to the projected local field, with polarization fractions reaching up to about $20\%$~\citep{planck_xix_2015}. 

The residual Galactic dust power spectra in temperature and polarization are modeled with a power-law in multipole $\ell$, while the frequency dependence follows a modified blackbody emission law:
\begin{equation}\label{eq:dust_pw}
    D_{\ell,g}^{X_i Y_j} = a_g^{XY}\left[ \frac{\ell}{\ell_0} \right]^{\alpha_g^{XY}}\frac{\mu(\nu_i; \beta_d, T_d^\mathrm{eff})\mu(\nu_j; \beta_d, T_d^\mathrm{eff})}{\mu^2(\nu_0; \beta_d, T_d^\mathrm{eff})} \,.
\end{equation}

In the baseline DR6 model, the power-law index describing the angular scale dependence is fixed to $\alpha_g^\mathrm{TT} = -0.6$ in temperature and to $\alpha_g^\mathrm{TE/EE}=-0.4$ for polarized emission, motivated by observations from \emph{Planck}~\citep{planck_poldust:2018}. The effective dust temperature is fixed to $T_d^\mathrm{eff}=19.6~\mathrm{K}$ and the modified blackbody spectral index is fixed to $\beta_d=1.5$. Separate amplitudes are fit in TT, TE, and EE, defined at a pivot scale $\ell_0=500$ (note that this pivot scale differs from that used for the other components in the sky model --- $\ell_0 = 3000$ --- shifting the fit to scales where the dust contribution is more relevant). As a possible extension of the baseline model, fixing the power-law indices $\alpha_g^\mathrm{TT}$, $\alpha_g^\mathrm{TE/EE}$ to different values than those assumed here, or marginalizing over them, is considered. When marginalizing over them, a uniform prior $[-2.0, 0.5]$ is imposed.

Because atmospheric fluctuations limit our ability to accurately measure the large-scale emission from Galactic dust using ACT alone, data from \emph{Planck} at $353~{\rm GHz}$ and $143~{\rm GHz}$ (with the \emph{Planck} passbands) are used to estimate the amplitude of the expected contamination. Details of the estimation procedure are given in L25; we briefly summarize them here. The following residual spectra are computed, by analysing the {\it Planck} maps in the DR6 footprint:
\begin{align}
    \Delta D^{XY, \rm data}_\ell &= D^{XY, \ 353 \rm GHz \times 353 \rm GHz}_{\ell, \ \rm planck} \nonumber \\
    &+  D^{XY, \ 143  \rm GHz \times 143  \rm GHz}_{\ell,  \ \rm planck} \nonumber \\
&- 2  D^{XY, \ 143 \rm GHz \times 353 \rm GHz}_{\ell,  \ \rm planck} .
\end{align}
Since the CMB signal cancels in this construction, we expect $\Delta D^{XY, \rm data}_\ell$ to be dust-dominated (dust here also refers to the CIB). These spectra are fit using different models in temperature and polarization. For TE and EE, the residuals above are assumed to be dominated by Galactic dust, whereas in TT, the CIB is also included:
\begin{align}
    \Delta D^{TT, \rm model}_\ell &= a^\mathrm{TT}_{g}(\ell / \ell_{0})^{\alpha_g^\mathrm{TT}} \Delta^{\rm dust}_{353, 143} +  \Delta D^{\rm CIB}_\ell (a_{c}, a_{p}), \nonumber \\
    \Delta D^{TE, \rm model}_\ell &= a^\mathrm{TE}_{g}(\ell / \ell_{0})^{\alpha_g^\mathrm{TE/EE}}\Delta^{\rm dust}_{353, 143}, \nonumber\\
    \Delta D^{EE, \rm model}_\ell &= a^\mathrm{EE}_{g}(\ell / \ell_{0})^{\alpha_g^\mathrm{TE/EE}}\Delta^{\rm dust}_{353, 143}.
\end{align}
Here $\Delta^{\rm dust}_{353, 143}$ gives the expected frequency scaling of the residual assuming an MBB, as in Equation~\eqref{eq:dust_pw}, integrated in the {\it Planck} measured passbands, defined in~\citet{planck_2013_ix}. Similarly,  $\Delta D^{\rm CIB}_\ell (a_{c}, a_{p})$ gives the frequency and $\ell$ scaling of the corresponding CIB residuals from Equations~\eqref{eqn:cib_p} and~\eqref{eqn:cib_c} as specified in \S\ref{subsec.cib}, where $\beta_c$ and $\beta_p$ have been fixed to their {\it Planck} best-fitting values, marginalizing over $a_c$ and $a_p$ while fitting for $a_g^\mathrm{TT}$. The $\ell$-power law provides an excellent fit to the data and the amplitudes (normalized at $150~{\rm GHz}$ and $\ell_0=500$) are estimated as (from L25)
\begin{align}
    \label{eq:dust_amps}
    a^\mathrm{TT}_{g} &=  8.0 \pm 0.2  \ \ ( {\rm PTE}: 20\%), \nonumber \\ 
    a^\mathrm{TE}_{g} &=  0.42 \pm 0.015  \ \ ( {\rm PTE}: 69\%), \nonumber \\
    a^\mathrm{EE}_{g} &=  0.17 \pm 0.01  \ \ ( {\rm PTE}: 38\%).
\end{align}
These best-fitting values are used as the priors for the dust amplitudes, and the Gaussian prior standard deviations are twice the $1\sigma$ uncertainties in Equation~\eqref{eq:dust_amps}. These fits have been performed for a range of values of dust power-law indices $\alpha_g^\mathrm{TT}$ and $\alpha_g^\mathrm{TE/EE}$, allowing us to vary the value of these parameters while maintaining consistent priors for the amplitudes.

\begin{figure*}[!t]
\center
\includegraphics[width=0.99\textwidth]{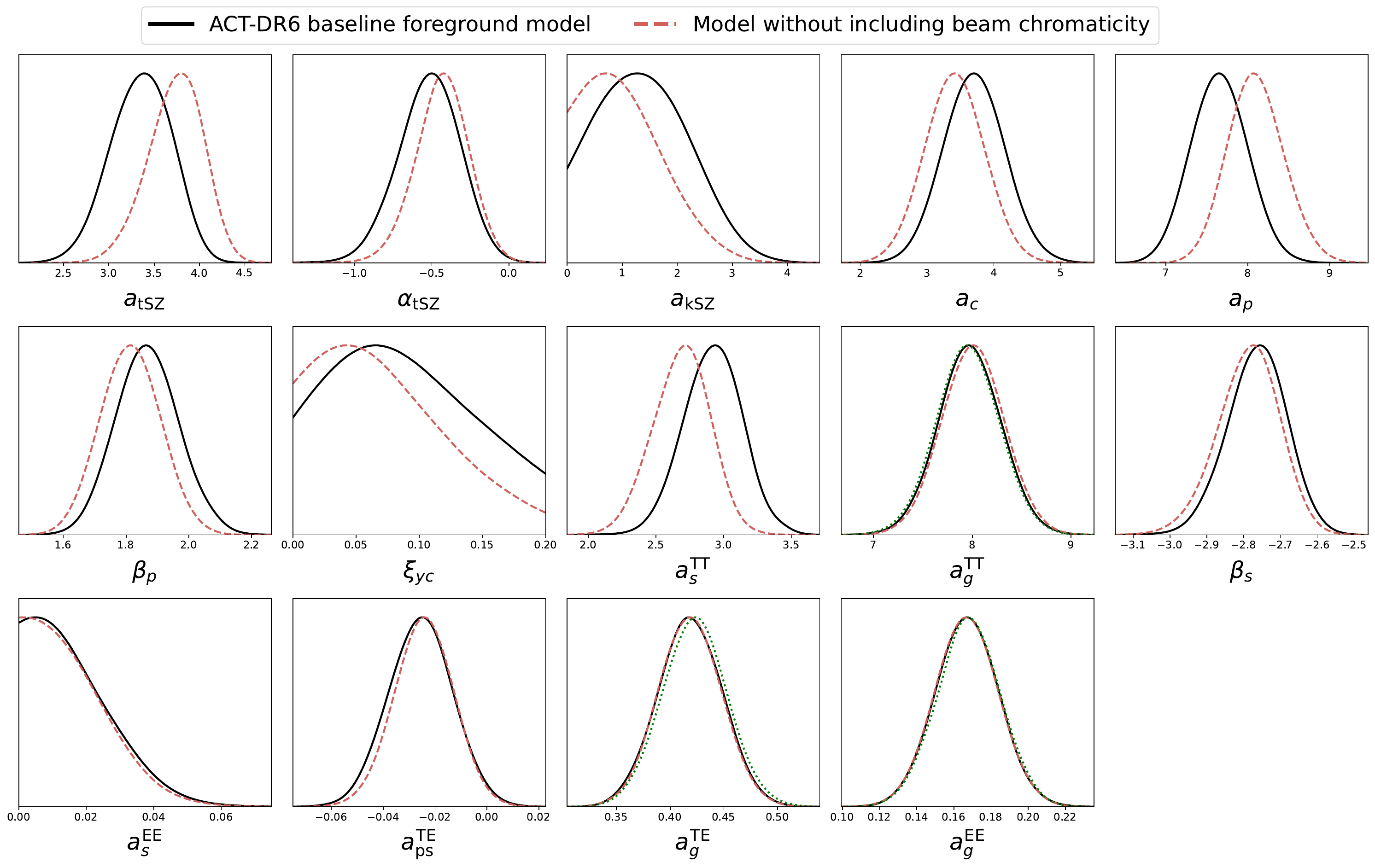}
\caption{Posterior distribution of the foreground parameters for the ACT DR6 baseline foreground model. The first two rows show the TT-related foreground parameters and last row reports the TE/EE foregrounds. Note that $\beta_s$ is assumed to be the same in temperature and polarization. The red dashed lines show the effect of neglecting beam chromaticity (the frequency dependence of the beams across the passbands). The green dotted lines show the priors imposed on the Galactic dust amplitudes.\\}
\label{fig:fg_dr6_bf}
\end{figure*}

\subsection{Parameter Summary}
Table~\ref{tab:fg_par_summary} summarizes the parameters of the DR6 baseline model, as well as the model extensions and associated additional free parameters (discussed further in \S \ref{sec:fg_tests}). In summary, the baseline DR6 foreground model includes Galactic dust (with free parameters $a_g^\mathrm{TT}$, $a_g^\mathrm{TE}$, and $a_g^\mathrm{EE}$), the tSZ effect (with free parameters $a_{\rm tSZ}$, $\alpha_{\rm tSZ}$), the late-time kSZ effect (with free parameter $a_{\rm kSZ}$), the CIB (with free parameters $a_c$, $a_p$, and $\beta_p$), radio emission (with free parameters $\beta_s$, $a_s^\mathrm{TT}$, $a_s^\mathrm{TE}$, and $a_s^\mathrm{EE}$), and the tSZ--CIB cross-correlation (with free parameter $\xi_{yc}$), for a total of 14 parameters.


\section{Results from ACT DR6 and Consistency with Other Data}

\label{sec:existing_data}
\subsection{ACT DR6 Foreground Estimates}
In this subsection, we briefly review the best-fitting foreground parameters from ACT DR6 for the baseline model discussed above, which were presented in L25. These results are shown in Figure~\ref{fig:fg_dr6_bf}. Unless otherwise stated, constraints are quoted at the $68\%$ confidence level (CL) and amplitudes are defined at $\nu_0 = 150~{\rm GHz}$ and $\ell_0=3000$.

\subsubsection{Thermal Sunyaev-Zel'dovich Effect}

As discussed in L25, previous works did not vary the shape of the tSZ power spectrum template, using a model with $\alpha_{\rm tSZ}=0$. The sensitivity of ACT DR6, however, requires the addition of this new parameter, which is measured at $>3\sigma$ when combining ACT and \emph{Planck} (L25) and leads to cosmological parameter shifts at more than $0.5\sigma$ when varying $\alpha_{\rm tSZ}$ (thus altering the shape of the inferred tSZ template). This is discussed further in \S \ref{sec:fg_tests}. A model that varies $\alpha_{\rm tSZ}$ as the baseline is thus adopted in L25.

The two tSZ-related parameters for the nominal model are then found in L25 for ACT-DR6-only to be (repeated here for completeness)
\begin{align}
    a_{\rm tSZ} &= 3.35^{+0.36}_{-0.32}~{\rm \mu K^2}, \nonumber \\
    \alpha_{\rm tSZ} &= -0.53^{+0.22}_{-0.19}.
\end{align}
The tSZ power spectrum amplitude is detected at high significance and the sensitivity of the data allows us to also constrain its $\ell$-shape, captured by the $\alpha_{\rm tSZ}$ parameter. As discussed in L25, the negative sign of $\alpha_{\rm tSZ}$ steepens the slope of the tSZ power spectrum towards larger scales, which is consistent with simulation-based predictions of stronger AGN feedback. Schematically, a higher AGN gas heating temperature leads to a more efficient ejection of gas from the cluster, which boosts the contribution from gas to tSZ power on large angular scales ($\ell \sim 1000$) and suppresses it on small angular scales~\citep[e.g.,][]{McCarthy2014}. This steepening towards large scales is consistent with recent results from~\citet{Efstathiou:2025ckq}, who reconstructed a binned version of the tSZ power spectrum using the 90-100 GHz data from \textit{Planck}, ACT DR4, and SPT.

\subsubsection{Kinematic Sunyaev-Zel'dovich Effect}
\label{sec:ksz}

The blackbody (kinematic) SZ component, measured by $a_{\rm kSZ}$, is not significantly detected, as reported in L25 for ACT-DR6-only:

\begin{equation}
    a_\mathrm{kSZ} = 1.48^{+0.71}_{-1.10}~\mathrm{\mu K^2}.
\end{equation}

The best-fitting value is consistent with previous limits. In particular, in their joint multi-frequency analysis of SPT-SZ and SPTPol data, ~\cite{Reichardt_SPT:2020} measured $a_\mathrm{kSZ} = 3.0 \pm 1.0~\mathrm{\mu K^2}$ for their fiducial kSZ template --- also reported in Figure~\ref{fig:fg_params_combined} below.

\begin{figure*}
\includegraphics[width = .5\linewidth]{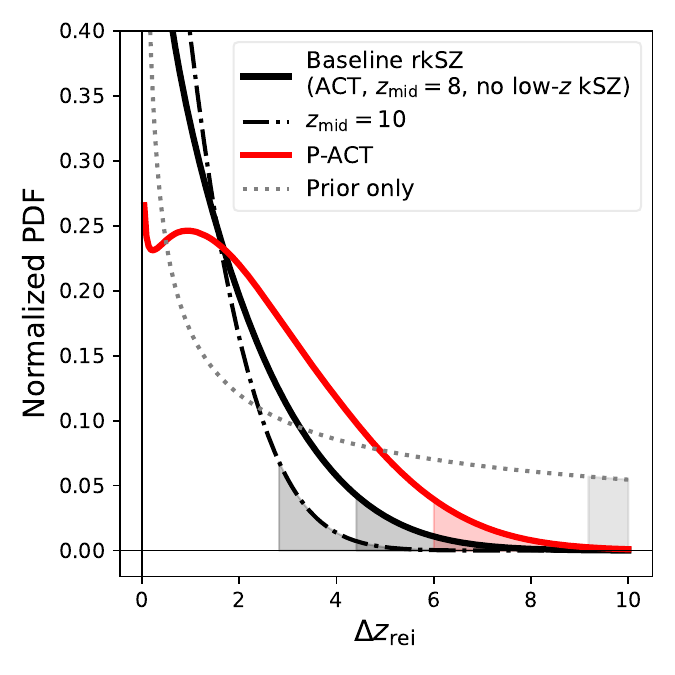}
\includegraphics[width = .5\linewidth]{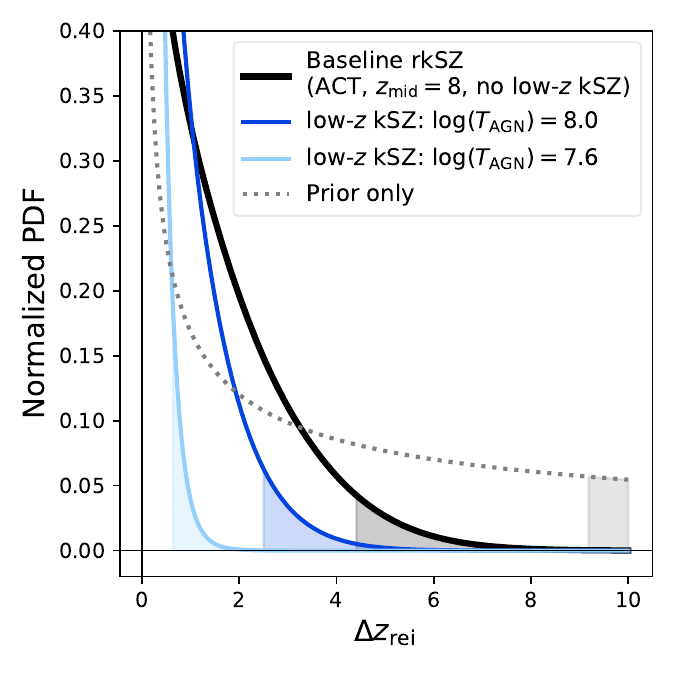}
\caption{Constraints on the duration of reionization, obtained by interpretation of the $a_\mathrm{kSZ}$ posteriors for ACT DR6 and P-ACT DR6 using the B13 scaling, under various assumptions. The shaded regions represent the 95\% exclusion limits. The baseline reionization kSZ treatment is shown in black, using ACT data with an assumed $z_\mathrm{mid} = 8$ and the most conservative treatment of the low-$z$ kSZ, in which all the signal is ascribed to the kSZ contribution from reionization. Left:  Small variations to the inferred limit are apparent for the P-ACT data (red) and for the ACT data when assuming $z_\mathrm{mid} = 10$  rather than 8. Right: allowing for a non-zero low-$z$ kSZ signal with the given amount of feedback (parametrized by the AGN heating temperatures), as implemented in the \texttt{AGORA} model, yields tighter constraints on the reionization duration. Note that all models considered here are informative over the prior (dotted grey).}
 \label{fig:deltaz_pdfs}
\end{figure*}

The kSZ signal is sourced by ionized gas with non-zero line-of-sight velocity from two different epochs. The component at $z \gtrsim 6$ from the reionization epoch is called the ``reionization kSZ'' (rkSZ) or ``patchy'' kSZ signal, while the kSZ contribution at $z\lesssim3$ arising from the circumgalactic medium associated with massive halos is known as the ``homogeneous'' kSZ or ``low-redshift'' kSZ signal. The amplitude of the reionization kSZ power spectrum is particularly sensitive to the duration of reionization \citep[e.g.,][]{Zahn2005,McQuinn2005,Iliev2007,Battaglia_2013_kszpatchy,2014JCAP...08..010C,Chen2023}. 
Although the ACT DR6 constraint on $a_{\rm kSZ}$ does not correspond to a detection of the kSZ power spectrum, it can be used to derive a limit on the duration of reionization, $\dzreion$.    

Together with both an assumed model for the low-redshift component and a prior on the tSZ power spectrum amplitude based on the tSZ bispectrum measured with SPT, \cite{Reichardt_SPT:2020} transformed their kSZ measurement into a limit on the duration of reionization, $\dzreion < 4.1$ at 95\% confidence level for their baseline modeling choice \citep[see also ][for results based on earlier measurements.]{Zahn2012,George_SPT:2014,2016A&A...596A.108P}. 

Here, a similar procedure is followed to transform the L25 constraint on $a_\mathrm{kSZ}$ into constraints on the duration of reionization. 
Specifically, starting with the ACT posterior for $a_{\rm kSZ}$ from Figure~\ref{fig:fg_dr6_bf}, the \cite{Battaglia_2013_kszpatchy} parametrization (B13 hereafter) is implemented. This parametrization is based on a semi-analytic model \citep{Battaglia_2013_model} calibrated with radiation-hydrodynamical simulations \citep{trac_cen_2007}.  It yields the dependence of $D_{\ell=3000}^{\rm rkSZ}$ on the midpoint of reionization $z_\mathrm{mid}$ and the duration $\dzreion$ (defined as the redshift interval between a spatially-averaged ionization fraction of 25\% and 75\%),
\begin{equation}
    a_{\rm kSZ} = D_{\ell=3000}^\mathrm{rkSZ}  = 2.03 \left[\frac{1 + z_\mathrm{mid}}{11} - 0.12\right] \left(\frac{\dzreion}{1.05}\right)^{0.51}. \label{eq:batt_model}
\end{equation}
Given that the rkSZ signal is most sensitive to $\dzreion$, and for simplicity, a fixed value of $z_\mathrm{mid}$ is assumed here in order to map the one-dimensional posterior distribution for $a_\mathrm{kSZ}$ to a distribution on $\dzreion$. Note that the optical depth measurement from the low-$\ell$ \emph{Planck} data constrains $z_\mathrm{mid}$ to a relatively narrow range of redshifts, $z_\mathrm{mid} = 8.14 \pm 0.61$~\citep{pagano/etal:2020}. Thus, a fiducial value of $z_\mathrm{mid}=8$ is set for the baseline analysis shown here, and the implications of other choices are also explored. A more involved analysis, with both parameters allowed to vary, is left to future work.

To transform probability distributions for $a_{\rm kSZ}$ to distributions for  $\dzreion$, the non-linear relationship between them in Equation~\eqref{eq:batt_model} must be accounted for:
\begin{equation}
\begin{aligned}
    P(\dzreion) & =  P(a_\mathrm{kSZ}) {\partial D_\mathrm{\ell=3000}^\mathrm{rkSZ} \over \partial \dzreion} \\
    & \propto  P(a_\mathrm{kSZ}) (\dzreion)^{-0.49}. \label{eq:deltaz_pdf}
\end{aligned}
\end{equation}
This applies to priors as well as posteriors; in particular, since the analysis chains were performed with a flat prior on $a_\mathrm{kSZ}$ (see Table~\ref{tab:fg_par_summary}), the prior on $\dzreion$ is $\propto (\dzreion)^{-0.49}$, yielding  a prior preference for low values of $\dzreion$. The degree to which the data prefer limits more stringent than those arising from  this prior will be determined. A maximum $\dzreion$ value of 10 was chosen for this work.

To obtain the most conservative limits on the reionization kSZ, it is assumed for the baseline choice that the low-redshift kSZ has no appreciable contribution, and its amplitude is set to zero. This yields a conservative 95\% upper limit of $\dzreion < 4.4$, for an assumed $z_\mathrm{mid} = 8.0.$

This analysis is performed on other cases as well: the P-ACT (referring to a combination of ACT DR6 and \emph{Planck} data, defined in~\S\ref{subsec:alpha_tsz}) result leads to a weaker upper limit, and a higher assumed $z_\mathrm{mid} =10$ yields a slightly tighter limit. Both of these cases are shown in the left-hand panel of Figure~\ref{fig:deltaz_pdfs} and in Table~\ref{tab:deltaz_constraints}.

The dependence of $a_{\rm kSZ}$ on $z_\mathrm{mid}$ and $\dzreion$ based on the assumptions used in the semi-numerical Abundance Matching Box for the Epoch of Reionization \citep[\texttt{AMBER}]{Trac2022} is also explored, as an alternative to B13. This code creates simulations based on a set of reionization parameters, which can be controlled more directly than with the hydrodynamical simulations of B13.  The additional ability to generate \texttt{AMBER} simulations quickly enables studies of the dependence of reionization observables on these parameters \citep{Chen2023,kramer2025}. Using the power-law scalings from Table 4 of \citet{kramer2025}, which were obtained from the \texttt{AMBER} simulations, an effective model for $D_{\ell=3000}^{\rm rkSZ} $ can be written as:
\begin{equation}
    D_{\ell=3000}^{\rm rkSZ} = 1.75 \left(\frac{z_\mathrm{mid}}{8.0}\right)^{1.4} \left( \frac{\Delta z_\mathrm{rei, 90}}{4.0}\right)^{0.75}.
\end{equation}
These are obtained for fixed values of the other parameters that can be varied within \texttt{AMBER}, namely, the radiation mean free path, $l_\mathrm{mfp} = 3.0\ \mathrm{Mpc}\ h^{-1}$, the minimum halo mass to host ionizing sources, $M_\mathrm{min} = 10^8 \, M_{\odot}$, and the asymmetry parameter, $A_z = 3.0$. These parameters have subdominant effects on $D_{\ell=3000}^{\rm rkSZ}$ \citep{Chen2023, kramer2025}. Note that this equation contains a different $\dzreion$ than used by B13: \texttt{AMBER} defines $\Delta z_\mathrm{rei,90}$ as the duration of reionization between when the Universe was 5\% ionized and 95\% ionized. Thus, the results obtained from using this parametrization make a different statement about reionization than those found using the B13 fit. The constraint on $\Delta z_\mathrm{rei, 90}$ from this \texttt{AMBER} model is $\Delta z_\mathrm{rei, 90} < 7.7$, much higher in value than $\dzreion$ from the middle 50\% in the B13 model, due to the larger ionization fraction range included.

Another set of $\dzreion$ constraints are obtained by accounting for a contribution to $a_\mathrm{kSZ}$ from the low-redshift kSZ signal (which is generically expected to be non-zero). The shape of the low-redshift kSZ template is not significantly different from that from reionization within the precision of the ACT data, as is confirmed with the test described in \S \ref{sec.ng_sims}. 
The amplitude of the low-redshift kSZ is sensitive both to cosmological parameters and to parameters that describe the distribution of ionized gas around halos \citep[e.g.,][]{Shaw2012}, with the largest uncertainty coming from the latter. Thus, the low-redshift amplitudes from the \texttt{BAHAMAS} 7.6 and 8.0 models in the \texttt{AGORA} kSZ simulations are added to the $a_\mathrm{kSZ}$ model in Equation~\eqref{eq:batt_model}. These choices lead to tighter limits of $\dzreion < 2.5$ and $0.7$ respectively, for the B13 parameterization and for a 50\% duration. The inferred reionization limits thus depend strongly on the assumed astrophysical feedback processes at low redshifts.

\begin{table}[]
    \centering
    \begin{tabular}{|r|c|}
          \hline
          Model & 95\% Upper Limit \\
          \hline
          \hline
          Baseline rkSZ &   \\
          (ACT,   B13 param., & $\dzreion < 4.4$ \\
         $z_\mathrm{mid}=8$, no low-$z$ kSZ) & \\
          \hline
          \hline
          $z_\mathrm{mid}=10$  & $\dzreion < 2.9$ \\
          \hline
          P-ACT & $\dzreion < 6.0$ \\
          \hline
          \hline
          low-$z$ kSZ: $\log(T_\mathrm{AGN}) = 8.0$ & $\dzreion < 2.5$ \\
          \hline
          low-$z$ kSZ: $\log(T_\mathrm{AGN}) = 7.6$ & $\dzreion < 0.7$ \\
          \hline
          \hline
          Prior only & $\dzreion < 9.1$ \\
          \hline
          \hline
         \texttt{AMBER} param. & $\Delta z_\mathrm{rei,90} < 7.7$ \\
          \hline
          \texttt{AMBER} param., $z_\mathrm{mid}=10$ & $\Delta z_\mathrm{rei,90} < 5.2$ \\
          \hline
    \end{tabular}
    \caption{Constraints on the duration of reionization, $\dzreion$, inferred from the fit to the kSZ power spectrum amplitude $a_\mathrm{kSZ}$; values are as described in  Fig.~\ref{fig:deltaz_pdfs}, apart from the \texttt{AMBER}-based results. Most results are obtained in terms of $\dzreion$, which here refers to the 50\% duration used by B13. The \texttt{AMBER} results are limits on a 90\% duration \citep{Trac2022, Chen2023}.} 
    \label{tab:deltaz_constraints}
\end{table}

It is worth noting that within our fits, $a_{\rm kSZ}$ is strongly (negatively) correlated with the thermal SZ component (to both $a_{\rm tSZ}$ and $\alpha_{\rm tSZ}$). Thus, the inferred constraints on $a_{\rm kSZ}$ (and hence $\dzreion$) are sensitive to the foreground model adopted in the analysis; see \S \ref{sec:fg_tests} for further exploration.

\vfill\null
\subsubsection{Cosmic Infrared Background}
The best-fitting CIB parameters for the baseline model in L25 are (included here for completeness)
\begin{align}
    a_c &= 3.69\pm 0.47~{\rm \mu K^2}, \nonumber \\
    a_p & = 7.65\pm 0.34~{\rm \mu K^2}, \nonumber \\
    \beta_p &= 1.87\pm 0.10,
\end{align}
consistent with previous results from ACT, SPT, and \emph{Planck}. We explore in greater detail the impact of allowing $\beta_p\neq \beta_c$ in \S \ref{sec:fg_tests}, but we note here that the inferred $\beta_c$ is also consistent with the value inferred from {\it Planck}~\citep{planck_collaboration_planck_2014}, where the estimated value for $\beta_{\rm CIB}$ was $1.75\pm0.06$, assuming that the clustered and Poisson components had the same frequency dependence. Given that the CIB is the dominant signal at high frequencies, the MBB spectral index $\beta_c$ strongly correlates with the bandpass shift $\Delta_\nu^{\rm f220}$. 

\subsubsection{tSZ -- CIB Correlation}

The tSZ--CIB correlation parameter is found in L25 to be
\begin{equation}
    \xi_{yc} = 0.091^{+0.041}_{-0.075}, 
\end{equation}
which is correlated with the SZ and clustered CIB amplitudes.  This result is compatible with 0 at $1.2\sigma$. However, modeling this correlation is challenging and the baseline template is taken from~\citet{Addison_2012} to allow for a direct comparison with~\citet{Dunkley_2013} and~\citet{choi_atacama_2020}. The constraint on $\xi_{yc}$ from L25 is consistent with previous ACT data, as well as predictions from simulations such as \texttt{AGORA}, which predicts $\xi_{yc} \sim 0.1$ at $\ell = 3000$.

\subsubsection{Radio Galaxies}
Power spectra (TT, TE, and EE) for radio emission from galaxies are characterized by four parameters in the baseline model (L25):
\begin{align}
    a_s^{\rm TT} &= 2.86\pm 0.21~{\rm \mu K^2}, \nonumber \\
    a_s^{\rm TE} & = -0.03\pm 0.01~{\rm \mu K^2}, \nonumber \\
    a_s^{\rm EE} &< 0.04~{\rm \mu K^2}~(95\%\:{\rm CL}), \nonumber \\
    \beta_s &= -2.78\pm{0.08}. 
\end{align} 
The amplitude in temperature is consistent with theoretical predictions from~\citet{Tucci_2011} of $a_s^{\rm TT} = 3.01~{\rm \mu K^2}$, given ACT's frequency coverage as well as flux-cut. There is no evidence for significant radio source emission in EE, and while the data prefer a TE radio source spectrum that is negative, it is still consistent with zero at the $3\sigma$ level. This is partly due to the point source mask (with a flux density cut of $15\:{\rm mJy}$ at $150\:{\rm GHz}$ in intensity), which removes most of the bright polarized point sources. Prior to masking, there is significant power in polarized point sources. 

\subsection{Comparing the Model with Other Observations}
\label{subsec:other_data}
The foreground models introduced in cosmological analysis of multi-frequency CMB data~\citep[e.g.,][]{Dunkley_2013, Planck:2018_pslkl, choi_atacama_2020, Reichardt_SPT:2020} make their own assumptions (using different component templates or SEDs, for example), rendering a direct comparison of the best-fitting foreground parameters difficult. In this subsection, we use the ACT DR6 baseline foreground model described above to fit other existing multi-frequency microwave data from \emph{Planck} and SPT. We focus on assessing whether the model is flexible enough to describe multiple datasets, both individually and when combined. 

As mentioned above, the ACT DR6 baseline model uses the previous ACT DR4 model~\citep{choi_atacama_2020} as a starting point and introduces some refinements. Hence, some foreground parameters between the two releases fluctuate in mean value while remaining consistent. For example, the thermal SZ amplitude $a_{\rm tSZ}$ in ACT DR6 is found to be $2.4\sigma$ (where $\sigma$ is combined from DR6 and DR4) lower than the DR4 estimate. This difference arises primarily from two key changes in the modeling: the inclusion of beam chromaticity effects and the marginalization over the tSZ shape parameter $\alpha_{\rm tSZ}$ (i.e., relaxing the assumption that $\alpha_{\rm tSZ}=0$). As shown in Figures~\ref{fig:fg_dr6_bf} and~\ref{fig:fg_from_fg_tests}, both of these contribute significantly to lowering the inferred tSZ amplitude. Another major distinction is the treatment of the radio point source spectral index $\beta_s$, which was fixed to $-2.5$ in ACT DR4 but allowed to vary in ACT DR6. The ACT DR6 best-fitting value is $\beta_s = -2.78\pm 0.08$ and letting this parameter float leads to moderate shifts in the Poisson amplitudes: $a_p$ and $a_s$ differ by $2.4\sigma$ and $1.8\sigma$, respectively, between the two datasets. We also note that the inclusion of the f220 channel in ACT DR6 enables us to lift the prior on the clustered CIB amplitude that was necessary in DR4, allowing for more data-driven foreground constraints.

We consider two other publicly available datasets in addition to DR6: data from the SPT-SZ+SPTPol surveys presented in~\cite{Reichardt_SPT:2020} (hereafter R21), and the \emph{Planck} PR3/\texttt{plik} multi-frequency spectra~\citep[][hereafter PL20]{Planck:2018_pslkl}. In the following, we briefly describe these two datasets and highlight the main differences between the model adopted in their analyses and the baseline DR6 foreground model. 

\subsubsection{SPT}
The SPT dataset is extracted from $500~\deg^2$ surveyed by SPTpol combined with $2540~\deg^2$ from the SPT-SZ survey. The data vector consists of $88$ $\ell$-bins capturing $6$ cross-frequency spectra between the 90, 150, and 220 GHz frequency channels in temperature only. The model used in R21 to characterize these data uses the same formalism described above, but with different templates for the tSZ power spectrum, kSZ power spectrum, and tSZ--CIB correlation. It also differentiates between the 1- and 2-halo contribution to the clustered CIB, each modeled with its own template and amplitude, but sharing a common frequency dependence. We implement the publicly available SPT likelihood\footnote{https://pole.uchicago.edu/public/data/reichardt20/} in the \verb|MFLike| framework and verify that we reproduce the results of R21. We then modify the likelihood to fit the SPT data with components of the DR6 model --- when possible, see below --- and finally bring together ACT DR6 and SPT in a joint fit. The two SPT surveys do not overlap with the ACT DR6 footprint and we treat them as independent, simply adding the individual log-likelihoods and neglecting any covariances between the two data vectors. 

For the modified SPT likelihood, we model the tSZ power spectrum, kSZ power spectrum, clustered CIB, and tSZ--CIB correlation using the same templates and frequency dependence as for the DR6 baseline model. For the Poisson-like terms (CIB and radio), we use separate amplitudes for ACT and SPT to account for the more aggressive flux-cut applied to the SPT temperature maps ($\sim 7.6~{\rm mJy}$ versus $15~{\rm mJy}$ at $150\:{\rm GHz}$). However, we assume a common frequency dependence for both experiments. To validate this assumption, we perform a joint fit allowing separate spectral indices, $\beta_s$ and $\beta_p$, for ACT and SPT. With this assumption, we find a $3.4\sigma$ discrepancy in the recovered radio spectral indices, while the difference in the CIB Poisson index $\beta_p$ is less than $1\sigma$. We interpret this result as a need to also relax other assumptions in the model. For example, introducing experiment-specific spectral indices necessitates relaxing the baseline condition $\beta_c = \beta_p$, but this would open up the model further, undermining the goal of testing the baseline DR6 foreground model on the combined dataset. For this reason, we opt to retain a common frequency scaling for Poisson sources in both ACT and SPT, which, as discussed in~\S\ref{subsec:results_joint}, gives a good fit to the data.  We defer a more detailed modeling of experiment-specific frequency behavior to future work.

The dust residual power spectra are also maintained with their respective models and priors. 

\subsubsection{\textit{Planck}}

{\it Planck} data are sensitive to secondary anisotropies in different multipole and frequency regimes compared to ACT and SPT. For example, the Galactic dust contamination is a dominant component for \emph{Planck}, affecting its large-scale measurements; the extragalactic sources are more problematic at frequencies around 100 GHz which lack resolution at $\ell>1500$. The approach used to marginalize, remove, and minimize foreground contamination varied across the different {\it Planck} high-$\ell$ likelihoods (\verb|plik|, \verb|Camspec|, and \verb|Hillipop|), each defining its own foreground model and in some cases after steps introduced at the map level. The impact of these model choices was studied in detail in \cite{Planck:2013_pslkl, Planck:2015_pslkl}. Capitalizing on the implementation of the \verb|plik| likelihood in the SO \texttt{MFLike} framework presented in~\citet{Li_2023}, we only consider the legacy PR3 dataset accompanied by the \verb|plik| likelihood described in detail in PL20. The dataset consists of 6 frequency cross-spectra between the $100$, $143$, and $217~{\rm GHz}$ channels in TT, TE, and EE (the $100\times 143$ and $100\times 217$ spectra are not included in TT). This is the \emph{Planck} dataset and likelihood that most resembles the approach used by DR6, i.e., multi-frequency spectra corrected for foregrounds only at the power spectrum level with a parametric model. The full model is presented in PL20 and has some differences compared to the ACT DR6 baseline model, which we highlight below:

\begin{itemize}
\itemsep0em 
    \item The clustered CIB model is constructed from the power spectra and emission law based on the one-plus-two halo model described in~\citet{planck_xxx_2016}. This model is fit to {\it Planck} data to produce a template. A perfect correlation between the emission at the three frequencies is assumed and is rescaled by a single amplitude, not allowing for any modeling of the frequency dependence.
    \item The tSZ and kSZ models follow the formalism described in \S\ref{sec.fg_modeling} but with different templates, respectively from~\citet{Efstathiou_2012} and~\citet{trac_templates_2011}.
    \item The CIB and radio Poisson terms are modeled (in temperature only) using a free amplitude for each cross-spectrum that rescales a flat angular power spectrum (in $C_\ell$). This is necessary because PR3 adopted a frequency-dependent flux-cut in temperature. Polarized point sources are neglected and therefore not modeled. 
    \item  Since the point-source mask is different for each channel, the residuals in temperature are channel-specific and their power spectra must be modeled using a different template for each cross-spectrum. Since point sources are neglected in polarization, a common template can be used to model the $\ell$-dependence of polarized dust residuals: a power-law (in $D_\ell$) with a (fixed) index $-0.4$. As indicated in~\citet{Planck:2018_pslkl}, the EE amplitudes are fixed, while the TT and TE amplitudes are allowed to vary within specific priors.
\end{itemize}

When combining DR6 with {\it Planck}, we follow the approach taken by L25 and restrict the \verb|plik| spectra to $\ell <1000$ in TT/TE and $\ell<600$ in EE. Including higher multipoles would require modeling the ACT-\emph{Planck} covariance, which is beyond the scope of this work. We refer to this dataset as \texttt{plik}$_{cut}$. To model foregrounds in this combination, we use the DR6 baseline templates for the tSZ power spectrum, kSZ power spectrum, clustered CIB, and tSZ--CIB correlation. However, we keep the modeling of Poisson sources and Galactic dust in \verb|plik| spectra as defined in PL20. 

\subsubsection{Results}\label{subsec:results_joint}
\begin{figure*}[!t]
\center
\includegraphics[width=0.99\textwidth]{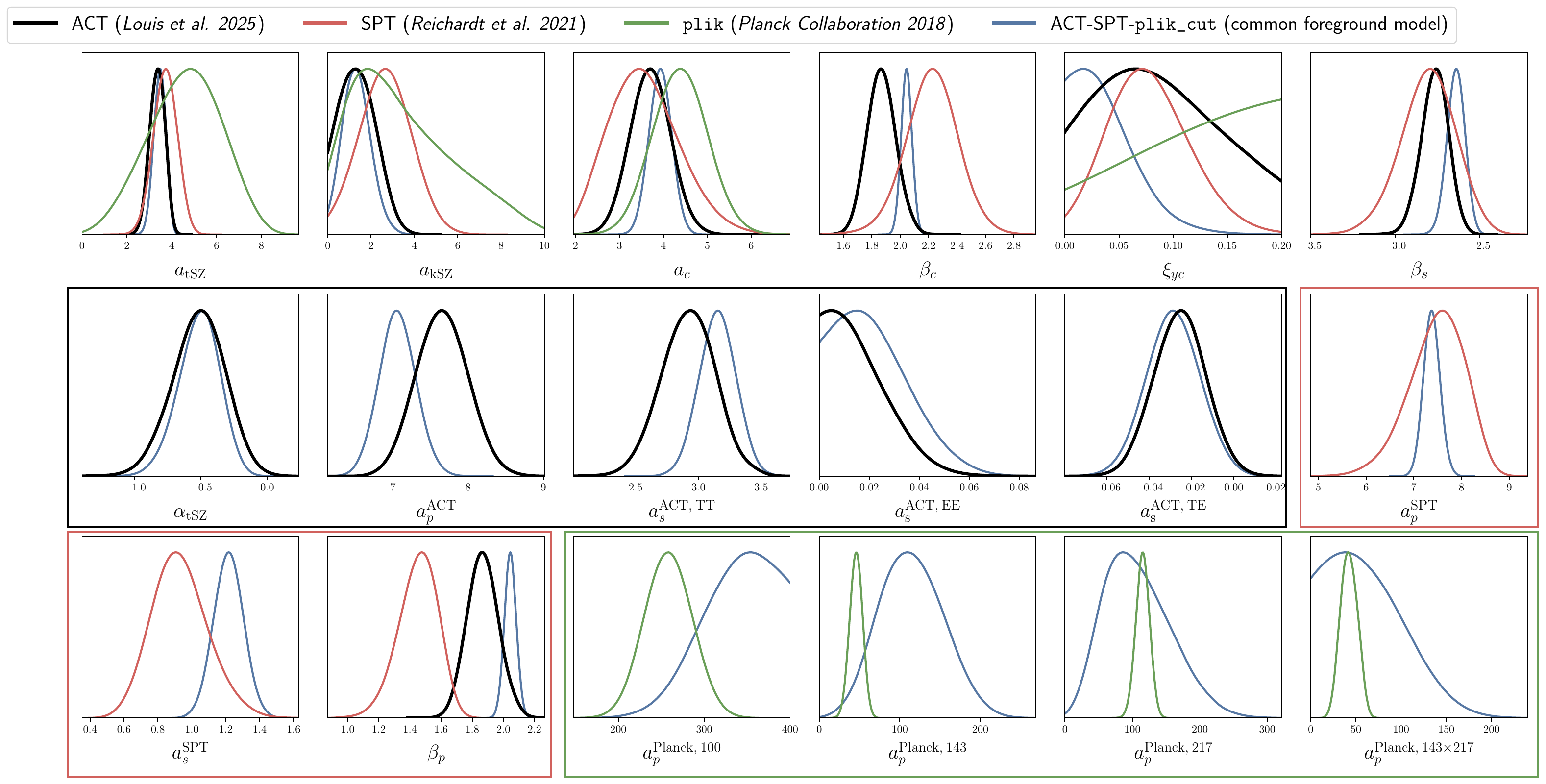}
\caption{Posterior distributions of the foreground parameters in the ACT DR6 baseline model, for various datasets. The top row shows the parameters used for a model common to ACT, SPT, and \textit{Planck}. The black, red, and \textcolor{red}{green} boxes in the second and third rows represent respectively the ACT-, SPT-, and \textit{Planck}-specific parameters. Note that the \textit{Planck} tSZ and clustered CIB amplitudes are rescaled to $150\:{\rm GHz}$. The SPT clustered CIB is the sum of the $1$- and $2$-halo contributions introduced in R21. Going from the \textcolor{red}{green} to the dark blue case removes the small-scale information in {\it Planck} data and hence we find a broadening of the \textit{Planck}-related point source parameters in the last row.\\}
\label{fig:fg_params_combined}
\end{figure*}

In Figure~\ref{fig:fg_params_combined} we present the posterior distributions of the foreground parameters for the different datasets\footnote{The SPT posteriors presented here vary slightly from those quoted in R21, which are for a fixed cosmology; here we allow cosmological parameters to vary with the same priors as in ACT DR6.} and the combination of ACT, SPT, and \emph{Planck} using a common foreground model built from the DR6 baseline. 

We find that the common foreground parameters (top row) are consistent to within $1\sigma$ for both the ACT DR6 and R21 baseline models, except for the CIB MBB parameters $\beta_p$ and $\beta_c$. This is entirely due to the ACT DR6 baseline model setting $\beta_p = \beta_c$ and using a CIB temperature $T_d = 9.6\:{\rm K}$, while the R21 model varies both $\beta_p$ and $\beta_c$ and sets $T_d=25\:{\rm K}$. Indeed, the model extension ``$\beta_c \neq \beta_p$, $T_d=25$'' (defined in the next section) brings all parameters in agreement to within $1\sigma$. As expected, the best-fitting values for the amplitude of the Poisson radio sources are different due to the different flux cuts applied to ACT and SPT data. However, the amplitude of the Poisson CIB sources is similar for both datasets: $7.65\pm 0.35~{\rm \mu K^2}$ (ACT) and $7.26^{+0.73}_{-0.57}~{\rm \mu K^2}$ (SPT).

We also note that the posteriors of the common foreground parameters (top row in Figure~\ref{fig:fg_params_combined}) are tightened when the three datasets are combined. In particular, the amplitude of the clustered CIB is constrained to about $6\%$ ($a_c = 3.93\pm 0.26~{\rm \mu K^2}$). The amplitudes of the Poisson terms (CIB and radio), even though modeled independently for the three datasets, are affected by the use of a common foreground model. This is due to both the common modeling of their frequency scaling in ACT and SPT (common $\beta_s$ and $\beta_p$) and the strong degeneracies between these amplitudes and $a_c$. The \textit{Planck} Poisson amplitudes are less constrained in this combination; this is entirely due to the limited $\ell$-range in the \verb|plik| spectra used when combining datasets. This range excludes the smaller angular scales where Poisson-like contributions, which scale approximately as $\ell^2$, are most prominent.

We find that the best-fitting model from the combined fit has a goodness-of-fit to each individual data vector as follows: $\chi^2_{\rm ACT} = 1608$ (1651 bins), $\chi^2_{\rm SPT} = 115$ (88 bins), and $\chi^2_{\rm Planck} = 1484$ (1470 bins). As a comparison, when fitting the baseline model to ACT DR6 only the goodness-of-fit is $\chi^2_{\rm ACT} = 1591$. Similarly, we fit the DR6 baseline model to SPT data only (while varying the cosmology) and find $\chi^2_{\rm SPT} = 93$. The fiducial SPT foreground model presented in R21 gives a goodness-of-fit $\chi^2_{\rm SPT} = 90$ (while varying the cosmology). Both models give similar probability to exceed (PTE): $9\%$ for the SPT fiducial model ($88$ bins $- 10$ foreground parameters $- 5$ cosmological parameters) and $7\%$ for the ACT DR6 baseline model fit to SPT data ($88 - 9 - 5$ degrees of freedom).

These results show that the ACT DR6 baseline foreground model provides a good fit to ACT, SPT, and \textit{Planck} data, enabling consistent joint analyses and confirming the flexibility of the model across experiments.


\section{Impact of Changing Foreground Models on ACT DR6 Cosmology}
\label{sec:fg_tests}
\begin{figure*}[!t]
    \centering
    \includegraphics[width=1.0\textwidth]{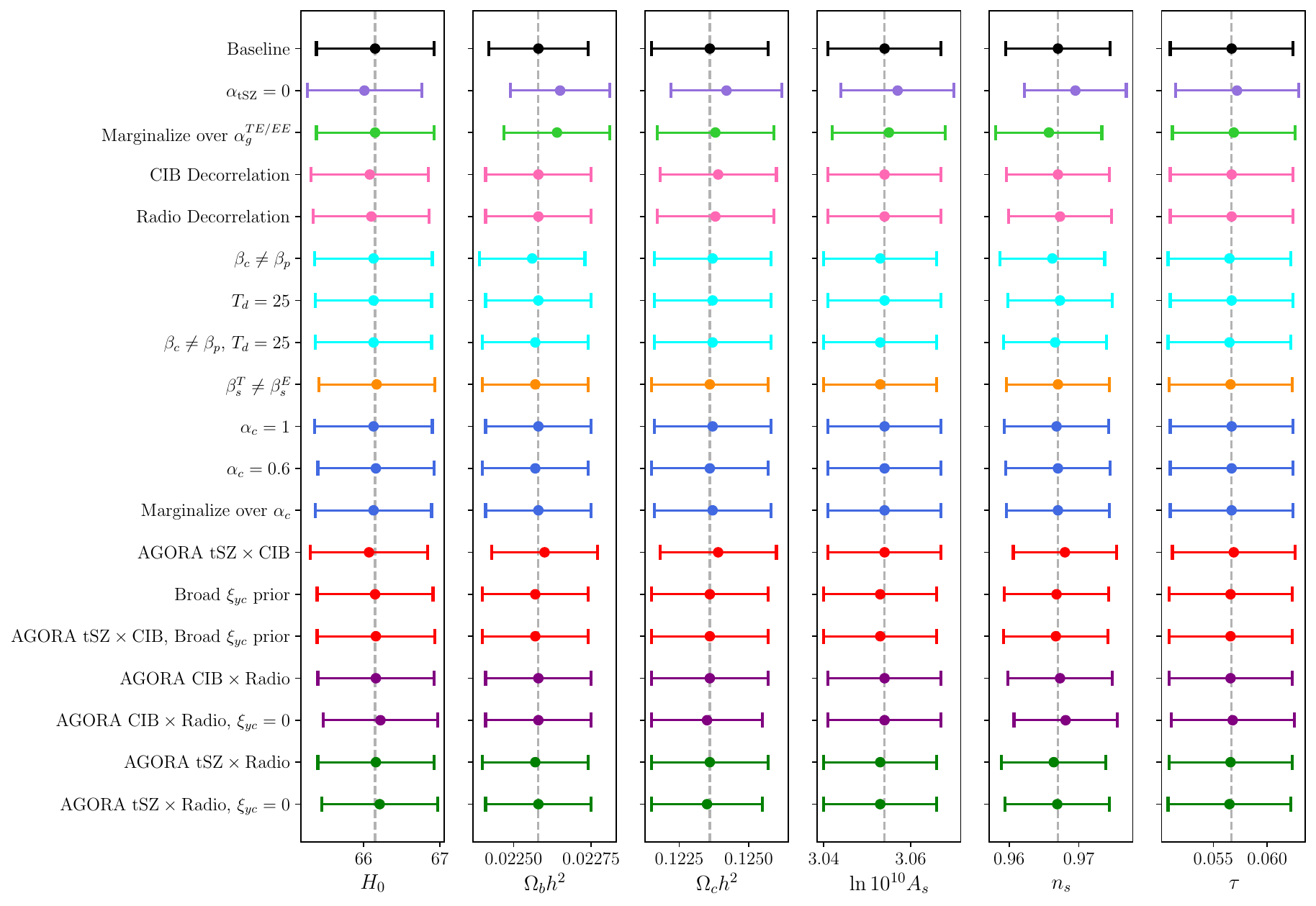}
    \caption{$\Lambda$CDM cosmological parameter constraints (with 68\% confidence intervals) from various foreground model modification tests. The black lines show the baseline results. The light purple, lime green, pink, cyan, orange, blue, red, purple, and green lines show the results involving modifications to the tSZ template shape, CIB and radio decorrelation, CIB SED, radio point source SED, CIB clustered template, tSZ $\times$ CIB template, CIB $\times$ radio inclusion, and tSZ $\times$ radio inclusion, respectively.\\} 
    \label{fig:cosmo_from_fg_tests}
\end{figure*}

\begin{figure*}[!ht]
    \centering
    \includegraphics[width=1.0\textwidth]{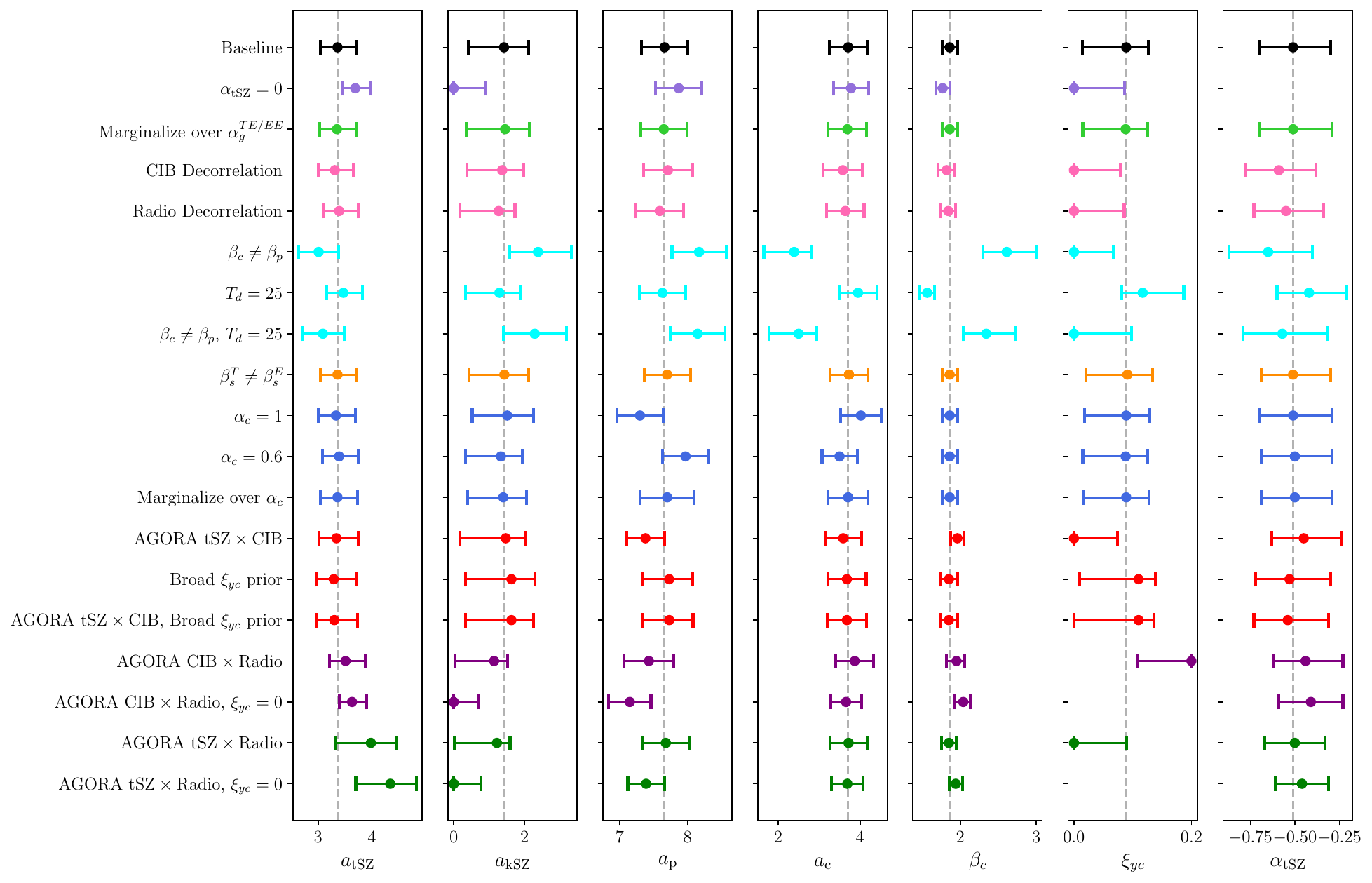}
    \caption{$\Lambda$CDM foreground parameter constraints (with 68\% confidence intervals) from various foreground model modification tests.  We show results for the tSZ, kSZ, and CIB-related foreground parameters.  The black lines show the baseline results. The light purple, lime green, pink, cyan, orange, blue, red, purple, and green lines show the results involving modifications to the tSZ template shape, CIB and radio decorrelation, CIB SED, radio point source indices, CIB clustered template, tSZ $\times$ CIB template, CIB $\times$ radio inclusion, and tSZ $\times$ radio inclusion, respectively. See the text for full descriptions of each test.\\} 
    \label{fig:fg_from_fg_tests}
\end{figure*}

In L25, it was verified that the baseline model explored above is indeed the minimal model required to characterize the multi-frequency measurements from ACT and to assure robustness of the extracted cosmological results. To assert this, several variations of the baseline DR6 foreground model were considered. Here we describe these variations and their results in greater detail.

The approach taken to identify the minimal model needed for DR6 involves checking the following criteria: 
\begin{itemize}
\itemsep0em
    \item A model extension causes a negligible ($<0.5\sigma$) shift in cosmological parameters;
    \item No significant preference for an additional foreground parameter is found ($>3\sigma$ deviation of the parameter from its baseline value);
    \item The extended model does not give a better fit to the data compared to the baseline case. (In the case of a model extension that involves merely adding model parameters, this is already captured in the previous point.) 
\end{itemize}

The extensions considered include altering (1) the shape of tSZ template, (2) the shape of the polarized dust template, and (3) the level of CIB and radio decorrelation; varying (4) separate CIB spectral indices and (5) radio point source spectral indices; using (6) different templates for clustered CIB and (7) different templates for tSZ--CIB; and including (8) CIB--radio templates, (9) tSZ--radio templates, and CO templates (the latter are shown in Appendix \ref{app:co}). 

Unless otherwise noted, these tests are all run using only the ACT DR6 data, without combining with any external datasets.

\subsection{Impact on $\Lambda$CDM Cosmology}

\begin{figure*}[t]
    \centering
    \includegraphics[width=0.9\textwidth]{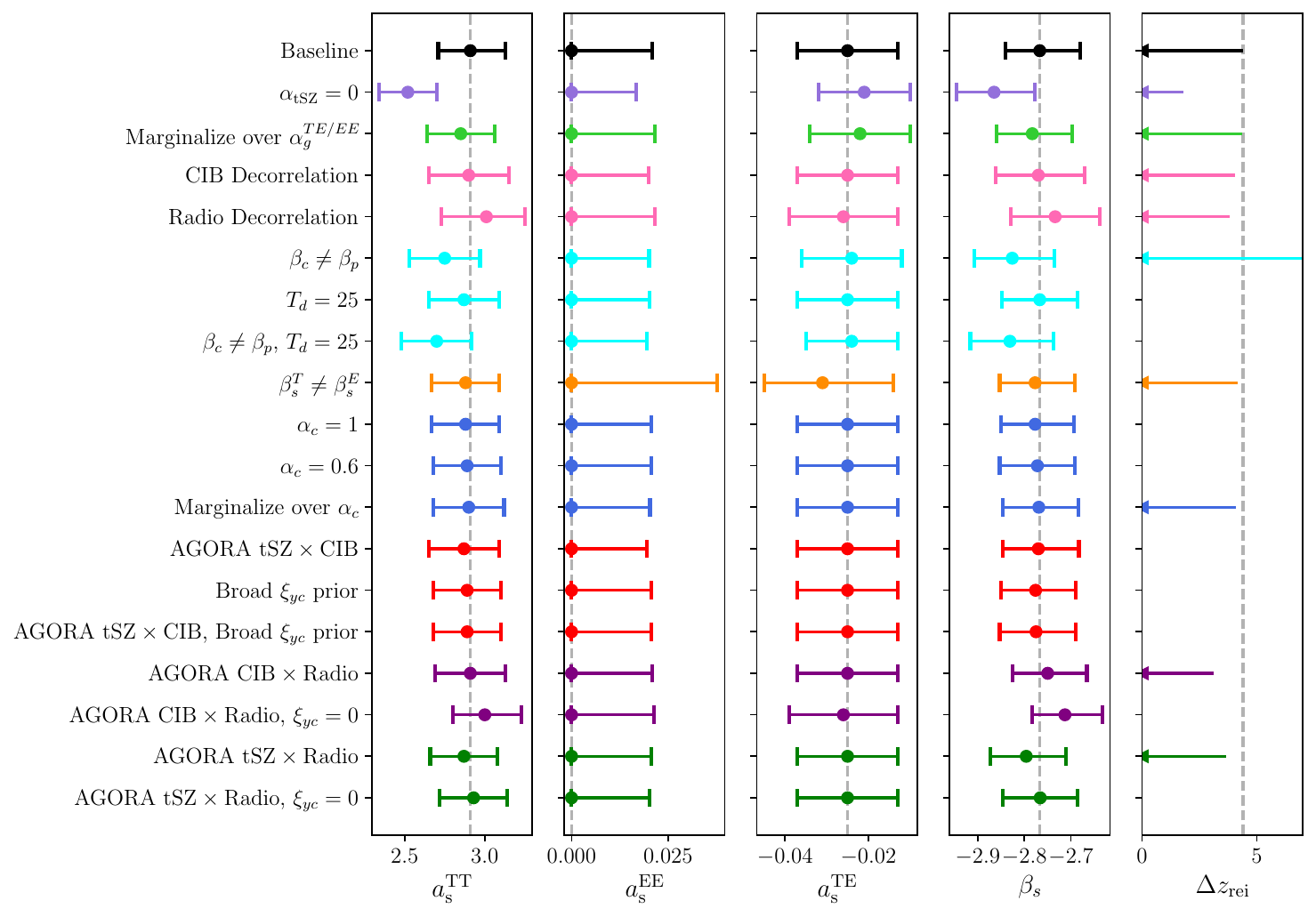}
    \caption{Radio point source parameter constraints and $\Delta z_{\rm rei}$ constraints from various foreground model modification tests. For radio point source parameters, all error bars are shown as 68\% confidence intervals or upper limits. The $\Delta z_{\rm rei}$ constraints are shown as 95\% upper limits, for a few foreground model variations of interest. The black lines show the baseline results. The light purple, lime green, pink, cyan, orange, blue, red, purple, and green lines show the results involving modifications to the tSZ template shape, CIB and radio decorrelation, CIB SED, radio point source indices, CIB clustered template, tSZ $\times$ CIB template, CIB $\times$ radio inclusion, and tSZ $\times$ radio inclusion, respectively. See the text for full descriptions of each test.\\} 
    \label{fig:extra_fg_from_fg_tests}
\end{figure*}

The baseline $\Lambda$CDM cosmology results presented in L25 are robust to all of these variations (see Figure~\ref{fig:cosmo_from_fg_tests}), and changes are absorbed by the foregrounds, shown in Figure~\ref{fig:fg_from_fg_tests} and Figure~\ref{fig:extra_fg_from_fg_tests}. All cosmological parameter shifts are less than $0.1\sigma$, with the exception of the $\alpha_{\rm tSZ}=0$ variation, discussed further in \S\ref{subsec:alpha_tsz}, and $\alpha_{g}^\mathrm{TE/EE}$ marginalization, discussed in \S\ref{subsec:alpha_dE}.

Table~\ref{table:extra_fg_par_detection_lcdm} reports the values of the additional foreground parameters for each model within a $\Lambda$CDM cosmology. None of the new parameters introduced are found to be $>3\sigma$ away from their baseline value.

When allowed to vary separately, $\beta_c$ and $\beta_p$ differ by $2.5\sigma$, with $\beta_c$ being shifted to larger values with broader uncertainties ($\beta_c = 2.6_{-0.3}^{+0.4}$) while $\beta_p$ shifts by less than $1\sigma$. Since the total CIB emission is dominated by the Poisson contribution at high-$\ell$ where ACT data are the most constraining, we expect $\beta_p$ to be better constrained. As shown in Figure~\ref{fig:fg_from_fg_tests}, the three baseline SZ parameters ($a_{\rm tSZ}$, $\alpha_{\rm tSZ}$, and $a_{\rm kSZ}$) as well as the clustered CIB amplitude $a_c$ are affected by allowing $\beta_c\neq\beta_p$. In particular, $a_{\rm kSZ}$ tends to prefer higher values in this case, as observed in R21. When additionally setting the dust temperature to $T_d=25\:{\rm K}$, $\beta_c$ and $\beta_p$ remain consistent with each other at $2.5\sigma$ and we recover the SPT spectral indices to within $2\sigma$. The lack of strong observational support for $\beta_p\neq\beta_c$ justifies the choice of keeping these parameters equal.

As expected from the lack of detection of polarized point sources, the parameter $\beta_s^E$ remains unconstrained when varied, while $\beta_s^T$ is consistent with the baseline $\beta_s$ value. Similarly, $\alpha_c$ is poorly constrained when allowed to vary, likely because the total CIB emission at high multipoles is dominated by the Poisson component. Even in models where $\beta_c$ and $\beta_p$ are allowed to differ, there is no clear preference for a particular value of $\alpha_c$. 

Cross-correlation parameters involving CIB and radio, as well as tSZ and radio, are either consistent with zero (within $1.3\sigma$) or only marginally constrained, and are typically detected as upper bounds. This holds even when setting the tSZ--CIB correlation to zero. We also note that the upper bounds are consistent with predictions from the \verb|Websky| or \verb|AGORA| simulations. Finally, there is no evidence for decorrelation in the CIB or radio spectra. 

\begin{table}[!htb]
\centering
\begin{tabular}{|c|c|}
\hline
\multirow{2}{8em}{Foreground Model}
   & New Parameter    \\  
  & (in $\Lambda$CDM Cosmology) \\
 \hline \hline
Baseline  & $\alpha_{\rm tSZ}=-0.53_{-0.19}^{+0.22}$  \\ 
(compare with $\alpha_{\rm tSZ}=0$) &  \\\hline

Marginalize over $\alpha_g^\mathrm{TE/EE}$ & $\alpha_g^\mathrm{TE/EE} = -0.98^{+0.64}_{-0.39}$\\
\hline

CIB  &$r^\mathrm{CIB}_\mathrm{90\times 220} > 0.896$ \\ 
 decorrelation&$r^\mathrm{CIB}_\mathrm{90\times 150} > 0.874$ \\ 
 &$r^\mathrm{CIB}_\mathrm{150\times 220} > 0.978$ \\ \hline
Radio  &$r^\mathrm{radio}_\mathrm{90\times 220} = --$ \\ 
 decorrelation&$r^\mathrm{radio}_\mathrm{90\times 150} = 0.970^{+0.027}_{-0.011}$ \\ 
&$r^\mathrm{radio}_\mathrm{150\times 220} = --$\\ \hline

\multirow{2}{5em}{$\beta_c \neq \beta_p$} &$\beta_c = 2.6_{-0.3}^{+0.4}$ \\
&$\beta_p=1.8 \pm 0.1$ \\ \hline
\multirow{2}{8em}{$\beta_c \neq \beta_p, T_d=25$} &$\beta_c=2.3_{-0.3}^{+0.4}$ \\ 
 &$\beta_p=1.5 \pm 0.1$ \\ \hline
\multirow{2}{5em}{$\beta_s^T \neq \beta_s^E$} &$\beta_s^T=-2.78^{+0.09}_{-0.08}$ \\ 
 &$\beta_s^E > -3.14$ \\ \hline
Marginalize over $\alpha_c$ &$\alpha_c = --$ \\ \hline

\verb|AGORA| CIB $\times$ radio&$\xi_{cs}=0.06_{-0.06}^{+0.02}$ \\ \hline
\verb|AGORA| CIB $\times$ radio, $\xi_{yc}=0$& $\xi_{cs}<0.05$ \\ \hline
\verb|AGORA| tSZ $\times$ radio& $\xi_{ys} < 0.13$ \\ \hline
\verb|AGORA| tSZ $\times$ radio, $\xi_{yc}=0$&$\xi_{ys}=0.08_{-0.06}^{+0.04}$\\ \hline

\end{tabular}
  \caption{1D marginalized posterior distributions (with 68\% confidence intervals or 95\% upper/lower bounds for undetected quantities) on additional foreground parameters for several foreground modifications of interest, in the $\Lambda$CDM scenario. These results are based on ACT DR6 alone.}
\label{table:extra_fg_par_detection_lcdm}
\end{table}

For each of the test cases (1)--(8) listed above, the maximum a posteriori (MAP) is found using the \texttt{cobaya} ``bobyqa''~\citep{Cartis_18a, Cartis_18b} minimizer, and the $\chi^2$ of the MAP from the modified model is compared to that of the baseline model. Table~\ref{table:chi2_fg_tests} shows $\Delta \chi^2 = \chi^2_{\rm{variation}}-\chi^2_{\rm baseline}$ for each test. For tests involving additional free parameters, we also compute the preference of that model relative to the baseline using the likelihood-ratio test statistic via Wilks's theorem. 

A few of the models improve the $\chi^2$ of the fit. The $\beta_c \neq \beta_p$ model, with one additional free parameter relative to the baseline model, improves the goodness-of-fit by $\Delta \chi^2 = 6.9$, corresponding to a $2.6\sigma$ preference. Nevertheless, this is not statistically significant enough --- as defined in the criteria listed above --- to trigger a model expansion and does not impact the cosmological parameters. Future observatories such as the Simons Observatory (SO)~\citep{2019JCAP...02..056A, SimonsObservatory:2025wwn} may be able to distinguish between $\beta_c$ and $\beta_p$.  Other models with additional free parameters, such as $\beta_s^T \neq \beta_s^E$ and letting $\alpha_c$ vary, do not improve the fit at all.

\begin{table}[htb]
\centering
\begin{tabular}{|c|c|c|c}
\hline
  Model ($\Lambda$CDM Cosmology) & $\Delta \chi^2$ & Pref. ($\sigma$) \\  
 \hline \hline
$\alpha_{\rm tSZ}=0$ &4.9  &$-2.2$   \\ \hline
Marginalize over $\alpha_g^\mathrm{TE/EE}$ &$-11.7$ &3.4 \\
\hline 
CIB decorrelation &$-1.0$ &0.3\\ \hline
Radio decorrelation &$-1.4$ &0.4\\ \hline
$\beta_c \neq \beta_p$ & $-6.9$ & 2.6  \\ \hline
$T_d=25$ & $-0.3$ & -- \\ \hline
$\beta_c \neq \beta_p, T_d=25$& $-6.9$  &2.6 \\ \hline 
$\beta_s^T \neq \beta_s^E$& 0.0 &0.0  \\ \hline 
Marginalize over $\alpha_c$&0.0 &0.0 \\ \hline 
\texttt{AGORA} tSZ $\times$ CIB& 0.3 & -- \\ \hline
Broad $\xi_{yc}$ prior&0.0 &-- \\ \hline
\texttt{AGORA} CIB $\times$ radio&$-0.9$ &0.9 \\ \hline
\texttt{AGORA} CIB $\times$ radio, $\xi_{yc}=0$&0.4 &-- \\ \hline 
\texttt{AGORA} tSZ $\times$ radio&$-1.5$ &1.2 \\ \hline
\texttt{AGORA} tSZ $\times$ radio, $\xi_{yc}=0$&$-1.4$ &-- \\ \hline
\end{tabular}
  \caption{$\Delta \chi^2 = \chi^2_{\rm{variation}}-\chi^2_{\rm baseline}$ for various foreground modification tests in the $\Lambda$CDM cosmological model and for ACT DR6 data only. The preference for the modified model (in $\sigma$), computed using the likelihood-ratio test statistic, is also shown for modifications involving additional free parameters. A negative value of the $\sigma$ preference indicates that the baseline model is preferred over the modification.}
\label{table:chi2_fg_tests}
\end{table}

\subsection{Impact on $\Lambda{\rm CDM}+N_{\rm eff}$ Cosmology}
\label{subsec:fg_tests_neff}

\begin{figure*}[t]
    \centering
    \includegraphics[width=1.0\textwidth]{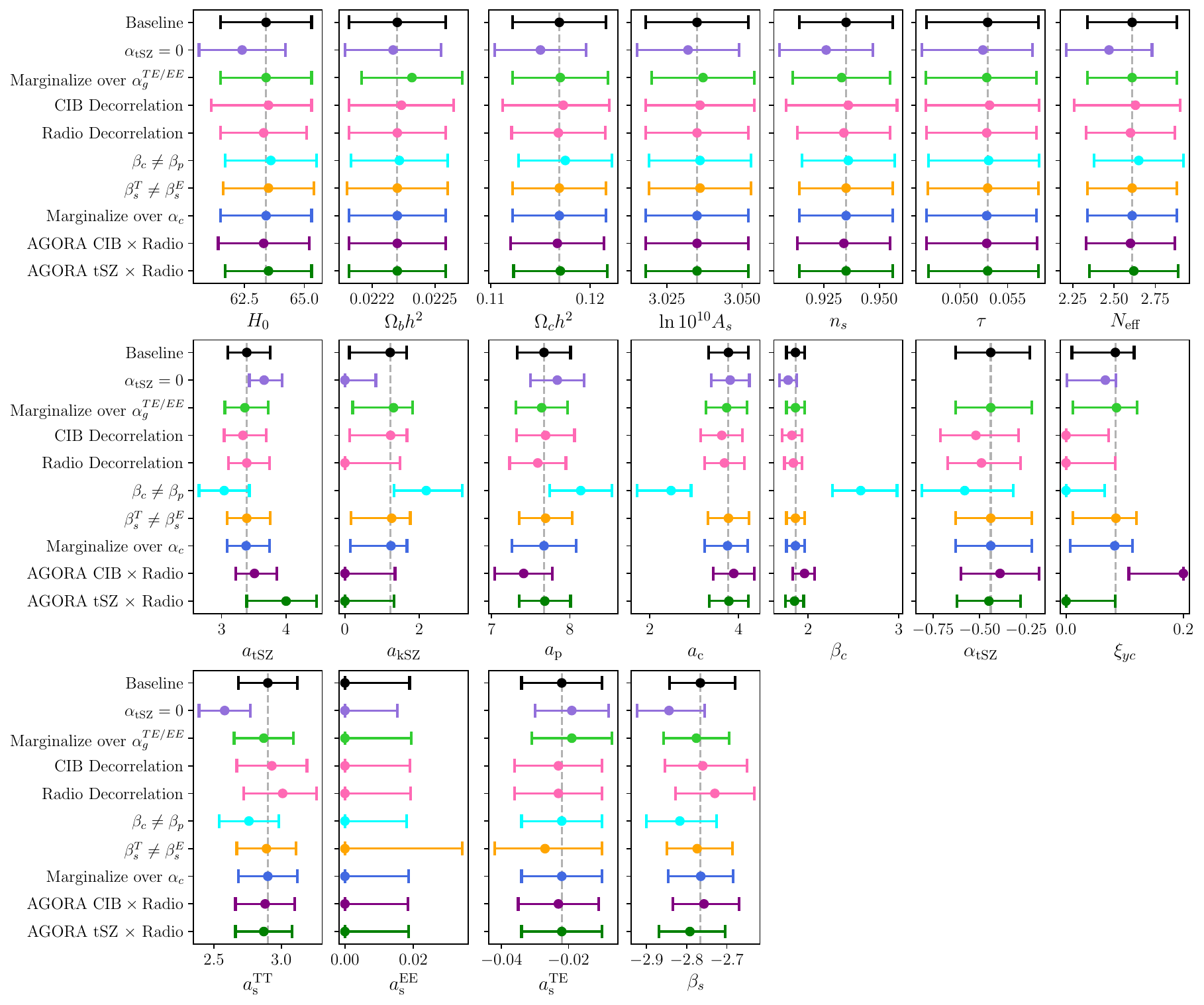}
    \caption{$\Lambda$CDM+$N_{\rm eff}$ cosmological and foreground parameter constraints from various foreground model modification tests. The black lines show the baseline results. The light purple, pink, cyan, orange, blue, purple, and green lines show the results involving modifications to the tSZ template shape, CIB and radio decorrelation, CIB SED, radio point source indices, CIB clustered template, CIB $\times$ radio inclusion, and tSZ $\times$ radio inclusion, respectively. See the text for full descriptions of each test. \\} 
    \label{fig:neff_fg_tests}
\end{figure*}

A key extension of the standard cosmological model is to allow the effective number of relativistic species, $N_\mathrm{eff}$, to vary. $N_\mathrm{eff}$ quantifies the total energy density in light, relativistic particles — such as neutrinos — in the early Universe. It reflects their contribution to the radiation density during their relativistic phase which includes Big Bang Nucleosynthesis and recombination, and is of interest because theories of particle physics beyond the standard model often predict the existence of additional light particles. We repeat a subset of the foreground extension tests for the $\Lambda$CDM+$N_{\rm eff}$ cosmological model, with  Figure~\ref{fig:neff_fg_tests} showing parameter constraints for various tests, Table \ref{table:extra_fg_par_detection_neff} reporting constraints for additional foreground parameters in tests of interest, and Table \ref{table:chi2_fg_tests_neff} showing $\Delta \chi^2 = \chi^2_{\rm{variation}}-\chi^2_{\rm baseline}$ for each test. 

As in the $\Lambda$CDM case, modifications to the foreground model do not lead to significant shifts in the cosmological parameters (with the exception of the $\alpha_{\rm tSZ}$ case, detailed below). No additional foreground parameters are found to deviate from their baseline values at a statistically significant level, and no alternative model provides a better fit than the baseline. These results reinforce the robustness of the baseline foreground model and its compatibility with the $\Lambda$CDM+$N_{\rm eff}$ cosmological framework.

\begin{table}[!h]
\centering
\begin{tabular}{|c|c|}
\hline
  Foreground Model &  New Parameter \\ 
  & ($\Lambda$CDM+$N_{\rm eff}$ Cosmology) \\
 \hline \hline
Baseline  &$\alpha_{\rm tSZ}=-0.44^{+0.21}_{-0.19}$   \\
(compare with $\alpha_{\rm tSZ}=0$) &  \\\hline

Marginalize over $\alpha_g^\mathrm{TE/EE}$ & $\alpha_g^\mathrm{TE/EE} = -0.94^{+0.65}_{-0.37} $\\
\hline 

CIB  &$r^\mathrm{CIB}_\mathrm{90\times 220} > 0.898$ \\ 
 decorrelation&$r^\mathrm{CIB}_\mathrm{90\times 150} > 0.874 $ \\ 
 &$r^\mathrm{CIB}_\mathrm{150\times 220} > 0.978 $ \\ \hline

Radio  &$r^\mathrm{radio}_\mathrm{90\times 220} = -- $ \\ 
decorrelation&$r^\mathrm{radio}_\mathrm{90\times 150} = 0.970^{+0.026}_{-0.011}$ \\ 
&$r^\mathrm{radio}_\mathrm{150\times 220} = -- $\\ \hline

$\beta_c \neq \beta_p$   &$\beta_c=2.6^{+0.4}_{-0.3}$\\
 &$\beta_p=1.8\pm 0.1$\\ \hline

$\beta_s^T \neq \beta_s^E$ & $\beta_s^T = -2.77^{+0.09}_{-0.08}$ \\ 
& $\beta_s^E > -3.15$\\ \hline
Marginalize over $\alpha_c$  &$\alpha_c = --$\\ \hline

\verb|AGORA| CIB $\times$ radio & $\xi_{cs} = 0.06^{+0.02}_{-0.06}$\\ \hline

\verb|AGORA| tSZ $\times$ radio & $\xi_{ys} < 0.13$\\ \hline
\end{tabular}
  \caption{1D marginalized posterior distributions (with 68\% confidence intervals or 95\% upper/lower bounds) on additional foreground parameters for several foreground modifications of interest, in the $\Lambda$CDM +$N_{\rm eff}$ scenario. These results are based on ACT DR6 alone.}
\label{table:extra_fg_par_detection_neff}
\end{table}

\subsection{Discussion of $\alpha_{\rm tSZ}$}
\label{subsec:alpha_tsz}

As discussed in L25 and above, the shape of the thermal SZ power spectrum component has not been explicitly varied before ACT DR6. Different templates, with somewhat different shapes affected by the exact modeling of the small-scale tSZ spectrum, have, however, been adopted in different analyses. These can map into different values of the $\alpha_{\rm tSZ}$ parameter. 

For instance, ACT DR4 used $\alpha_{\rm tSZ} =0$, but the R21 tSZ template (from~\citealp{Shaw2010}) is well approximated by $\alpha_{\rm tSZ} =-0.2$, while \verb|AGORA| predicts a range of values for $\alpha_{\rm tSZ}$ (from $-0.3$ to $-0.5$) depending on the AGN feedback model.\footnote{Figure 23 of L25 explicitly shows how the tSZ power spectra for different \texttt{BAHAMAS} AGN gas heating temperature models in \texttt{AGORA} are described by different values of $\alpha_{\rm tSZ}$.}  Analyses of the sensitive ACT DR6 small-scale temperature data show that marginalizing over the exact shape of the tSZ spectrum is now necessary to extract robust cosmological constraints.

As can be seen from Table~\ref{table:alpha_tsz}, the ACT $\Lambda$CDM parameters shift by up to $0.3\sigma$ when allowing $\alpha_{\rm tSZ}$ to vary, with the most affected parameter being $\Omega_bh^2$. When including $N_{\rm eff}$, and because of its strong correlations with densities and $H_0$, the shift is largest for $H_0$, $\Omega_ch^2$, and $N_{\rm eff}$, all moving by $0.3\sigma$. Even though these are smaller than the $0.5\sigma$ limit set above, the number of parameters affected led us to further investigate this issue and consider the impact of $\alpha_{\rm tSZ}$ on the {\it Planck} + ACT DR6 combination, denoted P-ACT in L25 and used extensively as the most constraining CMB dataset in L25 and C25. We should note that this combination differs from the one introduced in \S\ref{subsec:other_data} as it combines a foreground-marginalized “$Planck_{cut}$” dataset using \textit{Planck} high-$\ell$ data at $\ell<1000$ in TT and $\ell<600$ in TE/EE from the PR3 likelihood~\citep{Planck:2018_pslkl}, as well as the low-$\ell$ \textit{Planck} temperature likelihood and substituting in the \verb|Sroll2| likelihood for low-$\ell$ polarization. 
Table~\ref{table:alpha_tsz} summarizes the P-ACT shifts in cosmological parameters when fixing $\alpha_{\rm tSZ}=0$. The tighter constraints on cosmological parameters offered by P-ACT causes all $\Lambda$CDM parameters to shift by more than $0.2\sigma$, with two parameters crossing the $0.5\sigma$ threshold: $\Omega_bh^2\:(0.6\sigma)$ and $n_s\:(0.8\sigma)$. For a $\Lambda$CDM$ + N_{\rm eff}$ cosmology, both $\Omega_ch^2$ and $N_{\rm eff}$ shift down by $0.5\sigma$. 

Additionally, we find evidence for non-zero $\alpha_{\rm tSZ}$ in certain configurations, reaching $3.4\sigma$ for P-ACT $\Lambda$CDM:  
\begin{align}
    \text{ACT, $\Lambda$CDM}: \alpha_{\rm tSZ}&=-0.53^{+0.22}_{-0.19}  &(2.5\sigma) \nonumber \\
    \text{ACT, $\Lambda$CDM + $N_{\rm eff}$}: \alpha_{\rm tSZ}&=-0.44^{+0.21}_{-0.19}  &(2.2\sigma) \nonumber \\
    \text{P-ACT, $\Lambda$CDM}: \alpha_{\rm tSZ}&=-0.64\pm 0.19  &(3.4\sigma) \nonumber \\
    \text{P-ACT, $\Lambda$CDM + $N_{\rm eff}$}: \alpha_{\rm tSZ}&=-0.53^{+0.21}_{-0.19}  &(2.7\sigma) \nonumber \\
\end{align}

\begin{figure}[!h]
    \centering
    \includegraphics[width=0.95\linewidth]{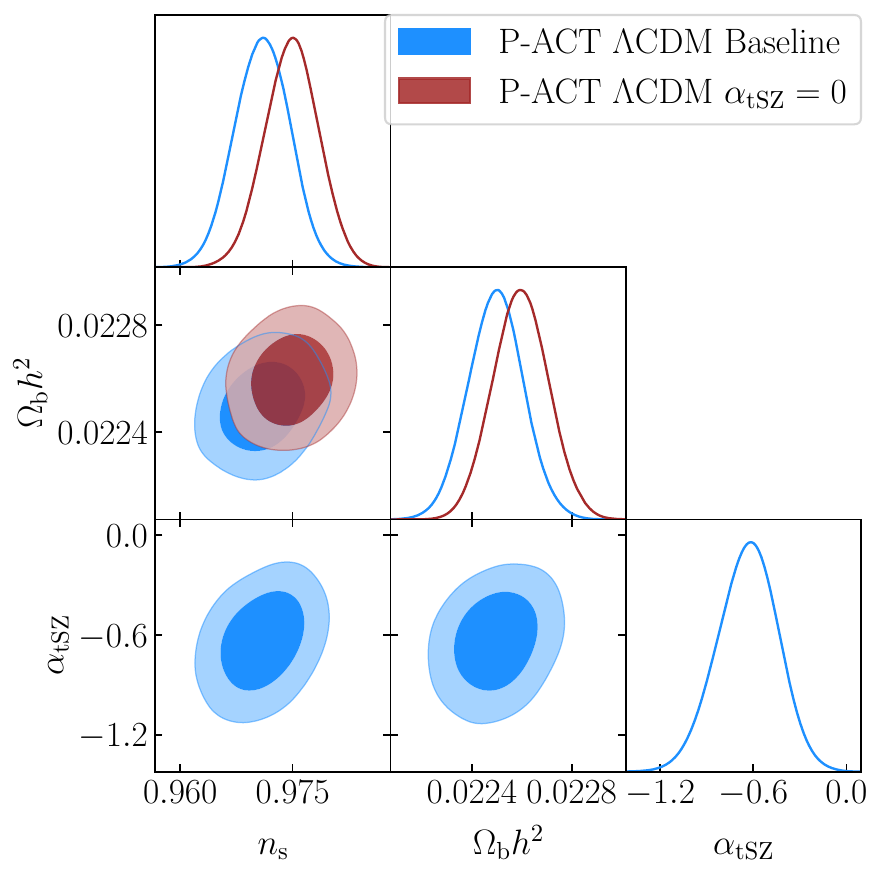}
    \caption{2D and 1D marginalized posterior distributions in the $\Lambda$CDM model using the P-ACT dataset combination for the baseline foreground model (blue), as compared to the foreground model variation with $\alpha_{\rm tSZ}=0$ (red). Results are shown for the $\alpha_{\rm tSZ}$ parameter, as well as $n_s$ and $\Omega_b h^2$, the two cosmological parameters having the highest correlation with $\alpha_{\rm tSZ}$.}
    \label{fig:pact_alphasz_2d}
\end{figure}

Figure~\ref{fig:pact_alphasz_2d} shows the 2D and 1D marginalized posterior distributions in the P-ACT $\Lambda$CDM case for both the baseline and $\alpha_{\rm tSZ}=0$ foreground models, in particular highlighting the positive correlations among $\alpha_{\rm tSZ}$, $\Omega_b h^2$, and $n_s$. Since $\alpha_{\rm tSZ}$ is negative in the baseline model, the $\alpha_{\rm tSZ}=0$ variant results in an increase in these cosmological parameters. Appendix~\ref{app:alphasz} provides the full set of constraints on cosmological parameters in the $\alpha_{\rm tSZ}=0$ case for both the $\Lambda$CDM and $\Lambda$CDM+$N_{\rm eff}$ cosmological models, with both the ACT and P-ACT datasets. It also compares spectra of individual components evaluated at the MAP from the $\alpha_{\rm tSZ}=0$ case to the baseline case in Figure~\ref{fig:alphasz0_vs_baseline_output}. Specifically, setting $\alpha_{\rm tSZ}=0$ forces the inferred tSZ spectrum to increase at high $\ell$. Because the tSZ effect and radio emission are the two brightest foregrounds at 90 GHz (on small scales relevant to ACT), if one is changed then the other should probably compensate; but due to the rigid Poisson shape of the radio component, it cannot correctly absorb the full effect on the total power spectrum shape, so $n_s$ increases to compensate for this over-compensation of the radio emission (and other damping-tail parameters also change, e.g., $\Omega_b h^2$).

\begin{table}[!t]
\centering
\begin{tabular}{|c|c|c|c}
\hline
  Model ($\Lambda$CDM+$N_{\rm eff}$ Cosmology) & $\Delta \chi^2$ & Pref. ($\sigma$) \\  
 \hline \hline
$\alpha_{\rm tSZ}=0$ & 2.5 &$-1.6$   \\ \hline
Marginalize over $\alpha_g^\mathrm{TE/EE}$ &$-11.9$ &3.4 \\
\hline 
CIB decorrelation &$-1.7$ &0.5\\ \hline
Radio decorrelation &$-2.0$ &0.6\\ \hline
$\beta_c \neq \beta_p$ &$-5.9$ &2.4\\ \hline
$\beta_s^{T} \neq \beta_s^{E}$ &0.0 &0.0\\ \hline
Marginalize over $\alpha_c$ &0.0 &0.0\\ \hline
\verb|AGORA| CIB $\times$ radio &$-0.7$ &0.8\\ \hline
\verb|AGORA| tSZ $\times$ radio &$-1.3$ &1.1\\ \hline
\end{tabular}
  \caption{$\Delta \chi^2 = \chi^2_{\rm{variation}}-\chi^2_{\rm baseline}$ for various foreground modification tests in the $\Lambda$CDM+$N_{\rm eff}$ cosmological model. The preference for the modified model (in $\sigma$), computed using the likelihood-ratio test statistic, is also shown for modifications involving additional free parameters. A negative value of the $\sigma$ preference indicates that the baseline model is preferred over the modification.}
\label{table:chi2_fg_tests_neff}
\end{table}

It is worth noting that the parametrization adopted for the tSZ power spectrum in Equation~\eqref{eq:tSZ} is empirical, chosen for its ability to capture a wide range of results from both simulations and previous analyses, while retaining a simple form. Alternative parametrizations of the tSZ power spectrum could potentially lead to additional shifts in the recovered parameters, although such shifts are expected to be comparable in magnitude to those observed here.

Finally, as shown in Table \ref{table:chi2_fg_tests}, the DR6 baseline model is a moderately better fit to the DR6 data than the $\alpha_{\rm tSZ}=0$ model, with a $\chi^2$ improvement of 4.9 in the $\Lambda$CDM scenario and 2.7 in the $\Lambda$CDM+$N_{\rm eff}$ scenario, both for one additional degree of freedom. These correspond to a preference of $2.2\sigma$ and $1.6\sigma$ for the baseline model, respectively.

\begin{table}[t]
\centering
\begin{tabular}{|c|c|c|c|c|c}
\hline
  Parameter & ACT  & ACT  & P-ACT & P-ACT \\  
    &$\Lambda$CDM & $N_{\rm eff}$ &  $\Lambda$CDM &  $N_{\rm eff}$\\  
 \hline \hline
$H_0$ &$-0.1$ &$-0.3$ &0.3 &$-0.4$ \\ \hline
$\Omega_b h^2$ &0.3 &0.0 &0.6 &$-0.2$ \\ \hline
$\Omega_c h^2$ &0.2 &$-0.3$ &$-0.2$ &$-0.5$ \\ \hline
$\ln{10^{10}A_s}$ &0.2 &0.0 &0.3 &$-0.1$ \\ \hline
$n_s$ &0.2 &$-0.2$ &0.8 &$-0.2$ \\ \hline
$\tau$ &0.1 &0.2 &0.3 &0.1 \\ \hline
$N_{\rm eff}$ &-- &$-0.3$ &-- &$-0.5$ \\ \hline
\end{tabular}
  \caption{Shifts (in $\sigma$) in cosmological parameters in the $\alpha_{\rm tSZ}=0$ foreground model relative to the baseline model, for both $\Lambda$CDM and $\Lambda$CDM+$N_{\rm eff}$, using both the ACT DR6 (``ACT'') and ACT DR6 + \emph{Planck} (``P-ACT'') dataset combinations. }
\label{table:alpha_tsz}
\end{table}

\subsection{Discussion of $\alpha_g^\mathrm{TE/EE}$}
\label{subsec:alpha_dE}

The second largest shift in cosmological parameters is caused by letting $\alpha_g^\mathrm{TE/EE}$ (defined in Equation~\eqref{eq:dust_pw}) vary (while updating the priors on the amplitude consistently), with $\Omega_bh^2$ increasing by $0.3\sigma$. With dust being most relevant on large scales, the slope of its power spectrum impacts the determination of the height of the acoustic peaks in the EE spectrum, which is controlled by $\Omega_bh^2$. With $\ell_{\rm min}$ being smaller in polarization than in temperature ($600$ versus $1000$), $\alpha_g^\mathrm{TE/EE}$ has a larger impact than $\alpha_g^\mathrm{TT}$. Indeed, there are no shifts on the cosmology resulting from setting $\alpha_g^\mathrm{TT}$ to a different value. 

\begin{figure}[!t]
    \centering
    \includegraphics[width=0.8\linewidth]{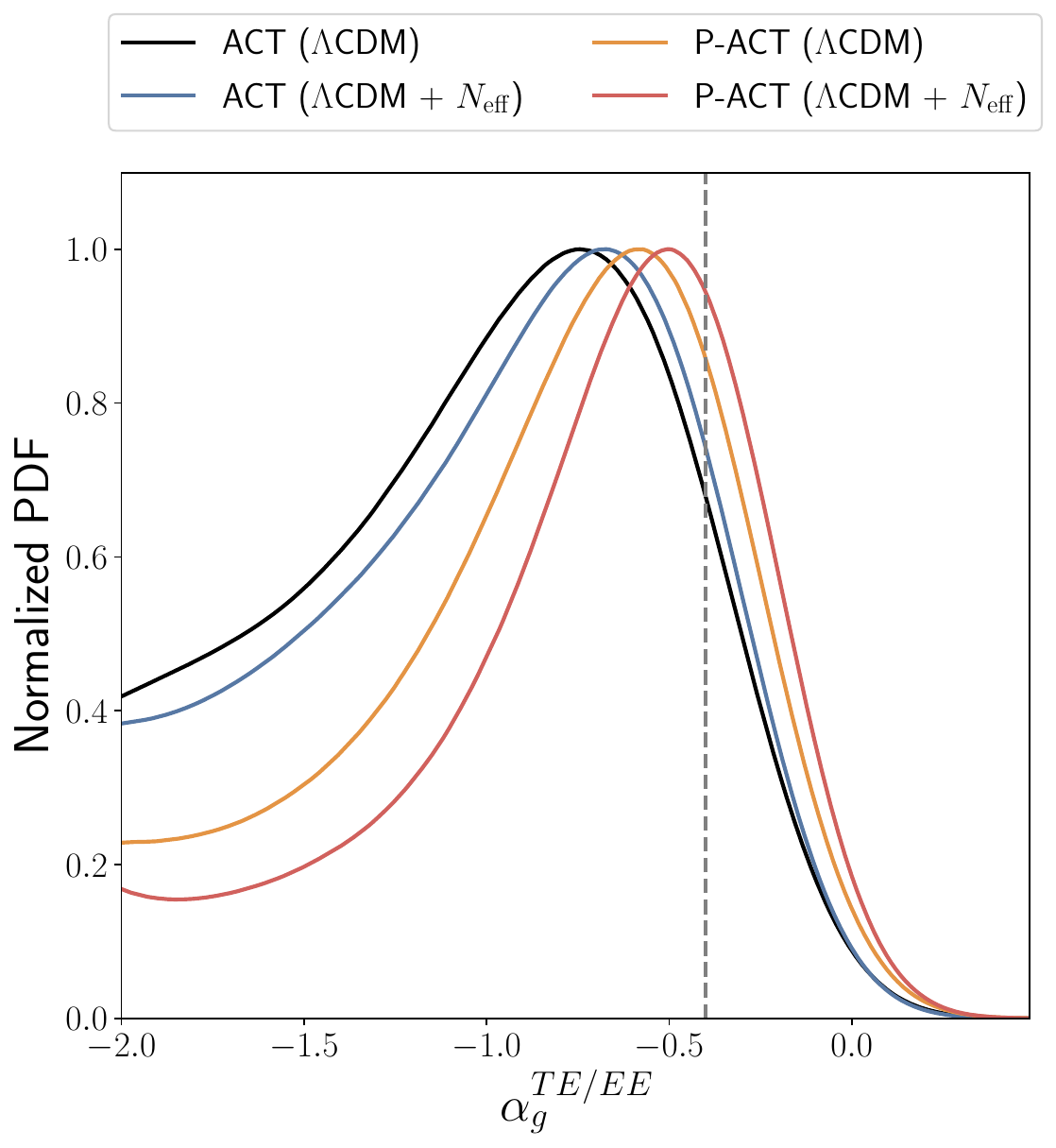}
    \caption{Posterior distributions of $\alpha_g^\mathrm{TE/EE}$ for ACT and P-ACT, shown for both $\Lambda$CDM and $\Lambda$CDM + $N_\mathrm{eff}$ cosmologies, when the parameter is allowed to vary. The dashed line indicates the baseline fixed value of $\alpha_g^\mathrm{TE/EE}=-0.4$. }
    \label{fig:alpha_gTEEE}
\end{figure}

To further assess the impact of varying $\alpha_g^\mathrm{TE/EE}$ on cosmological parameters, the same tests are repeated as for $\alpha_{\rm tSZ}$. For ACT only, in $\Lambda$CDM+$N_{\rm eff}$ cosmology, $\Omega_bh^2$ shifts by $0.2\sigma$. For P-ACT, the shifts are even smaller, at $0.1\sigma$ for both $\Lambda$CDM and $\Lambda$CDM+$N_{\rm eff}$. In addition, as shown in Figure~\ref{fig:alpha_gTEEE}, there is no statistically significant deviation of $\alpha_{g}^\mathrm{TE/EE}$ from its baseline value of $-0.4$ for any of these configurations, and it is therefore kept fixed at this value in L25 and C25.

Even though marginalizing over $\alpha_{g}^\mathrm{TE/EE}$ yields a $3.4\sigma$ improvement in the fit, this test reflects an extreme and conservative scenario for the foreground model. Dust emission is relatively faint in the 90 and 150 GHz ACT channels, and its influence is strongest on large angular scales that ACT alone does not probe. Lacking this large-scale sensitivity, ACT cannot effectively constrain $\alpha_{g}^\mathrm{TE/EE}$, as illustrated by the broad posterior in Figure~\ref{fig:alpha_gTEEE}. In contrast, \citet{planck_poldust:2018} derived a tight constraint of $\alpha_g^\mathrm{TE/EE} = -0.42 \pm 0.02$ using full-sky measurements at 353 GHz. When restricted to a smaller patch in the southern hemisphere, \citet{planck_poldust:2018} reports an even smaller value: $\alpha_g^\mathrm{EE} = -0.3\pm0.08$. The broad prior on the slope of the dust power spectrum adopted in this test is therefore very conservative relative to the \emph{Planck} constraints. This situation differs from the case of $\alpha_{\rm tSZ}$, for which no comparable prior information exists. When $\alpha_g^\mathrm{TE/EE}$ is allowed to vary freely, the posterior peaks near $-0.8$, a value clearly inconsistent with the \emph{Planck} constraint. Nonetheless, even with such an extreme value, the impact on cosmological parameters remains minimal, further justifying the decision to fix $\alpha_g^\mathrm{TE/EE}$ in the baseline model.

\subsection{Systematic Uncertainty in DR6 $\Lambda$CDM Foreground Parameter Constraints}
\label{sec.sys_uncertainty}
The extensive foreground modeling tests conducted in this work not only demonstrate the robustness of the cosmological constraints derived from ACT DR6, but also underscore the uncertainty around some of these components, and the challenges involved in consistently modeling foregrounds and comparing parameter values across different experiments. Here, we estimate systematic uncertainties in the DR6 $\Lambda$CDM foreground parameter constraints by assessing shifts in the parameters with different foreground models, as in Figure~\ref{fig:fg_from_fg_tests}. For each foreground parameter, we first consider each of the tests that have been performed. Then we narrow down the tests to those that result in a preference of $\geq 1\sigma$ over the baseline model (as shown in Table \ref{table:chi2_fg_tests}). Of the tests that pass that mild threshold (these tests are: allowing variation of the dust template shape; allowing $\beta_c \neq \beta_p$; allowing $\beta_c \neq \beta_p$ with $T_d=25.0$ K; and inclusion of a tSZ $\times$ radio template), we take the maximum of the differences between the central value of the foreground parameter constraint in each of the model variations and the central value of the constraint in the baseline model as an overall systematic uncertainty on that parameter. This results in foreground parameter constraints with statistical and systematic uncertainties shown in Table~\ref{tab:fg_model_sys}. 

\begin{table*}[htb]
\centering
\begin{tabular}{lcccc}
\hline
Parameter & Central Value & Statistical Uncertainty & FG Model Systematic Uncertainty & (FG Sys.)/Stat. Ratio \\
\hline
$a_{\rm tSZ}$   &3.36    & ${+0.36}{-0.32}$       & ${+0.6}{-0.4}$     & $+1.67\ -1.25$ \\ \hline
$a_{\rm kSZ}$    &1.42   & ${+0.69}{-1.00}$       & ${+1.0}{-0.2}$     & $+1.45\ -0.20$ \\ \hline
$a_p$    & 7.66          & $\pm 0.34$             & ${+0.5}{-0.0}$     & $+1.47\ -0.00$ \\ \hline
$a_c$  &   3.71          & $\pm 0.46$             & ${+0.0}{-1.3}$     & $+0.00\ -2.83$ \\ \hline
$\beta_c$   &  1.86      & $\pm 0.10$             & ${+0.7}{-0.0}$     & $+7.00\ -0.00$ \\ \hline
$\xi_{yc}$     & 0.089    & ${+0.038}{-0.075}$     & ${+0.00}{-0.09}$   & $+0.00\ -1.20$ \\ \hline
$\alpha_{\rm tSZ}$ & -0.51 & ${+0.21}{-0.19}$       & ${+0.0}{-0.1}$     & $+0.00\ -0.53$ \\
\hline
\end{tabular}
\caption{Summary of parameter uncertainties, both statistical and from the foreground model systematic uncertainty. The final column shows the ratio of systematic to statistical uncertainty, separated for positive and negative errors.}
\label{tab:fg_model_sys}
\end{table*}

We thus find that the systematic uncertainty is on the order of, or slightly larger than, the statistical uncertainty on each of these parameters. For $a_{\rm tSZ}$, the upper systematic error bar is driven by the \verb|AGORA| tSZ $\times$ radio template inclusion, while the lower systematic error bar is driven by allowing $\beta_c \neq \beta_p$. For $a_{\rm kSZ}$ and $\beta_c$, the upper systematic error bar is driven by allowing $\beta_c \neq \beta_p$. The lower systematic error on $a_{\rm kSZ}$ is driven by the inclusion of the \verb|AGORA| tSZ $\times$ radio template. This underscores the strong sensitivity of SZ constraints to the assumed foreground model and highlights the difficulty of comparing individual parameter values across different analyses without accounting for the specifics of the modeling framework.

For $a_c$, $\xi_{yc}$, and $\alpha_{\rm tSZ}$, the lower systematic error bar is driven by allowing $\beta_c \neq \beta_p$. For $a_p$, allowing $\beta_c \neq \beta_p$ drives the upper systematic error. We note in passing that the preference for a nonzero value of $\alpha_{\rm tSZ}$ is relatively robust to the choice of foreground model, with the systematic error being smaller than the statistical error.

From this assessment, we see that the foreground parameters can be quite sensitive to the choice of foreground model. In particular, the inclusion of a tSZ $\times$ radio template is not adopted in our baseline analysis or in the analysis for any other CMB experiment to date, but as seen here, its inclusion can decrease the inferred value of $a_{\rm kSZ}$.

The impact of these foreground models on the $95\%$ confidence upper limit placed on $\dzreion$ is not negligible, but does not increase it significantly except for one case. The foreground model that loosens the upper limit on $\dzreion$ the most is $\beta_c \neq \beta_p$, increasing the limit to $\dzreion < 7.0$. The model that pushes the upper limit the lowest is the inclusion of CIB$\times$radio, where the limit goes to $\dzreion < 3.1$. The other foreground model variations yield upper limits between these values, most sitting in a tight range just below the baseline ACT upper limit of $\dzreion < 4.4$; see Figure~\ref{fig:extra_fg_from_fg_tests} for a depiction of these results. The impact of foreground modeling on reionization constraints stems from its influence on the allowed kSZ amplitude: models that absorb or add small-scale power directly affect the room left for a reionization kSZ signal, tightening or loosening $\dzreion$ bounds.


\section{Correlated, Non-Gaussian Sky Simulations}
\label{sec.ng_sims}

Defining a model for the auto- and cross-frequency power spectra as done in~\S\ref{subsec:fg_model} involves making general assumptions about the form of the different foreground power spectra, with associated free parameters for each component. If the data differ significantly from the fiducial models, the free parameters of the sky model must be flexible enough to absorb any discrepancies, while still yielding unbiased cosmological parameters from the inferred CMB component of the TT, TE, and EE power spectra. In previous CMB analyses \citep[e.g.,][]{SPT:2017, Planck:2018_pslkl, choi_atacama_2020}, tests of the power spectrum model generally involved simulating all of the foregrounds as Gaussian random fields, usually generated with the same models as used in the inference process, and often uncorrelated with one another.\footnote{One important exception was the Planck Sky Model developed to test various \emph{Planck} analysis pipelines, which included some, though not all, of the expected correlations amongst various sky components~\citep{Delabrouille:2009, Delabrouille:2011}.}  These are significantly limiting assumptions, since in reality, many of the foreground fields are highly non-Gaussian and correlated.  Furthermore, given our imperfect knowledge of the microwave sky, it is crucial to test analysis pipelines on sky models that differ from those being used in the inference process.

In this and the following section, we test the ACT DR6 power spectrum and cosmological parameter inference pipeline on realistic non-Gaussian simulations that feature correlations amongst the sky components.  We replicate the entire DR6 data analysis pipeline on the simulations to show that we can recover the input cosmological parameters of the CMB component in the simulations, in spite of having foregrounds that were constructed independently of the models used in the ACT DR6 analysis. While non-Gaussian extragalactic simulations were used to test for biases in delensed spectra in ACT power spectrum analyses in~\cite{ACT:2020goa}, this is the first end-to-end test of parameter recovery of its kind performed for a modern CMB experiment.

We construct these simulations by synthesizing sky maps from models of the relevant Galactic and extragalactic components,  which are evaluated with the ACT passbands, beams, and systematic effects.  The Galactic components are generated with \verb|PySM| \citep{Thorne_2017, Zonca_2021, Pan-ExperimentGalacticScienceGroup:2025vcd}, and include thermal dust, synchrotron, and AME. The extragalactic components are built from the \verb|AGORA| simulations \citep{Omori:2022uox}, and include the CMB, CIB, radio source emission, tSZ, and kSZ (all gravitationally lensed). Additionally, we add a realization of the reionization kSZ effect from \cite{Battaglia_2013_kszpatchy}. The Galactic and extragalactic component models are described in \S \ref{subsubsec.gal_comps} and \S \ref{subsubsec.extragal_comps}, respectively. 

This section focuses on a description of the simulated maps as well as a description of how the simulations are processed and the power spectra are computed. \S \ref{sec.ng_sims_params} discusses parameter recovery using the simulations. 

\subsection{Simulation Pipeline Overview}
Maps of each individual Galactic and extragalactic component are produced with the DR6 passbands using the \verb|HEALPix| pixelization scheme \citep{Healpix, Healpy} in Galactic coordinates, at resolution parameter $N_{\mathrm{side}}=8192$ ($\approx 0.43$ arcmin pixels). To prevent aliasing effects, prior to transforming to harmonic space, pixels with flux density greater than 100 mJy in temperature at 150 GHz (34 pixels for the CIB map and 2276 pixels for the radio map) and their immediate neighbors are set to zero. The maps are then transformed to harmonic space, where they are convolved with the beams. The maps are then rotated into equatorial coordinates, and then reprojected to the CAR pixelization scheme using \verb|pixell|\footnote{\url{https://github.com/simonsobs/pixell}} via the DUCC Spherical Harmonics Transforms tools\footnote{\url{https://gitlab.mpcdf.mpg.de/mtr/ducc}}. Noise is added using the DR6 \verb|mnms| noise simulations \citep{Atkins:2023yzu}. Finally, sources are subtracted following the procedure described in N25. 

We generate 10 different sets of simulations, each with a different CMB realization, but with the same foregrounds and noise realizations. While the 10 unlensed CMB maps are independent from one another and the rest of the foregrounds, they are each lensed by the same lensing convergence map. From hereafter, one ``set" will refer to one of these 10 sets of simulations. For each set of simulations, we have 20 maps. The 20 maps consist of 5 detector array / frequency pairs (PA4 220, PA5 90, PA5 150, PA6 90, and PA6 150), with 4 different noise splits for each. Here PA stands for Polarimeter Array. Finally, we compute all auto- and cross-power spectra for each set using \verb|PSpipe|.\footnote{ \url{https://github.com/simonsobs/PSpipe}} The full power spectrum computation pipeline is described in \S \ref{subsubsec.ps}.

\subsection{Galactic Components}
\label{subsubsec.gal_comps}

For the Galactic component models from \verb|PySM|, the \textbf{d10}, \textbf{s5}, and \textbf{a1} models are adopted for dust, synchrotron emission, and AME, respectively. These are components of the ``medium complexity" model proposed by \citet{Pan-ExperimentGalacticScienceGroup:2025vcd}, described by small-scale extrapolation in both amplitude and spectral parameters. We do not include free-free or CO line emission in the simulations. For more detail on the \textbf{d10} and \textbf{s5} components, see \citet{Pan-ExperimentGalacticScienceGroup:2025vcd}, and for more detail on the \textbf{a1} component, see \citet{Thorne_2017}.

The \textbf{d10} dust map is modeled as an MBB, based on \emph{Planck} generalized needlet internal linear combination (GNILC) maps in intensity \citep{Planck:2016_dust}. The \textbf{d10} model adds random small-scale ($\ell > 100$) fluctuations to the templates, which are computed using a polarization fraction tensor formalism, such that they are moderately non-Gaussian. The fluctuations are modulated by the large-scale emission before being added to the large-scale template, such that they inherit some of the larger-scale amplitude structure of the maps. In terms of frequency scaling, an MBB spectrum model is assumed, with $\beta_d$ and $T_d$ on large scales fixed to the values from the PR2 GNILC maps \citep{Planck:2016_dust}. On small scales, the frequency scaling is modulated by the large scales.

The \textbf{s5} synchrotron model uses a power-law scaling with a spatially variable spectral index. For temperature, the template used is from \citet{Remazeilles:2014mba}, derived from the reprocessed Haslam 408 MHz map \citep{Haslam}, and for polarization, the template is from the WMAP 9-year 23 GHz map \citep{Bennett2013}. Small-scale ($\ell > 38$) fluctuations are added to both in a similar way as done for the dust. In terms of frequency scaling, a power-law model is assumed, with the large-scale frequency scaling having a fixed power-law index $\beta_s$.  It is based on the spectral index map of \citet{Miville-Deschenes:2008lza} (produced by combining the \citet{Haslam} 408 MHz intensity map with WMAP 3-year K-band data \citep{WMAP:2007}), rescaled in variance to match the S-PASS power spectrum \citep{Krachmalnicoff:2018imw}, then extrapolated to small scales. 

The \textbf{a1} AME model uses the Commander code \citep{Planck:2015mvg} to model the emission as coming from two spinning dust populations. For each population, the SpDust2 code \citep{Ali-Haimoud:2008kyz} is used to generate the emission law. One population is assumed to have a spatially varying peak frequency, while the other population is assumed to have a spatially constant peak frequency but nonetheless with an overall spatially varying emission template. Small scales are added in temperature. The model assumes that there is no polarized AME. 

\subsection{Extragalactic Components}
\label{subsubsec.extragal_comps}
The \verb|AGORA| simulation suite is used for the extragalactic components, including the lensed CMB, CIB, radio emission, tSZ effect, and late-time kSZ effect. For more detail on each component, see \citet{Omori:2022uox}.

For the lensed primary CMB, unlensed CMB maps are first generated following \citet{Giannantonio2016}. \verb|LensPix| \citep{lenspix} is then used to deflect the maps with the lensing convergence field $\kappa$, where the lensing field is generated either via the Born approximation or multi-plane ray-tracing. Here we use simulations generated via the ray-tracing approach.

The CIB is modeled in detail on a galaxy-by-galaxy basis (see~\citet{Omori:2022uox}), with each galaxy's infrared SED modeled as an MBB with a power-law transition, meaning that the MBB is multiplied by a power-law at high frequencies. The galaxy field itself is created using the \verb|UniverseMachine| catalogs \citep{universemachine}, with the CIB model parameters coming from the best-fitting to the auto- and cross-power spectra from \citet{Lenz:2019ugy}. 

The tSZ maps are generated by inferring gas pressure and density profiles from a hydrodynamic simulation, and pasting the profiles onto halos with mass $M_h > 10^{12}h^{-1}M_{\odot}$ in the lightcone. The thermodynamic profile fits are made to the \verb|BAHAMAS| simulations \citep{Mccarthy:2017yqf}\footnote{\url{https://www.astro.ljmu.ac.uk/~igm/BAHAMAS/}} using the formalism from~\citet{Mead:2020vgs}.  The map included here is that derived from the pressure profile model obtained with AGN gas heating temperature $10^{8.0}$ K in \verb|BAHAMAS|.\footnote{We note that the tSZ power spectra computed from the \texttt{AGORA}-based \texttt{BAHAMAS} models have mild differences with respect to the tSZ power spectra computed directly from the \texttt{BAHAMAS} simulations, which affect the maps on large scales; this does not impact our results, as the \texttt{AGORA} maps still represent a plausible model of the tSZ field, suitable for testing our CMB parameter inference pipeline.}

For the kSZ maps, the ionized gas density profiles inferred from fits to \verb|BAHAMAS| are used to estimate the electron number density, which is then multiplied by the pixelized particle velocity maps to obtain the kSZ field in a given density shell. The differential kSZ temperature maps are integrated for shells up to $z=3$ to obtain the full late-time kSZ map.

Importantly, as all of the above components are computed from the same large-scale structure simulation, they are all realistically correlated.

In addition to the above components from the \verb|AGORA| simulation suite, a Gaussian realization of the reionization kSZ power spectrum is included, computed using the kSZ power spectrum from \citet{Battaglia_2013_kszpatchy}.

\subsection{Passbands}
Maps of each individual Galactic and extragalactic component are produced using the truncated DR6 passbands \citep[]{passbands_inprep}. We use the PA4 f220 passband, the PA5 f150 passband, and the PA6 f090 passband (note that PA4 f220 is the only 220 GHz channel in the ACT DR6 dataset, but there are other 90 and 150 GHz channels; for simplicity we adopt a single f090 and a single f150 passband here). Thus, after integrating individual component signals over the passbands and summing all the components, we have a total of five maps per set. We note that we do not use chromatic beams in the passband integration here for simplicity; the lack of chromatic beam use is accounted for in the downstream analysis.

\subsection{Beams}
We apply coadded beams to each map \citep[]{beams_inprep}. In order to mimic the real data as closely as possible, we use all detector array and frequency pairs for the beams: PA4 f220, PA5 f090, PA5 f150, PA6 f090, and PA6 f150. However, we only have three bandpassed maps from before the beam application (PA4 f220, PA5 f150, and PA6 f090). To produce the full set of five maps, we apply the beams for each array to the corresponding frequency map. For example, for the 150 GHz bandpassed map, we apply the PA5 f150 beam to obtain the final PA5 f150 map, and the PA6 f150 beam to obtain the final PA6 f150 map (even though the map itself was generated using the PA5 f150 passband). Note that this may lead to a mismatch in the applied beam and the passband used, but we account for this difference when computing the spectra. 

\begin{figure}[t]
    \centering
    \includegraphics[width=0.4\textwidth]{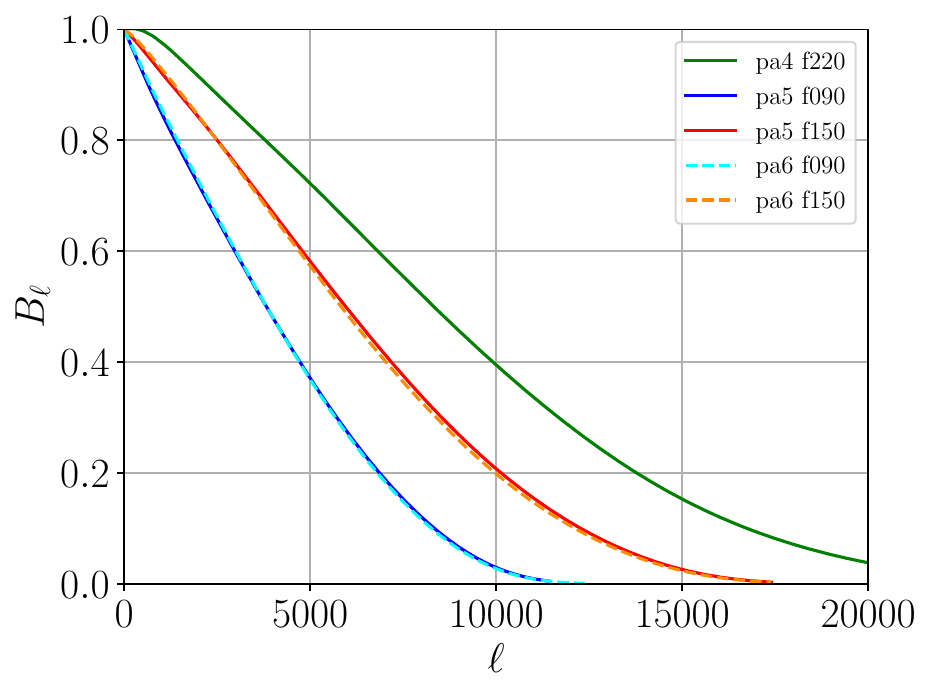}
    \caption{Coadded beams used for production of the simulations, for each detector array/frequency.} 
    \label{fig:beams}
\end{figure}

We note that we apply the coadded beams instead of the per-split beams for computational efficiency since the differences in the per-split and coadded beams are extremely small \citep{beams_inprep}. Figure~\ref{fig:beams} shows the normalized beams used in the simulations. 

\subsection{Noise}
Noise simulations are generated with \verb|mnms| \citep{Atkins:2023yzu} (also see \citealp{Atkins:2024jlo} for updates to the methodology). We employ a tiled noise model based on \citet{Naess:2020wgi}, which uses interleaved tiles on the sky. The model is still Gaussian in the sense that it is fully defined by a covariance matrix, but it is more complicated than typical Gaussian distributions encountered in CMB analysis. For instance, rather than being diagonal in the spherical harmonic basis --- as is the case for the primary CMB --- or the pixel basis --- as is the case for white noise --- the model is diagonal in the \textit{tiled, two-dimensional Fourier basis}. As a result, the simulated noise contains non-trivial properties: its correlation structure is directional or ``stripy,'' and the character of this structure is a function of position on the sky. In addition, the overall noise power also varies with sky position on both large and small spatial scales. The complicated structure of the noise is due to the interplay between the spatial variations of the scanning pattern of the telescope projected on the sky, the instrumental noise and fluctuations in the atmospheric emission \citep{Morris2022, Morris2025}, which act as the dominant source of noise at large angular scales in the maps. Since the \verb|mnms| simulations are band-limited to $\ell=10800$, we fill in smaller scales with white-noise realizations drawn using the DR6 inverse-variance maps.

Noise simulations are generated for each array-band, but the noise realizations are consistent across the 10 sets of simulations, i.e., each set of simulations has a separate CMB realization but the same noise realizations as each of the other sets. The noise realization is kept the same across sets for computational simplicity, since changing the noise in each set would require rerunning the source subtraction procedure on each set independently. 

\subsection{Source Subtraction}
Point sources are subtracted in the same way as in the DR6 data pipeline (N25, L25). We subtract point sources with a flux density cut of 15 mJy at 150 GHz.

\subsection{Power Spectra Computation}
\label{subsubsec.ps}

\begin{figure*}[t]
    \centering
    \includegraphics[width=0.9\textwidth]{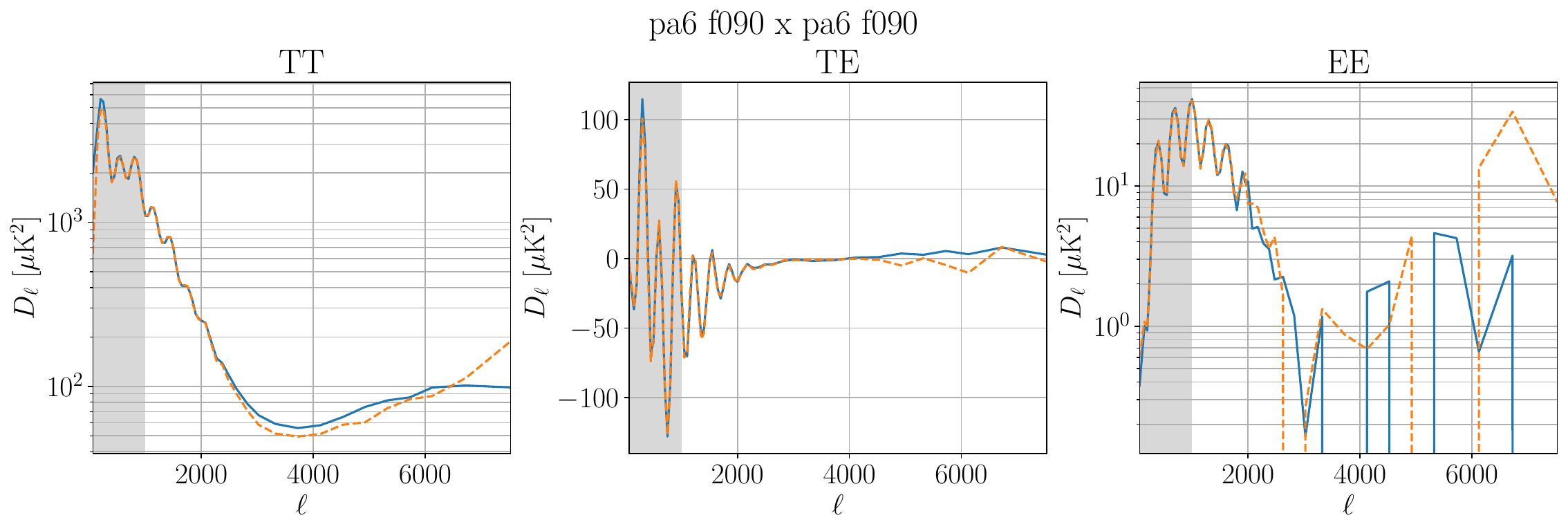}
    \includegraphics[width=0.9\textwidth]{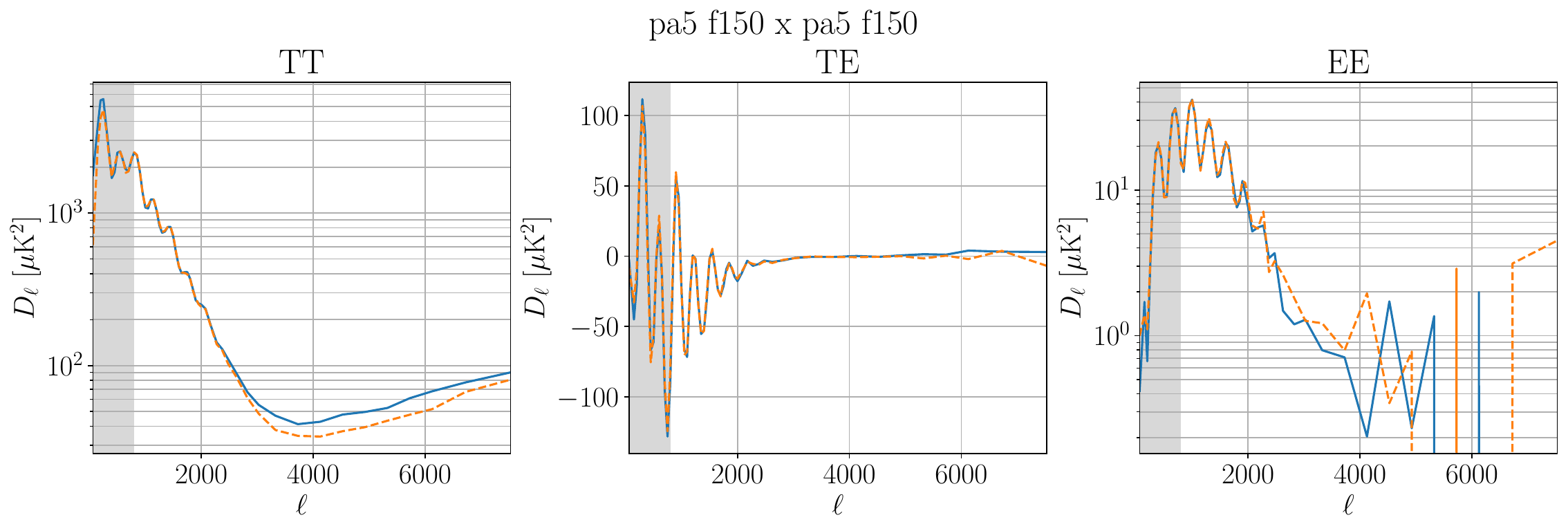}
    \includegraphics[width=0.33\textwidth]{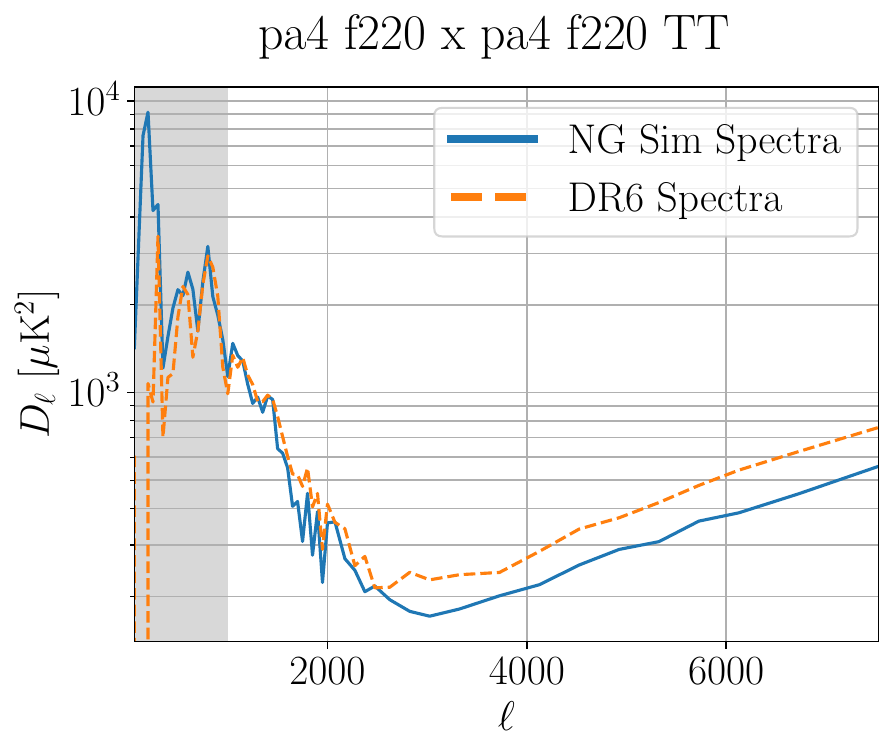}
    \caption{Example comparisons of spectra from the full simulations (blue) with the DR6 data (orange), shown for three example array-band combinations. Each of the spectra is computed using cross-spectra for noise-debiasing (thus why the EE spectrum is negative at some multipoles). The shaded gray band denotes multipoles that are not used in the likelihood analysis. The blue and orange curves are not expected to match since the simulations are constructed independently of the DR6 data and foreground model.\\ } 
    \label{fig:ng_sim_ps}
\end{figure*}

Spectra are computed with \verb|PSPipe|, in the same way as in the DR6 data pipeline (see L25), which follows the general approach from  \cite{choi_atacama_2020}. Window functions used for temperature and polarization are the same, based on the edges of the survey, a Galactic mask, and point source mask obtained from the source subtraction procedure. Windows and beams are deconvolved using the standard MASTER approach \citep{Hivon2002}. We do not apply any leakage or aberration corrections since these effects were not included when producing the simulations. To avoid noise bias, power spectra are computed from independent splits of the simulations. This results in fifteen spectra for TT,
using all five array-bands; ten for EE, using four array-bands (all excluding PA4 220); and sixteen for TE/ET, again using four array-bands (all excluding PA4 220). For the power spectrum covariance, we use the same covariance matrix that is used for the DR6 data \citep{Atkins:2024jlo}. For more details on the power spectrum computation pipeline, see L25.

Figure~\ref{fig:ng_sim_ps} shows examples of spectra computed from the simulations, compared with the true DR6 data power spectra. We do not expect these spectra to be identical, since the simulations were constructed independently of the DR6 data and foreground models.  

\section{Parameter Recovery on Correlated, Non-Gaussian Simulated Maps}
\label{sec.ng_sims_params}

\subsection{Parameter Constraints from Non-Gaussian Extragalactic Simulations}

\begin{table}[t]
\centering
\begin{tabular}{|c|c|c|c}
\hline
  Parameter & $\sigma$ Shift from Input & $\sigma$ Shift from Input  \\ 
   &(Extragalactic-Only Sims) & (Full Sims) \\ 
 \hline \hline
$\Omega_c h^2$& $-0.25$ & $-0.27$  \\ \hline
$\Omega_b h^2$ & $-1.72$ & $-1.86$ \\ \hline
$\ln 10^{10}A_s$ &  $-1.39$ & $-1.45$ \\ \hline
$n_s$ & $-0.20$ & $-0.02$ \\ \hline
$H_0$ & $-0.40$ & $-0.43$ \\ \hline
$\tau_{\mathrm{reio}}$ & $-0.38$ & $-0.38$ \\ \hline
\end{tabular}
  \caption{Average shift (in $\sigma$) of central values for cosmological parameters from the input true parameter values. The average is computed over the 10 simulation sets, each of which have different CMB realizations but the same foreground and noise realization. Results are shown for both extragalactic-only simulations and full simulations (containing Galactic components as well).}
\label{table:ng_sim_sigma_shifts}
\end{table}

In this subsection, we demonstrate recovery of the input cosmological parameters on non-Gaussian simulations including extragalactic foregrounds (tSZ effect, patchy (reionization) kSZ effect, late-time kSZ effect, CIB, and radio emission) additionally to the CMB. We note that the DR6 baseline foreground model does not contain a template for the patchy kSZ effect, so this test serves also as validation of that assumption.

\begin{figure}[t]
    \centering
    \includegraphics[width=0.47\textwidth]{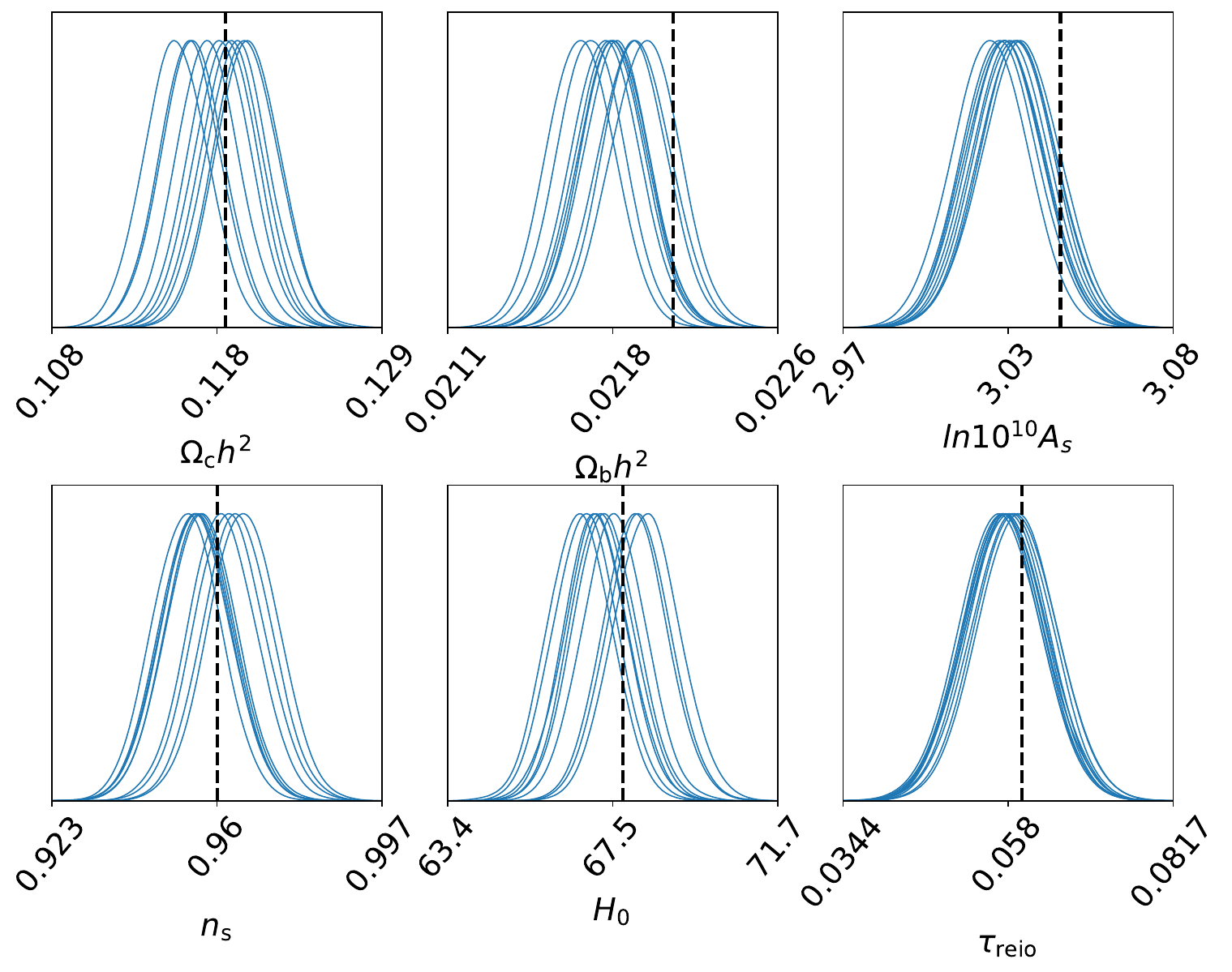}
    \caption{1D marginalized cosmological parameter constraints (with a linearly scaled $y$-axis) for simulations containing only the CMB and non-Gaussian extragalactic foregrounds. Results are shown for 10 simulations (blue), each of which have independent CMB realizations but the same foreground and noise realizations. The dashed black line shows the true cosmological parameters that are inputs to the simulation. This collection of parameter shifts is consistent with expectations for a single realization of the simulated data, implying that the single realization of the foregrounds and noise, rather than the cosmic variance, dominates the shifts in the output parameters from their input values.}
    \label{fig:extragal_10sims_cosmo}
\end{figure}

Parameter constraints are determined using the same input priors and settings as with the actual DR6 data and as done for the results of~\S\ref{sec:fg_tests}. We use each of the 10 different simulation sets, noting that each set contains a different CMB realization but the same foreground and noise realization. Thus, the use of 10 simulation sets is just an assessment of the cosmic variance of the CMB. Figure~\ref{fig:extragal_10sims_cosmo} shows the 1D posterior distributions for cosmological parameters for each of the 10 sets of simulations. The average (over the 10 sets) shifts in $\sigma$ from the input cosmological parameters are shown in Table~\ref{table:ng_sim_sigma_shifts}. Overall, we find that all of the parameters agree with the inputs within $2\sigma$. This is a strong test of our parameter recovery from the non-Gaussian simulations, particularly since the simulations include the exact same noise and foreground realizations, and hence this estimate does not include stochastic variations in these components. 

\subsection{Parameter Constraints from Full Non-Gaussian Simulations}

\begin{figure}[t]
    \centering
    \includegraphics[width=0.47\textwidth]{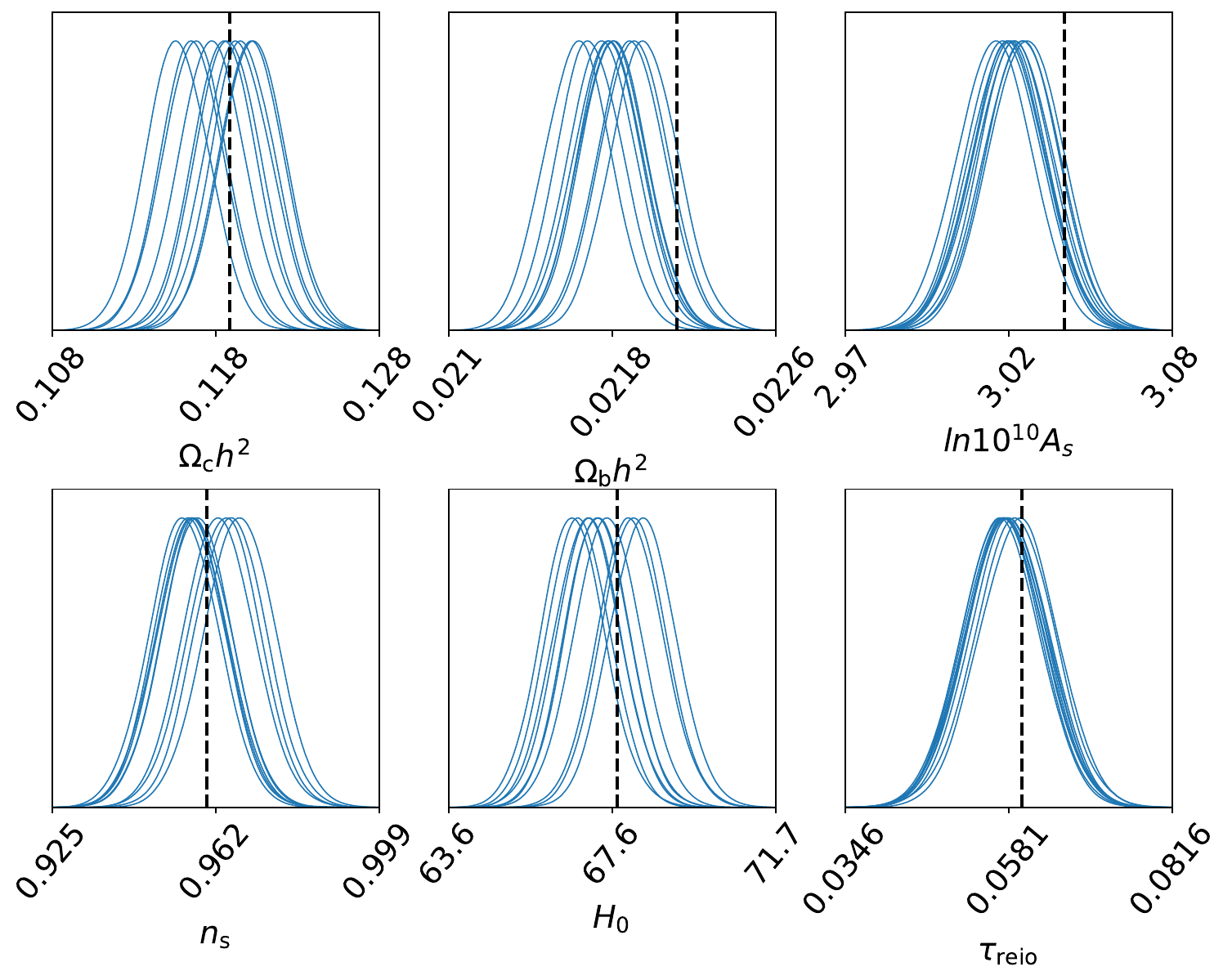}
    \caption{Same as Figure~\ref{fig:extragal_10sims_cosmo} but from simulations including both extragalactic and Galactic foregrounds.} 
    \label{fig:full_10sims_cosmo}
\end{figure}

\begin{figure*}[t]
    \centering
    \includegraphics[width=0.9\textwidth]{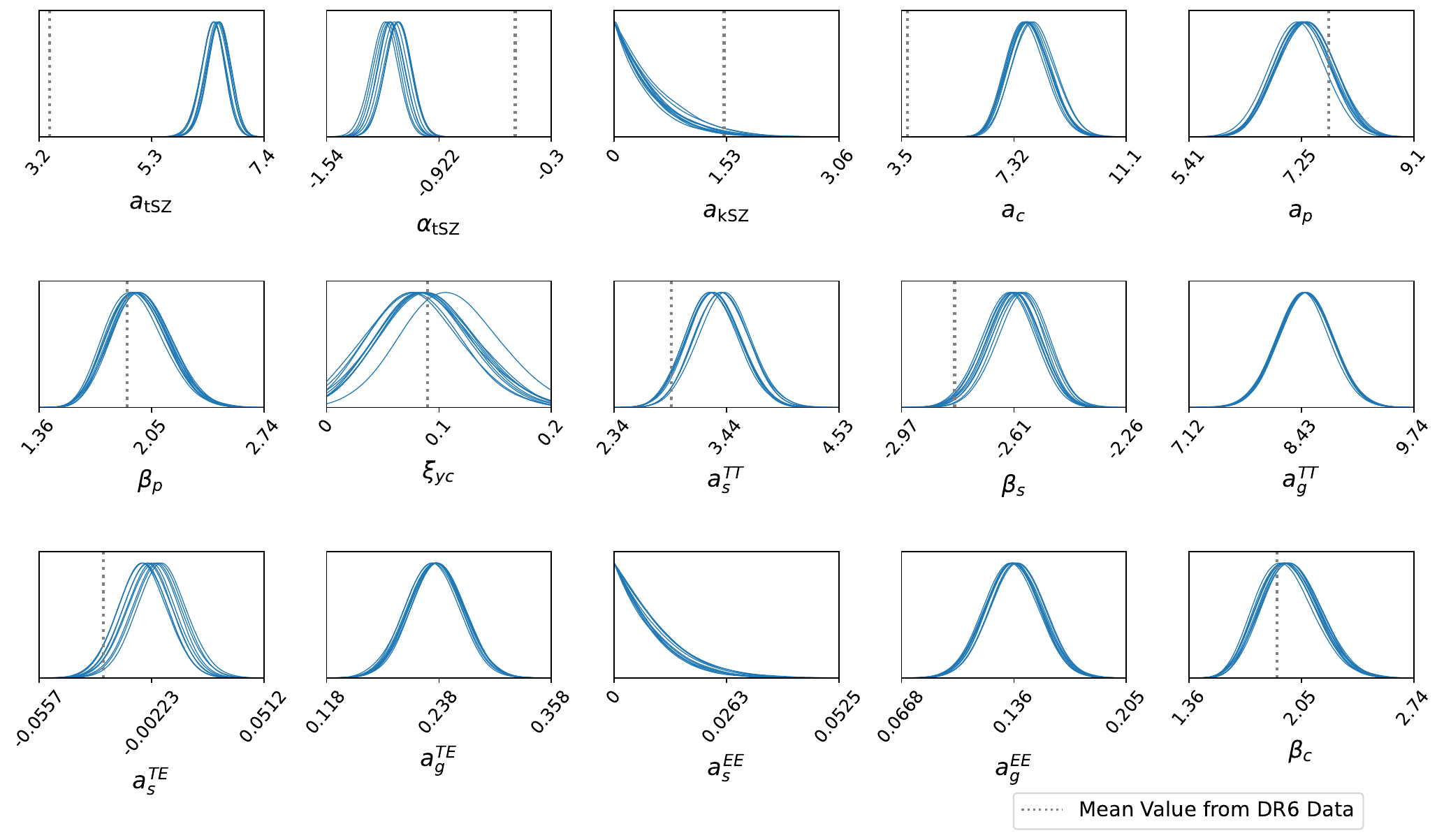}
    \caption{1D marginalized foreground parameter constraints for simulations containing the CMB with both non-Gaussian extragalactic and Galactic foregrounds. Results are shown for 10 simulations (blue), each of which have independent CMB realizations but the same foreground and noise realizations. Here, $\mathrm{a}_{\rm tSZ}$, $\alpha_{\rm tSZ}$, and $\mathrm{a}_{\rm kSZ}$ are SZ parameters; $a_p$, $\beta_p$, $a_c$, and $\beta_c$ are CIB parameters; a$_s$ and $\beta_s$ are radio parameters; a$^{\rm TT}_{\rm dust}$, a$^{\rm TE}_{\rm dust}$, and a$^{\rm EE}_{\rm dust}$ are dust parameters; and $\xi$ is the tSZ--CIB correlation. Dotted gray lines show the mean values from the DR6 data. Importantly, these are \emph{not} the true inputs to the simulations. These are here to show how much variation is possible in these parameters --- the model we fit in power spectrum analyses and current state-of-the-art inputs in complex simulations can be very different. Dotted lines are not shown for the dust parameters, which are prior-dominated in the ACT DR6 inference, and a$_{\rm ps}^{\rm EE}$, where there is only an upper bound from the DR6 data. \\ } 
    \label{fig:full_10sims_fg}
\end{figure*}

Next, we consider non-Gaussian simulations containing the CMB, as well as both extragalactic and Galactic components. 

In this case we change the prior on dust amplitudes from that used in the DR6 baseline model, in order to more closely match the procedure used in the real analysis.  We use the same standard deviation in the dust prior as in the runs with actual data. The central value, however, is determined as follows. We generate a 353 GHz dust-only map (using the \verb|PySM| \textbf{d10} model, evaluated with the \emph{Planck} passband). We compute the spectra using the same mask (survey edges + Galactic mask + point source mask) that is used in the simulations. We set the central value of the dust priors to the dust amplitudes $a_g^{XY}$ from Equation~\eqref{eq:dust_pw} in TT, TE, and EE, at the pivot multipole $\ell_0=500$. We obtain the following best-fitting values in $\mu$K${}^2$ for TT, TE, and EE, respectively: 8.3, 0.24, and 0.14 (for comparison, the DR6 data runs set these central values to 8.0, 0.42, and 0.17 from Equation~\ref{eq:dust_amps}). 

We note that the only Galactic component included in the baseline DR6 foreground model is the dust. Thus, including synchrotron emission and AME in the simulations is a non-trivial test that our parameter constraints are insensitive to the presence of these components. 

The resulting cosmological parameter constraints are shown in Figure~\ref{fig:full_10sims_cosmo}. The average (over the 10 sets) shifts in $\sigma$ from the input cosmological parameters is shown in Table~\ref{table:ng_sim_sigma_shifts}. Again, we find that all of the parameters agree with the inputs within $2\sigma$, demonstrating good agreement with the inputs. 

Figure~\ref{fig:full_10sims_fg} shows the foreground parameters obtained from the full simulations. We find that the $\alpha_{\rm tSZ}$ parameter is critical to allowing enough flexibility for this agreement, as it is observed to be robustly nonzero. From constraints with the simulations, we see that the mean $\alpha_{\rm tSZ}$ value is found to be $-1.18$, averaged across simulations. This is a larger magnitude than what is found from the DR6 data, or from fitting the \verb|AGORA| tSZ power spectrum alone, but likely arises due to degeneracies with other foreground parameters in the simulations (Figure~\ref{fig:pact_alphasz_2d_fg} shows such degeneracies from the DR6 data).

We emphasize that appearance of deviations from the input CMB cosmological parameters are likely due to fluctuation rather than bias. This is because of the use of a single realization of each foreground field and noise. Simple statistical tests show that the observed parameter shifts from their input values are consistent with expectations for a single realization of the simulated data. In theory, the foreground and noise realization could be changed too, but we do not explore that avenue here. It is computationally expensive to produce new foreground realizations, and new noise realizations would require redoing the source subtraction procedure on each set of simulations, as the noise impacts detection of sources.

\subsection{Understanding Fits to Foregrounds}

\begin{figure*}[htb]
    \centering
    \includegraphics[width=0.81\textwidth]{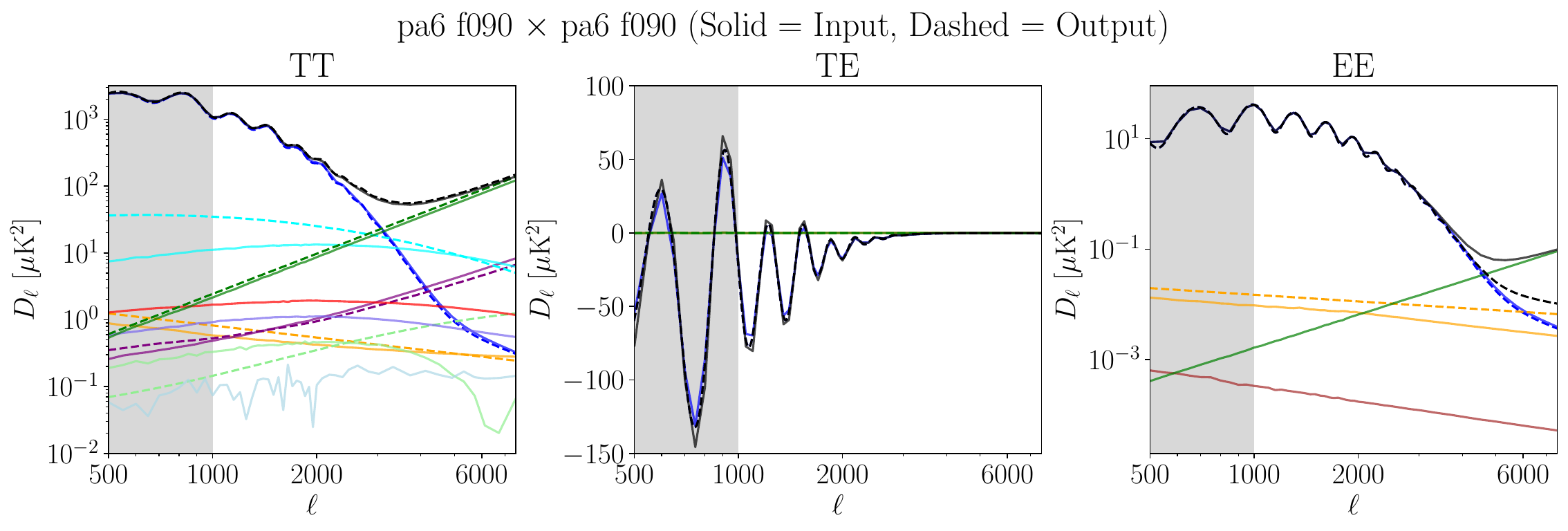}
    \includegraphics[width=0.81\textwidth]{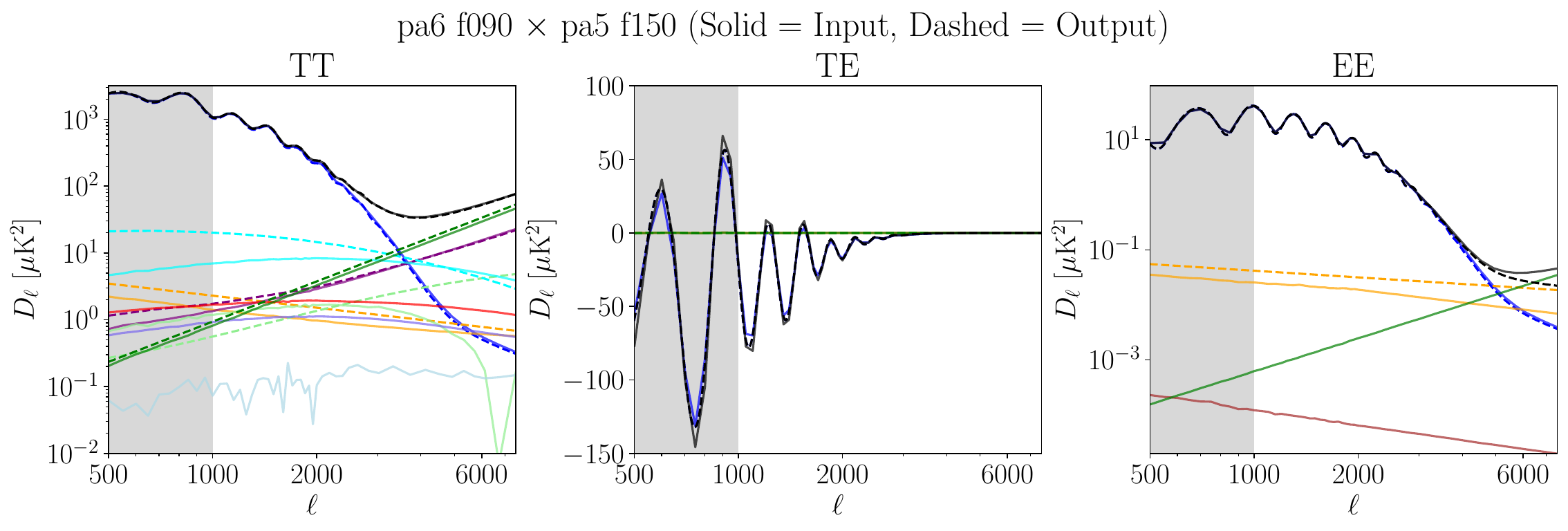}
    \includegraphics[width=0.81\textwidth]{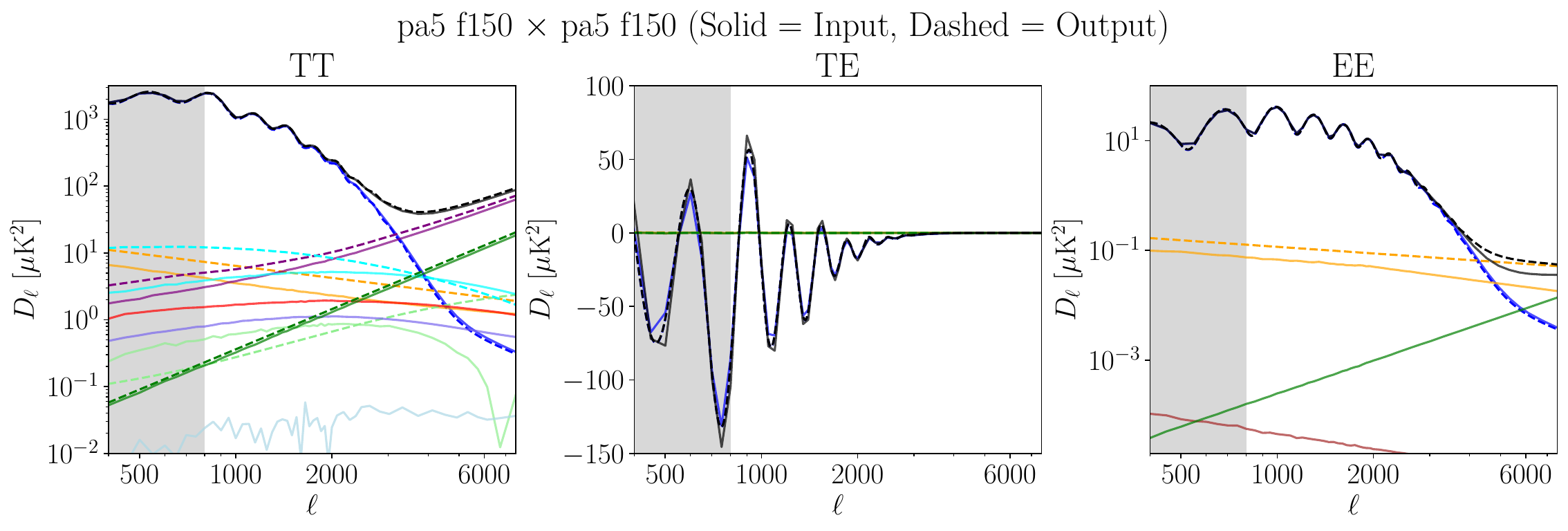}
    \includegraphics[width=0.50\textwidth]{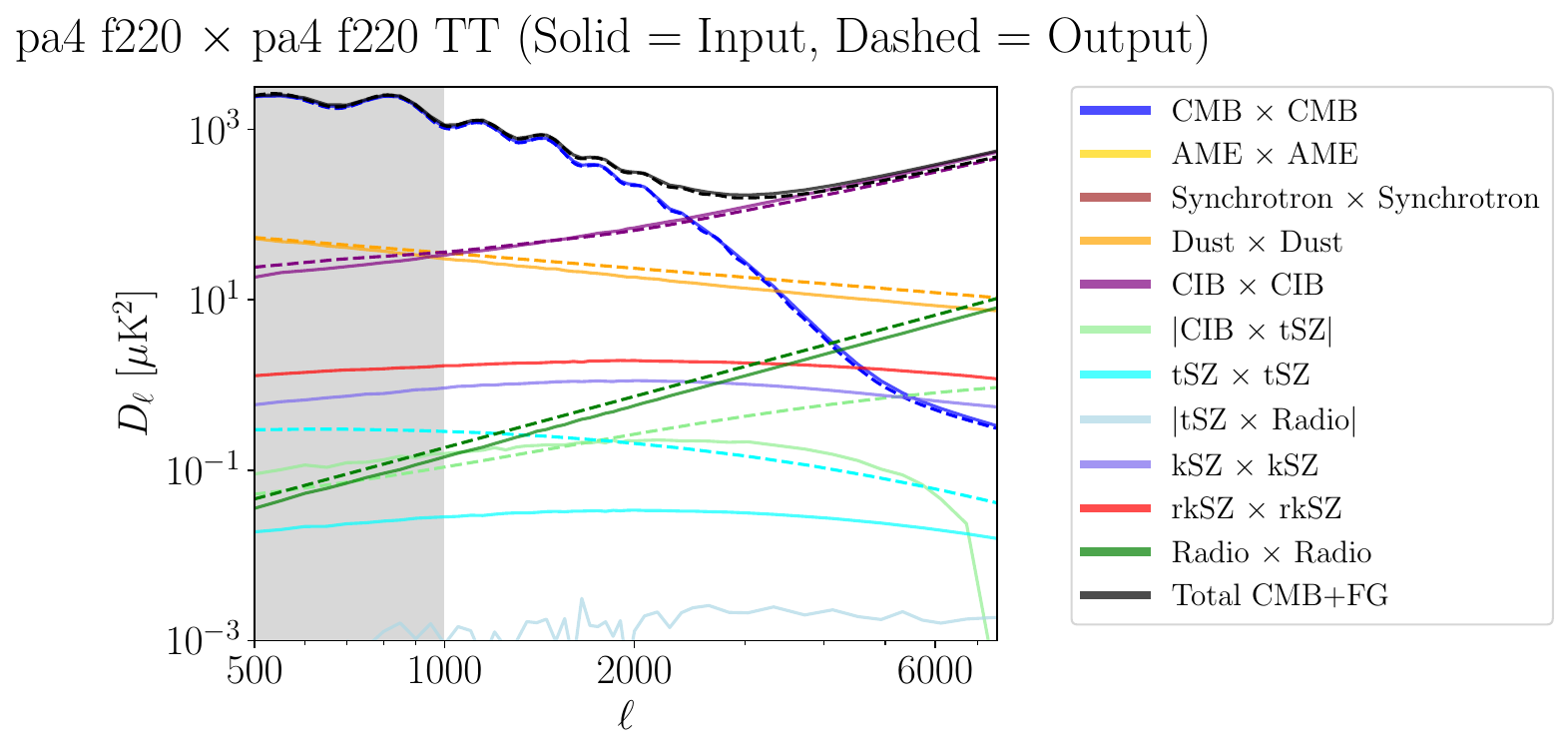}
    \caption{Example power spectra of individual component fields in the non-Gaussian simulations, for array-band pairs pa6 f090 $\times$ pa6 f090 (top row), pa6 f090 $\times$ pa5 f150 (second row), pa5 f150 $\times$ pa5 f150 (third row), and pa4 f220 $\times$ pa4 f220 (bottom row). Solid lines show ``inputs" to the simulations (processing individual foregrounds through the full power spectrum pipeline). Dashed lines show the model ``outputs," the foreground models in the DR6 likelihood evaluated at the MAP of the fit to the simulations. The black line shows the total power spectrum of the component fields (CMB plus foregrounds, not including noise). For the CIB $\times$ tSZ component, the absolute value is shown. The sign of the CIB $\times$ tSZ cross-correlation changes at high $\ell$ in the inputs, a feature of the \texttt{AGORA} simulations. Here rkSZ is the reionization kSZ effect. The shaded gray band indicates multipole values that are not used in the likelihood analysis. In TT, the input AME and synchrotron emission lie well below the dynamic range of the plot shown and are omitted for clarity of the other components.} 
    \label{fig:ng_sim_comp_ps}
\end{figure*}

In this subsection, we compare the power spectra of the input foreground simulations to the output foreground models evaluated at the MAP parameter values. The input power spectra are computed by processing individual foreground fields (which have been beam-convolved) through the full power spectrum pipeline, including effects such as masking. As done above, to obtain the MAP of the fits we use a \texttt{cobaya} minimization routine and the resulting parameters are used to evaluate the output foreground models.  Figure~\ref{fig:ng_sim_comp_ps} shows spectra of individual component fields that went into the simulations (``inputs'') and the MAP model of each component (``outputs'').

\begin{figure*}[htb]
    \centering
    \includegraphics[width=0.35\textwidth]
    {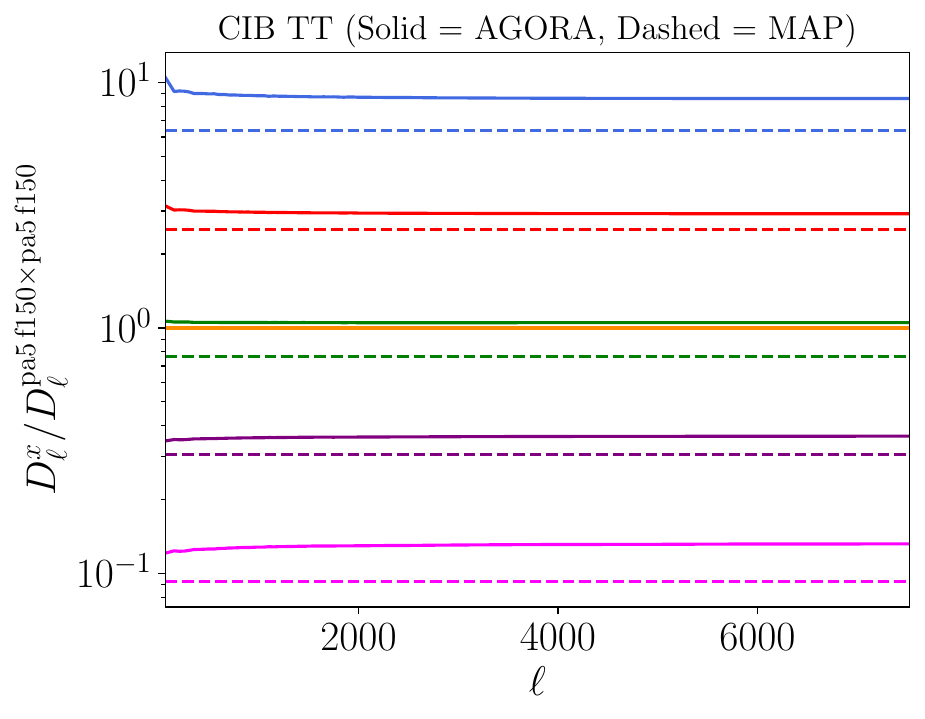}
    \includegraphics[width=0.35\textwidth]{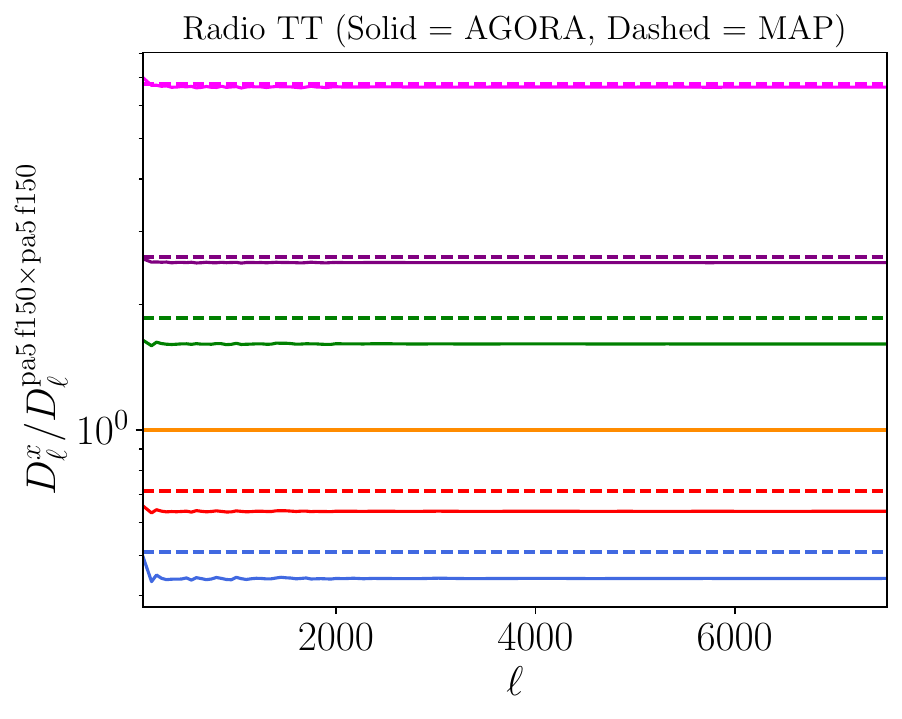}
    \includegraphics[width=0.35\textwidth]{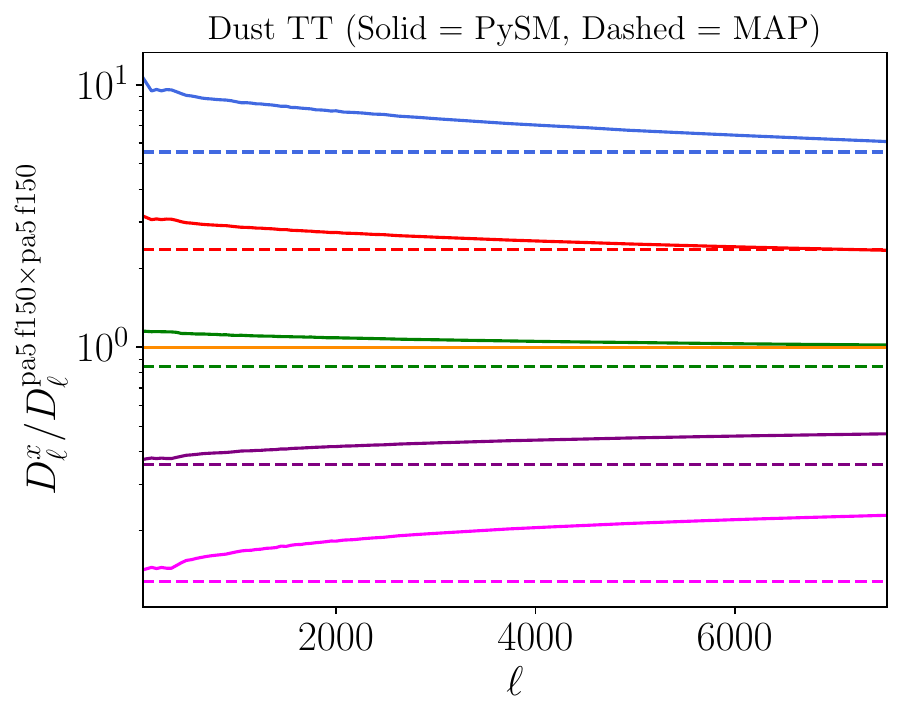}
    \includegraphics[width=0.35\textwidth]{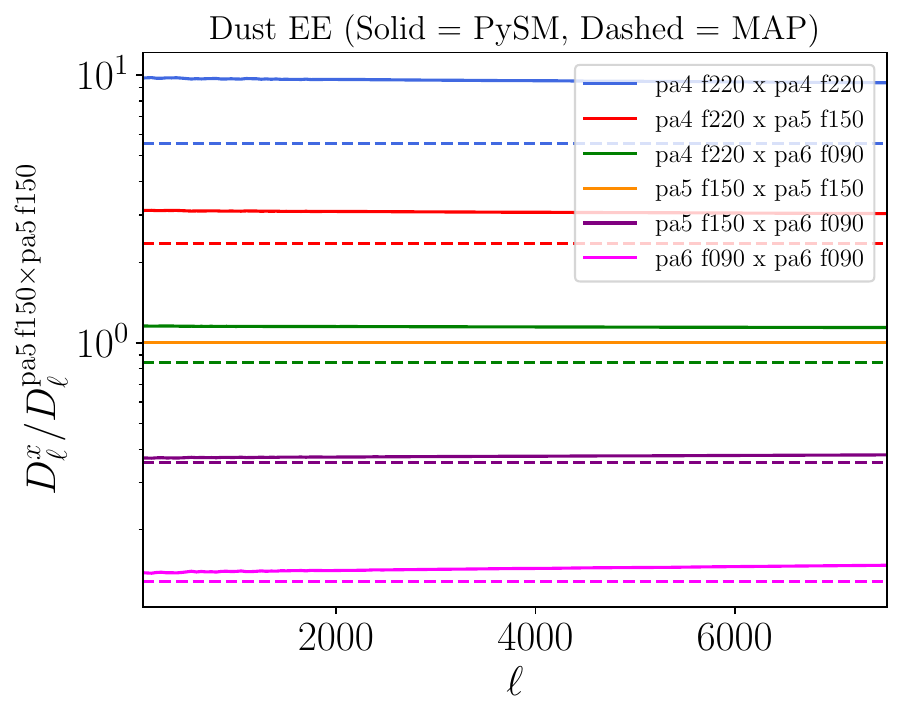}
    \caption{Frequency scaling of foreground fields in the non-Gaussian simulations: the CIB, radio, and dust in TT, as well as dust in EE. Plotted are the ratios of $D_\ell$ at various array-bands to $D_\ell$ evaluated at the PA5 f150 array-band. Solid lines show input spectra to the simulations (\texttt{AGORA} for CIB and radio and \texttt{PySM} for dust) normalized by the PA5 f150 simulation input. Dashed lines show the recovered output foreground models, evaluated at the MAP, normalized by the PA5 f150 MAP output. We note that there is no free SED parameter in the DR6 foreground model for dust.\\ } 
    \label{fig:freq_scaling}
\end{figure*}

We note that, formally, our best-fitting model is not actually a good fit to the simulations, with the total CMB + foreground power spectra being mildly underestimated at some frequencies and overestimated at others. However, we first note that only the MAP is shown, while the actual predictions can have significant scatter. Moreover, as shown in L25, the model fits the actual DR6 data very well, indicating that the worse fit here is likely due to the simulations rather than the model itself. The worse fit to the simulations can be attributed to various differences between the DR6 foreground models and the components of the simulations. First, various foreground fields used in the simulations have nontrivial SEDs that differ from the fiducial models, and may even vary as functions of $\ell$ in the case of the dust field. Figure~\ref{fig:freq_scaling} shows the frequency scaling of some of the foreground fields in the simulations, versus the inferred frequency scaling based on the output foreground models evaluated at the MAP. There is also (a small amount of) decorrelation across frequencies for the CIB, radio, and dust components, which is not accounted for in the foreground model (though tests from \S \ref{sec:fg_tests} show that including parameters for these decorrelations results in negligible shifts in cosmological parameters). For the CIB input to the simulations, the decorrelation is only on the order of half a percent. For radio, it is around 2-6\%. For dust, it is around 2-9\% in TT and 1-2\% in EE. We note that these decorrelation percentages are only for the ACT frequencies from 90 to 220 GHz and may be larger for experiments with larger frequency coverage. Here the largest decorrelation is between the 90 and 220 GHz channels. Overall, it is difficult to find an MBB (power-law) that is a good fit to the CIB (radio emission) at all array-band pairs.

There are a few interesting features in Figure~\ref{fig:ng_sim_comp_ps}. First, the DR6 model does not include foreground models for the reionization kSZ effect, AME, or synchrotron emission, and thus, dashed output lines are not shown for those components. For the late-time kSZ effect, the MAP of the model has $a_{\rm kSZ}=0$, and so no dashed output line is shown for that component either. However, we note the full posterior distribution on $a_{\rm kSZ}$ encompasses a range of nonzero values. 

Additionally, the overall inferred tSZ template lies above the input tSZ power spectrum. At low $\ell$, the CMB is so dominant that there is very little sensitivity to the tSZ effect. At scales beyond $\ell \approx 3000$, we have more sensitivity to the tSZ  effect, and in this region, the inferred model is a much better match to the input tSZ power spectrum. 

Another feature is the tSZ--CIB cross-correlation. In the \verb|AGORA| simulations, this cross-correlation changes sign at high $\ell$. Since this sign change is not physically motivated, the DR6 template does not account for such a change, and the recovered output differs from the input power spectrum. 

In the TE power spectrum, the CMB is dominant, with small contributions from polarized radio sources and dust and synchrotron emission. Finally, in the EE power spectrum, the MAP of our model has $a_{\rm ps}^{\rm EE}=0$ (the radio point source EE amplitude is not detected), so no dashed output line is shown for the EE radio component. As with the late-time kSZ effect, the full posterior distribution, however, encompasses a range of nonzero values for this amplitude. The actual input spectrum to our simulations has $a_{\rm ps}^{\rm EE}=0.03$, in the tail of the recovered distribution from the full fit. The EE point source amplitude in these simulations is thus not detectable at our current sensitivity. This is consistent with the non-detection in the real DR6 data, where the 95\% upper limit is 0.04. Thus, if the true EE point source power is at the \verb|AGORA|-predicted level, then it is likely that upcoming experiments like the Simons Observatory can detect it.\footnote{However, the point sources may also be suppressed via masking.}


\section{Conclusion}
\label{sec.conclusion}
In this work, we have described in greater detail the baseline ACT DR6 foreground model from L25, detailing the DR6 models for Galactic dust, the tSZ and kSZ effect, the CIB, radio galaxy emission, and cross-correlations between these components. We have summarized the L25 constraints on the model parameters from the ACT DR6 multi-frequency observations and further checked that the model (where meaningful comparisons are possible) is consistent with other data from \emph{Planck} and SPT. In particular, from our constraint on the kSZ power spectrum, we derived conservative limits on the duration of reionization of $\Delta z_{\rm rei} < 4.4$, but with some dependence on the assumed median reionization redshift and on the other foreground templates. We have further described tests from L25 on an extensive suite of foreground modifications, extending the baseline model with additional free parameters, new or modified foreground templates, and modified fiducial values. In all cases, the ACT DR6 cosmological parameter constraints shift by less than $0.5\sigma$, and typically significantly less. Finally, we have constructed realistic non-Gaussian microwave sky simulations, with correlations amongst the various fields. We have shown that we recover the input cosmological parameters of the CMB component within $1\sigma$ in most cases, with the largest deviation at $1.9\sigma$ (noting that the distribution of deviations is consistent with results expected from a single simulation, and indeed only one foreground realization is available, and that the same noise realization is used in each simulation).

From a foreground cleaning perspective, the current models describe the DR6 sky acceptably well, but in L25 and further described here, we have seen that the sensitivity of the DR6 data has necessitated adding the $\alpha_{\rm tSZ}$ parameter, which describes the scale dependence of the tSZ power spectrum. Future high-sensitivity experiments like the Simons Observatory~\citep{2019JCAP...02..056A,SimonsObservatory:2025wwn} may thus require even more complexity in features such as the CIB SED and the shape of the tSZ template in order to obtain unbiased cosmological parameters from the CMB.

Although the DR6 foreground model is sufficient for the goal of recovering cosmological parameters, Figure~\ref{fig:fg_from_fg_tests} shows that foreground parameter constraints can vary non-negligibly depending on the specific foreground model used. Thus, astrophysical interpretation of the foreground parameters must be done with care. For example, in some cases the kSZ power spectrum is detected at non-zero values, while in other cases it is only an upper bound. Similarly, the tentative evidence for a steepening of the tSZ spectrum slope at small scales warrants further investigation and future observations.

This work has primarily focused on foreground cleaning for the primary CMB, but there is also rich science to be done with the component foreground power spectra \citep[e.g.,][]{2022ApJ...929..166H, Efstathiou:2025ckq}. As an example, we have used our kSZ limits to obtain a limit on the duration of reionization. We expect significant gains in the signal-to-noise ratio for these measurements in SO \citep{SimonsObservatory:2025wwn}, CMB-S4 \citep{CMB_S4_Science_Book}, CCAT~\citep{CCAT-Prime:2021lly}, and CMB-HD \citep{CMBHD}. With these future datasets, we anticipate that degeneracies can be further broken, hopefully yielding a high signal-to-noise ratio detection of the kSZ power spectrum and improved constraints on other foreground parameters.

\section{Acknowledgments}
We thank Yuuki Omori for generation of and ongoing help with the \verb|AGORA| simulations. We thank Chris Cain for discussion around the kSZ-based inference of the duration of reionization in~\S\ref{sec:ksz}. Several software tools were used in the development and presentation of results shown in this paper, including \verb|HEALPix/healpy| \citep{Healpix, Healpy}, \verb|numpy| \citep{numpy}, \verb|scipy| \citep{scipy}, \verb|matplotlib| \citep{matplotlib}, and \verb|getdist| \citep{getdist}.

BB acknowledges funding from two Horizon 2020 ERC Starting Grants, CMBForward (PI: Erminia Calabrese, Grant agreement No 849169) and the SCIPOL project\footnote{\url{scipol.in2p3.fr}} (PI: Josquin Errard, Grant agreement No.~101044073). KS is supported by the National Science Foundation Graduate Research Fellowship Program under Grant No. DGE 2036197. JCH acknowledges support from NSF grant AST-2108536, the Sloan Foundation, and the Simons Foundation. ZA acknowledges support from NSF grant AST-2108126. EC and HJ acknowledge support from the Horizon 2020 ERC Starting Grant (Grant agreement No 849169). SG acknowledges support from STFC and  UKRI (grant numbers ST/W002892/1 and ST/X006360/1). SJG acknowledges support from NSF grant AST-2307727. This work was carried out in part at the Jet Propulsion Laboratory, California Institute of Technology, under a contract with the National Aeronautics and Space Administration. R.H. acknowledges funding through the NSERC Discovery Grant RGPIN-2025-06483. KM acknowledges support from the National Research Foundation of South Africa. This work was supported by a grant from the Simons Foundation (CCA 918271, PBL). NS acknowledges support from DOE award number DE-SC0025309. CS acknowledges support from the Agencia Nacional de Investigaci\'on y Desarrollo (ANID) through Basal project FB210003.

This research used resources of the Advanced Research Computing at Cardiff (ARCCA) as well as the National Energy Research Scientific Computing Center (NERSC), a U.S. Department of Energy Office of Science User Facility located at Lawrence Berkeley National Laboratory, operated under Contract No. DE-AC02-05CH11231. This research also used computing resources from Columbia University's Shared Research Computing Facility project, which is supported by NIH Research Facility Improvement Grant 1G20RR030893-01, and associated funds from the New York State Empire State Development, Division of Science Technology and Innovation (NYSTAR) Contract C090171, both awarded April 15, 2010. 

Support for ACT was through the U.S.~National Science Foundation through awards AST-0408698, AST-0965625, and AST-1440226 for the ACT project, as well as awards PHY-0355328, PHY-0855887 and PHY-1214379. Funding was also provided by Princeton University, the University of Pennsylvania, and a Canada Foundation for Innovation (CFI) award to UBC. ACT operated in the Parque Astron\'omico Atacama in northern Chile under the auspices of the Agencia Nacional de Investigaci\'on y Desarrollo (ANID). The development of multichroic detectors and lenses was supported by NASA grants NNX13AE56G and NNX14AB58G. Detector research at NIST was supported by the NIST Innovations in Measurement Science program. 

We thank the Republic of Chile for hosting ACT in the northern Atacama, and the local indigenous Licanantay communities whom we follow in observing and learning from the night sky.

\bibliographystyle{act_titles}
\bibliography{refs}

@misc{meta_CO_25,
  doi = {10.48550/ARXIV.2506.16028},
  url = {https://arxiv.org/abs/2506.16028},
  author = {Mehta,  Yogesh and Roy,  Anirban and Foreman,  Simon and van Engelen,  Alexander and Battaglia,  Nick},
  title = {The Modeling Landscape of Extragalactic CO in CMB Surveys},
  publisher = {arXiv},
  year = {2025},
  copyright = {Creative Commons Attribution 4.0 International}
}

@book{rybicki_1985,
  title = {Radiative Processes in Astrophysics},
  ISBN = {9783527618170},
  url = {http://dx.doi.org/10.1002/9783527618170},
  DOI = {10.1002/9783527618170},
  publisher = {Wiley},
  author = {Rybicki,  George B. and Lightman,  Alan P.},
  year = {1985},
  month = may 
}

@article{Tucci_2011,
  title = {High-frequency predictions for number counts and spectral properties of extragalactic radio sources. New evidence of a break at mm wavelengths in spectra of bright blazar sources},
  volume = {533},
  ISSN = {1432-0746},
  url = {http://dx.doi.org/10.1051/0004-6361/201116972},
  DOI = {10.1051/0004-6361/201116972},
  journal = {Astronomy \& Astrophysics},
  publisher = {EDP Sciences},
  author = {Tucci,  M. and Toffolatti,  L. and De Zotti,  G. and Martínez-González,  E.},
  year = {2011},
  month = aug,
  pages = {A57}
}

@misc{getdist_2019,
  doi = {10.48550/ARXIV.1910.13970},
  url = {https://arxiv.org/abs/1910.13970},
  author = {Lewis,  Antony},
  title = {GetDist: a Python package for analysing Monte Carlo samples},
  publisher = {arXiv},
  year = {2019},
  copyright = {arXiv.org perpetual,  non-exclusive license}
}

@article{Torrado_2021,
  title = {Cobaya: code for Bayesian analysis of hierarchical physical models},
  volume = {2021},
  ISSN = {1475-7516},
  url = {http://dx.doi.org/10.1088/1475-7516/2021/05/057},
  DOI = {10.1088/1475-7516/2021/05/057},
  number = {05},
  journal = {Journal of Cosmology and Astroparticle Physics},
  publisher = {IOP Publishing},
  author = {Torrado,  Jesús and Lewis,  Antony},
  year = {2021},
  month = may,
  pages = {057}
}

@article{Sharp_2010,
  title = {A MEASUREMENT OF ARCMINUTE ANISOTROPY IN THE COSMIC MICROWAVE BACKGROUND WITH THE SUNYAEV-ZEL’DOVICH ARRAY},
  volume = {713},
  ISSN = {1538-4357},
  url = {http://dx.doi.org/10.1088/0004-637X/713/1/82},
  DOI = {10.1088/0004-637x/713/1/82},
  number = {1},
  journal = {The Astrophysical Journal},
  publisher = {American Astronomical Society},
  author = {Sharp,  Matthew K. and Marrone,  Daniel P. and Carlstrom,  John E. and Culverhouse,  Thomas and Greer,  Christopher and Hawkins,  David and Hennessy,  Ryan and Joy,  Marshall and Lamb,  James W. and Leitch,  Erik M. and Loh,  Michael and Miller,  Amber and Mroczkowski,  Tony and Muchovej,  Stephen and Pryke,  Clem and Woody,  David},
  year = {2010},
  month = mar,
  pages = {82–89}
}

@article{Hall_2010,
  title = {ANGULAR POWER SPECTRA OF THE MILLIMETER-WAVELENGTH BACKGROUND LIGHT FROM DUSTY STAR-FORMING GALAXIES WITH THE SOUTH POLE TELESCOPE},
  volume = {718},
  ISSN = {1538-4357},
  url = {http://dx.doi.org/10.1088/0004-637X/718/2/632},
  DOI = {10.1088/0004-637x/718/2/632},
  number = {2},
  journal = {The Astrophysical Journal},
  publisher = {American Astronomical Society},
  author = {Hall,  N. R. and Keisler,  R. and Knox,  L. and Reichardt,  C. L. and Ade,  P. A. R. and Aird,  K. A. and Benson,  B. A. and Bleem,  L. E. and Carlstrom,  J. E. and Chang,  C. L. and Cho,  H.-M. and Crawford,  T. M. and Crites,  A. T. and de Haan,  T. and Dobbs,  M. A. and George,  E. M. and Halverson,  N. W. and Holder,  G. P. and Holzapfel,  W. L. and Hrubes,  J. D. and Joy,  M. and Lee,  A. T. and Leitch,  E. M. and Lueker,  M. and McMahon,  J. J. and Mehl,  J. and Meyer,  S. S. and Mohr,  J. J. and Montroy,  T. E. and Padin,  S. and Plagge,  T. and Pryke,  C. and Ruhl,  J. E. and Schaffer,  K. K. and Shaw,  L. and Shirokoff,  E. and Spieler,  H. G. and Stalder,  B. and Staniszewski,  Z. and Stark,  A. A. and Switzer,  E. R. and Vanderlinde,  K. and Vieira,  J. D. and Williamson,  R. and Zahn,  O.},
  year = {2010},
  month = jul,
  pages = {632–646}
}

@article{planck_2013_ix,
collaboration = {Planck},
  title = {Planck2013 results. IX. HFI spectral response},
  volume = {571},
  ISSN = {1432-0746},
  url = {http://dx.doi.org/10.1051/0004-6361/201321531},
  DOI = {10.1051/0004-6361/201321531},
  journal = {Astronomy \& Astrophysics},
  publisher = {EDP Sciences},
  author = {Ade,  P. A. R. and Aghanim,  N. and Armitage-Caplan,  C. and Arnaud,  M. and Ashdown,  M. and Atrio-Barandela,  F. and Aumont,  J. and Baccigalupi,  C. and Banday,  A. J. and Barreiro,  R. B. and Battaner,  E. and Benabed,  K. and Benoît,  A. and Benoit-Lévy,  A. and Bernard,  J.-P. and Bersanelli,  M. and Bielewicz,  P. and Bobin,  J. and Bock,  J. J. and Bond,  J. R. and Borrill,  J. and Bouchet,  F. R. and Boulanger,  F. and Bridges,  M. and Bucher,  M. and Burigana,  C. and Cardoso,  J.-F. and Catalano,  A. and Challinor,  A. and Chamballu,  A. and Chary,  R.-R. and Chen,  X. and Chiang,  H. C. and Chiang,  L.-Y and Christensen,  P. R. and Church,  S. and Clements,  D. L. and Colombi,  S. and Colombo,  L. P. L. and Combet,  C. and Comis,  B. and Couchot,  F. and Coulais,  A. and Crill,  B. P. and Curto,  A. and Cuttaia,  F. and Danese,  L. and Davies,  R. D. and de Bernardis,  P. and de Rosa,  A. and de Zotti,  G. and Delabrouille,  J. and Delouis,  J.-M. and Désert,  F.-X. and Dickinson,  C. and Diego,  J. M. and Dole,  H. and Donzelli,  S. and Doré,  O. and Douspis,  M. and Dupac,  X. and Efstathiou,  G. and Enßlin,  T. A. and Eriksen,  H. K. and Falgarone,  E. and Finelli,  F. and Forni,  O. and Frailis,  M. and Franceschi,  E. and Galeotta,  S. and Ganga,  K. and Giard,  M. and Giraud-Héraud,  Y. and González-Nuevo,  J. and Górski,  K. M. and Gratton,  S. and Gregorio,  A. and Gruppuso,  A. and Hansen,  F. K. and Hanson,  D. and Harrison,  D. and Henrot-Versillé,  S. and Hernández-Monteagudo,  C. and Herranz,  D. and Hildebrandt,  S. R. and Hivon,  E. and Hobson,  M. and Holmes,  W. A. and Hornstrup,  A. and Hovest,  W. and Huffenberger,  K. M. and Hurier,  G. and Jaffe,  A. H. and Jaffe,  T. R. and Jones,  W. C. and Juvela,  M. and Keih\"{a}nen,  E. and Keskitalo,  R. and Kisner,  T. S. and Kneissl,  R. and Knoche,  J. and Knox,  L. and Kunz,  M. and Kurki-Suonio,  H. and Lagache,  G. and Lamarre,  J.-M. and Lasenby,  A. and Laureijs,  R. J. and Lawrence,  C. R. and Leahy,  J. P. and Leonardi,  R. and Leroy,  C. and Lesgourgues,  J. and Liguori,  M. and Lilje,  P. B. and Linden-Vørnle,  M. and López-Caniego,  M. and Lubin,  P. M. and Macías-Pérez,  J. F. and Maffei,  B. and Mandolesi,  N. and Maris,  M. and Marshall,  D. J. and Martin,  P. G. and Martínez-González,  E. and Masi,  S. and Massardi,  M. and Matarrese,  S. and Matthai,  F. and Mazzotta,  P. and McGehee,  P. and Melchiorri,  A. and Mendes,  L. and Mennella,  A. and Migliaccio,  M. and Mitra,  S. and Miville-Desch\^enes,  M.-A. and Moneti,  A. and Montier,  L. and Morgante,  G. and Mortlock,  D. and Munshi,  D. and Murphy,  J. A. and Naselsky,  P. and Nati,  F. and Natoli,  P. and Netterfield,  C. B. and Nørgaard-Nielsen,  H. U. and North,  C. and Noviello,  F. and Novikov,  D. and Novikov,  I. and Osborne,  S. and Oxborrow,  C. A. and Paci,  F. and Pagano,  L. and Pajot,  F. and Paoletti,  D. and Pasian,  F. and Patanchon,  G. and Perdereau,  O. and Perotto,  L. and Perrotta,  F. and Piacentini,  F. and Piat,  M. and Pierpaoli,  E. and Pietrobon,  D. and Plaszczynski,  S. and Pointecouteau,  E. and Polenta,  G. and Ponthieu,  N. and Popa,  L. and Poutanen,  T. and Pratt,  G. W. and Prézeau,  G. and Prunet,  S. and Puget,  J.-L. and Rachen,  J. P. and Reinecke,  M. and Remazeilles,  M. and Renault,  C. and Ricciardi,  S. and Riller,  T. and Ristorcelli,  I. and Rocha,  G. and Rosset,  C. and Roudier,  G. and Rusholme,  B. and Santos,  D. and Savini,  G. and Scott,  D. and Shellard,  E. P. S. and Spencer,  L. D. and Starck,  J.-L. and Stolyarov,  V. and Stompor,  R. and Sudiwala,  R. and Sureau,  F. and Sutton,  D. and Suur-Uski,  A.-S. and Sygnet,  J.-F. and Tauber,  J. A. and Tavagnacco,  D. and Terenzi,  L. and Tomasi,  M. and Tristram,  M. and Tucci,  M. and Umana,  G. and Valenziano,  L. and Valiviita,  J. and Van Tent,  B. and Vielva,  P. and Villa,  F. and Vittorio,  N. and Wade,  L. A. and Wandelt,  B. D. and Yvon,  D. and Zacchei,  A. and Zonca,  A.},
  year = {2014},
  month = oct,
  pages = {A9},
  eprint = {1303.5070}
}

@misc{Cartis_18a,
  doi = {10.48550/ARXIV.1804.00154},
  url = {https://arxiv.org/abs/1804.00154},
  author = {Cartis,  Coralia and Fiala,  Jan and Marteau,  Benjamin and Roberts,  Lindon},
  title = {Improving the Flexibility and Robustness of Model-Based Derivative-Free Optimization Solvers},
  publisher = {arXiv},
  year = {2018},
  copyright = {arXiv.org perpetual,  non-exclusive license}
}

@misc{Cartis_18b,
  doi = {10.48550/ARXIV.1804.00154},
  url = {https://arxiv.org/abs/1804.00154},
  author = {Cartis,  Coralia and Fiala,  Jan and Marteau,  Benjamin and Roberts,  Lindon},
  title = {Improving the Flexibility and Robustness of Model-Based Derivative-Free Optimization Solvers},
  publisher = {arXiv},
  year = {2018},
  copyright = {arXiv.org perpetual,  non-exclusive license}
}

@article{McCarthy2014,
  title = {The thermal Sunyaev–Zel’dovich effect power spectrum in light of Planck},
  volume = {440},
  ISSN = {0035-8711},
  url = {http://dx.doi.org/10.1093/mnras/stu543},
  DOI = {10.1093/mnras/stu543},
  number = {4},
  journal = {Monthly Notices of the Royal Astronomical Society},
  publisher = {Oxford University Press (OUP)},
  author = {McCarthy,  I. G. and Le Brun,  A. M. C. and Schaye,  J. and Holder,  G. P.},
  year = {2014},
  month = apr,
  pages = {3645–3657}
}

@article{Maniyar_2023,
  title = {Extragalactic CO emission lines in the CMB experiments: A forgotten signal and a foreground},
  volume = {107},
  ISSN = {2470-0029},
  url = {http://dx.doi.org/10.1103/PhysRevD.107.123504},
  DOI = {10.1103/physrevd.107.123504},
  number = {12},
  journal = {Physical Review D},
  publisher = {American Physical Society (APS)},
  author = {Maniyar,  Abhishek S. and Gkogkou,  Athanasia and Coulton,  William R. and Li,  Zack and Lagache,  Guilaine and Pullen,  Anthony R.},
  year = {2023},
  month = jun 
}

@article{trac_templates_2011,
  title = {Templates for the Sunyaev--Zel'dovich Angular Power Spectrum},
  author = {Trac, Hy and Bode, Paul and Ostriker, Jeremiah P.},
  year = {2011},
  month = jan,
  journal = {The Astrophysical Journal},
  volume = {727},
  number = {2},
  pages = {94},
  publisher = {The American Astronomical Society},
  issn = {0004-637X},
  doi = {10.1088/0004-637X/727/2/94},
  urldate = {2025-03-16},
  langid = {english}
}

@article{Efstathiou_2012,
  title = {A simple empirically motivated template for the thermal Sunyaev-Zel’dovich effect: Sunyaev-Zel’dovich template},
  volume = {423},
  ISSN = {0035-8711},
  url = {http://dx.doi.org/10.1111/j.1365-2966.2012.21059.x},
  DOI = {10.1111/j.1365-2966.2012.21059.x},
  number = {3},
  journal = {Monthly Notices of the Royal Astronomical Society},
  publisher = {Oxford University Press (OUP)},
  author = {Efstathiou,  George and Migliaccio,  Marina},
  year = {2012},
  month = may,
  pages = {2492–2497}
}

@article{planck_xxx_2016,
collaboration = {Planck},
  title = {Planckintermediate results: XXX. The angular power spectrum of polarized dust emission at intermediate and high Galactic latitudes},
  volume = {586},
  ISSN = {1432-0746},
  url = {http://dx.doi.org/10.1051/0004-6361/201425034},
  DOI = {10.1051/0004-6361/201425034},
  journal = {Astronomy \& Astrophysics},
  publisher = {EDP Sciences},
  author = {Adam,  R. and Ade,  P. A. R. and Aghanim,  N. and Arnaud,  M. and Aumont,  J. and Baccigalupi,  C. and Banday,  A. J. and Barreiro,  R. B. and Bartlett,  J. G. and Bartolo,  N. and Battaner,  E. and Benabed,  K. and Benoit-Lévy,  A. and Bernard,  J.-P. and Bersanelli,  M. and Bielewicz,  P. and Bonaldi,  A. and Bonavera,  L. and Bond,  J. R. and Borrill,  J. and Bouchet,  F. R. and Boulanger,  F. and Bracco,  A. and Bucher,  M. and Burigana,  C. and Butler,  R. C. and Calabrese,  E. and Cardoso,  J.-F. and Catalano,  A. and Challinor,  A. and Chamballu,  A. and Chary,  R.-R. and Chiang,  H. C. and Christensen,  P. R. and Clements,  D. L. and Colombi,  S. and Colombo,  L. P. L. and Combet,  C. and Couchot,  F. and Coulais,  A. and Crill,  B. P. and Curto,  A. and Cuttaia,  F. and Danese,  L. and Davies,  R. D. and Davis,  R. J. and de Bernardis,  P. and de Zotti,  G. and Delabrouille,  J. and Delouis,  J.-M. and Désert,  F.-X. and Dickinson,  C. and Diego,  J. M. and Dolag,  K. and Dole,  H. and Donzelli,  S. and Doré,  O. and Douspis,  M. and Ducout,  A. and Dunkley,  J. and Dupac,  X. and Efstathiou,  G. and Elsner,  F. and Enßlin,  T. A. and Eriksen,  H. K. and Falgarone,  E. and Finelli,  F. and Forni,  O. and Frailis,  M. and Fraisse,  A. A. and Franceschi,  E. and Frejsel,  A. and Galeotta,  S. and Galli,  S. and Ganga,  K. and Ghosh,  T. and Giard,  M. and Giraud-Héraud,  Y. and Gjerløw,  E. and González-Nuevo,  J. and Górski,  K. M. and Gratton,  S. and Gregorio,  A. and Gruppuso,  A. and Guillet,  V. and Hansen,  F. K. and Hanson,  D. and Harrison,  D. L. and Helou,  G. and Henrot-Versillé,  S. and Hernández-Monteagudo,  C. and Herranz,  D. and Hivon,  E. and Hobson,  M. and Holmes,  W. A. and Huffenberger,  K. M. and Hurier,  G. and Jaffe,  A. H. and Jaffe,  T. R. and Jewell,  J. and Jones,  W. C. and Juvela,  M. and Keih\"{a}nen,  E. and Keskitalo,  R. and Kisner,  T. S. and Kneissl,  R. and Knoche,  J. and Knox,  L. and Krachmalnicoff,  N. and Kunz,  M. and Kurki-Suonio,  H. and Lagache,  G. and Lamarre,  J.-M. and Lasenby,  A. and Lattanzi,  M. and Lawrence,  C. R. and Leahy,  J. P. and Leonardi,  R. and Lesgourgues,  J. and Levrier,  F. and Liguori,  M. and Lilje,  P. B. and Linden-Vørnle,  M. and López-Caniego,  M. and Lubin,  P. M. and Macías-Pérez,  J. F. and Maffei,  B. and Maino,  D. and Mandolesi,  N. and Mangilli,  A. and Maris,  M. and Martin,  P. G. and Martínez-González,  E. and Masi,  S. and Matarrese,  S. and Mazzotta,  P. and Meinhold,  P. R. and Melchiorri,  A. and Mendes,  L. and Mennella,  A. and Migliaccio,  M. and Mitra,  S. and Miville-Desch\^enes,  M.-A. and Moneti,  A. and Montier,  L. and Morgante,  G. and Mortlock,  D. and Moss,  A. and Munshi,  D. and Murphy,  J. A. and Naselsky,  P. and Nati,  F. and Natoli,  P. and Netterfield,  C. B. and Nørgaard-Nielsen,  H. U. and Noviello,  F. and Novikov,  D. and Novikov,  I. and Pagano,  L. and Pajot,  F. and Paladini,  R. and Paoletti,  D. and Partridge,  B. and Pasian,  F. and Patanchon,  G. and Pearson,  T. J. and Perdereau,  O. and Perotto,  L. and Perrotta,  F. and Pettorino,  V. and Piacentini,  F. and Piat,  M. and Pierpaoli,  E. and Pietrobon,  D. and Plaszczynski,  S. and Pointecouteau,  E. and Polenta,  G. and Ponthieu,  N. and Popa,  L. and Pratt,  G. W. and Prunet,  S. and Puget,  J.-L. and Rachen,  J. P. and Reach,  W. T. and Rebolo,  R. and Remazeilles,  M. and Renault,  C. and Renzi,  A. and Ricciardi,  S. and Ristorcelli,  I. and Rocha,  G. and Rosset,  C. and Rossetti,  M. and Roudier,  G. and Rouillé d’Orfeuil,  B. and Rubiño-Martín,  J. A. and Rusholme,  B. and Sandri,  M. and Santos,  D. and Savelainen,  M. and Savini,  G. and Scott,  D. and Soler,  J. D. and Spencer,  L. D. and Stolyarov,  V. and Stompor,  R. and Sudiwala,  R. and Sunyaev,  R. and Sutton,  D. and Suur-Uski,  A.-S. and Sygnet,  J.-F. and Tauber,  J. A. and Terenzi,  L. and Toffolatti,  L. and Tomasi,  M. and Tristram,  M. and Tucci,  M. and Tuovinen,  J. and Valenziano,  L. and Valiviita,  J. and Van Tent,  B. and Vibert,  L. and Vielva,  P. and Villa,  F. and Wade,  L. A. and Wandelt,  B. D. and Watson,  R. and Wehus,  I. K. and White,  M. and White,  S. D. M. and Yvon,  D. and Zacchei,  A. and Zonca,  A.},
  year = {2016},
  month = feb,
  pages = {A133}
}

@article{Li_2023,
  title = {The Simons Observatory: a new open-source power spectrum pipeline applied to the Planck legacy data},
  volume = {2023},
  ISSN = {1475-7516},
  url = {http://dx.doi.org/10.1088/1475-7516/2023/09/048},
  DOI = {10.1088/1475-7516/2023/09/048},
  number = {09},
  journal = {Journal of Cosmology and Astroparticle Physics},
  publisher = {IOP Publishing},
  author = {Li,  Zack and Louis,  Thibaut and Calabrese,  Erminia and Jense,  Hidde and Alonso,  David and Atkins,  Zachary and Bond,  J. Richard and Choi,  Steve K. and Dunkley,  Jo and Fabbian,  Giulio and Garrido,  Xavier and H. Jaffe,  Andrew and Madhavacheril,  Mathew S. and Meerburg,  P. Daniel and Natale,  Umberto and Qu,  Frank J.},
  year = {2023},
  month = sep,
  pages = {048}
}

@article{pagano/etal:2020,
 adsurl = {https://ui.adsabs.harvard.edu/abs/2020A&A...635A..99P},
 archiveprefix = {arXiv},
 author = {{Pagano}, L. and {Delouis}, J. -M. and {Mottet}, S. and {Puget}, J. -L. and {Vibert}, L.},
 doi = {10.1051/0004-6361/201936630},
 eid = {A99},
 eprint = {1908.09856},
 journal = {\aap},
 month = {March},
 pages = {A99},
 primaryclass = {astro-ph.CO},
 title = {{Reionization optical depth determination from Planck HFI data with ten percent accuracy}},
 volume = {635},
 year = {2020}
}

@article{Lewis_1999,
      author         = "Lewis, Antony and Challinor, Anthony and Lasenby,
                        Anthony",
      title          = "{Efficient computation of CMB anisotropies in closed FRW
                        models}",
      journal        = "\apj",
      volume         = "538",
      year           = "2000",
      pages          = "473-476",
      doi            = "10.1086/309179",
      eprint         = "astro-ph/9911177",
      archivePrefix  = "arXiv",
      primaryClass   = "astro-ph",
      SLACcitation   = "%%CITATION = ASTRO-PH/9911177;%%"
}

@article{Jense_2025,
  title = {A complete framework for cosmological emulation and inference with CosmoPower},
  volume = {4},
  ISSN = {2752-8200},
  url = {http://dx.doi.org/10.1093/rasti/rzaf002},
  DOI = {10.1093/rasti/rzaf002},
  journal = {RAS Techniques and Instruments},
  publisher = {Oxford University Press (OUP)},
  author = {Jense,  H T and Harrison,  I and Calabrese,  E and Spurio Mancini,  A and Bolliet,  B and Dunkley,  J and Hill,  J C},
  year = {2025}
}

@article{Addison_2012,
  title = {Modelling the correlation between the thermal Sunyaev Zel’dovich effect and the cosmic infrared background: tSZ–CIB correlation},
  volume = {427},
  ISSN = {1365-2966},
  url = {http://dx.doi.org/10.1111/j.1365-2966.2012.21664.x},
  DOI = {10.1111/j.1365-2966.2012.21664.x},
  number = {2},
  journal = {Monthly Notices of the Royal Astronomical Society},
  publisher = {Oxford University Press (OUP)},
  author = {Addison,  G. E. and Dunkley,  J. and Spergel,  D. N.},
  year = {2012},
  month = nov,
  pages = {1741–1754}
}

@article{DeZotti_2009,
  title = {Radio and millimeter continuum surveys and their astrophysical implications},
  volume = {18},
  ISSN = {1432-0754},
  url = {http://dx.doi.org/10.1007/s00159-009-0026-0},
  DOI = {10.1007/s00159-009-0026-0},
  number = {1–2},
  journal = {The Astronomy and Astrophysics Review},
  publisher = {Springer Science and Business Media LLC},
  author = {De Zotti,  Gianfranco and Massardi,  Marcella and Negrello,  Mattia and Wall,  Jasper},
  year = {2009},
  month = oct,
  pages = {1–65}
}

@article{Battaglia_2010,
  title = {Simulations of the Sunyaev-Zel’dovich Power Spectrum with Active Galactic Nucleus Feedback},
  volume = {725},
  ISSN = {1538-4357},
  url = {http://dx.doi.org/10.1088/0004-637X/725/1/91},
  DOI = {10.1088/0004-637x/725/1/91},
  number = {1},
  journal = {The Astrophysical Journal},
  publisher = {American Astronomical Society},
  author = {Battaglia,  N. and Bond,  J. R. and Pfrommer,  C. and Sievers,  J. L. and Sijacki,  D.},
  year = {2010},
  month = nov,
  pages = {91–99}
}

@article{planck_xix_2015,
collaboration = {Planck},
  title = {Planck intermediate results. XIX. An overview of the polarized thermal emission from Galactic dust},
  volume = {576},
  ISSN = {1432-0746},
  url = {http://dx.doi.org/10.1051/0004-6361/201424082},
  DOI = {10.1051/0004-6361/201424082},
  journal = {Astronomy \& Astrophysics},
  publisher = {EDP Sciences},
  author = {{Planck Collaboraion XIX} and Ade,  P. A. R. and Aghanim,  N. and Alina,  D. and Alves,  M. I. R. and Armitage-Caplan,  C. and Arnaud,  M. and Arzoumanian,  D. and Ashdown,  M. and Atrio-Barandela,  F. and Aumont,  J. and Baccigalupi,  C. and Banday,  A. J. and Barreiro,  R. B. and Battaner,  E. and Benabed,  K. and Benoit-Lévy,  A. and Bernard,  J.-P. and Bersanelli,  M. and Bielewicz,  P. and Bock,  J. J. and Bond,  J. R. and Borrill,  J. and Bouchet,  F. R. and Boulanger,  F. and Bracco,  A. and Burigana,  C. and Butler,  R. C. and Cardoso,  J.-F. and Catalano,  A. and Chamballu,  A. and Chary,  R.-R. and Chiang,  H. C. and Christensen,  P. R. and Colombi,  S. and Colombo,  L. P. L. and Combet,  C. and Couchot,  F. and Coulais,  A. and Crill,  B. P. and Curto,  A. and Cuttaia,  F. and Danese,  L. and Davies,  R. D. and Davis,  R. J. and de Bernardis,  P. and de Gouveia Dal Pino,  E. M. and de Rosa,  A. and de Zotti,  G. and Delabrouille,  J. and Désert,  F.-X. and Dickinson,  C. and Diego,  J. M. and Donzelli,  S. and Doré,  O. and Douspis,  M. and Dunkley,  J. and Dupac,  X. and Efstathiou,  G. and Enßlin,  T. A. and Eriksen,  H. K. and Falgarone,  E. and Ferrière,  K. and Finelli,  F. and Forni,  O. and Frailis,  M. and Fraisse,  A. A. and Franceschi,  E. and Galeotta,  S. and Ganga,  K. and Ghosh,  T. and Giard,  M. and Giraud-Héraud,  Y. and González-Nuevo,  J. and Górski,  K. M. and Gregorio,  A. and Gruppuso,  A. and Guillet,  V. and Hansen,  F. K. and Harrison,  D. L. and Helou,  G. and Hernández-Monteagudo,  C. and Hildebrandt,  S. R. and Hivon,  E. and Hobson,  M. and Holmes,  W. A. and Hornstrup,  A. and Huffenberger,  K. M. and Jaffe,  A. H. and Jaffe,  T. R. and Jones,  W. C. and Juvela,  M. and Keih\"{a}nen,  E. and Keskitalo,  R. and Kisner,  T. S. and Kneissl,  R. and Knoche,  J. and Kunz,  M. and Kurki-Suonio,  H. and Lagache,  G. and L\"{a}hteenm\"{a}ki,  A. and Lamarre,  J.-M. and Lasenby,  A. and Lawrence,  C. R. and Leahy,  J. P. and Leonardi,  R. and Levrier,  F. and Liguori,  M. and Lilje,  P. B. and Linden-Vørnle,  M. and López-Caniego,  M. and Lubin,  P. M. and Macías-Pérez,  J. F. and Maffei,  B. and Magalhães,  A. M. and Maino,  D. and Mandolesi,  N. and Maris,  M. and Marshall,  D. J. and Martin,  P. G. and Martínez-González,  E. and Masi,  S. and Matarrese,  S. and Mazzotta,  P. and Melchiorri,  A. and Mendes,  L. and Mennella,  A. and Migliaccio,  M. and Miville-Desch\^enes,  M.-A. and Moneti,  A. and Montier,  L. and Morgante,  G. and Mortlock,  D. and Munshi,  D. and Murphy,  J. A. and Naselsky,  P. and Nati,  F. and Natoli,  P. and Netterfield,  C. B. and Noviello,  F. and Novikov,  D. and Novikov,  I. and Oxborrow,  C. A. and Pagano,  L. and Pajot,  F. and Paladini,  R. and Paoletti,  D. and Pasian,  F. and Pearson,  T. J. and Perdereau,  O. and Perotto,  L. and Perrotta,  F. and Piacentini,  F. and Piat,  M. and Pietrobon,  D. and Plaszczynski,  S. and Poidevin,  F. and Pointecouteau,  E. and Polenta,  G. and Popa,  L. and Pratt,  G. W. and Prunet,  S. and Puget,  J.-L. and Rachen,  J. P. and Reach,  W. T. and Rebolo,  R. and Reinecke,  M. and Remazeilles,  M. and Renault,  C. and Ricciardi,  S. and Riller,  T. and Ristorcelli,  I. and Rocha,  G. and Rosset,  C. and Roudier,  G. and Rubiño-Martín,  J. A. and Rusholme,  B. and Sandri,  M. and Savini,  G. and Scott,  D. and Spencer,  L. D. and Stolyarov,  V. and Stompor,  R. and Sudiwala,  R. and Sutton,  D. and Suur-Uski,  A.-S. and Sygnet,  J.-F. and Tauber,  J. A. and Terenzi,  L. and Toffolatti,  L. and Tomasi,  M. and Tristram,  M. and Tucci,  M. and Umana,  G. and Valenziano,  L. and Valiviita,  J. and Van Tent,  B. and Vielva,  P. and Villa,  F. and Wade,  L. A. and Wandelt,  B. D. and Zacchei,  A. and Zonca,  A.},
  year = {2015},
  month = apr,
  pages = {A104}
}

@article{Birkinshaw_1999,
  title = {The Sunyaev–Zel’dovich effect},
  volume = {310},
  ISSN = {0370-1573},
  url = {http://dx.doi.org/10.1016/S0370-1573(98)00080-5},
  DOI = {10.1016/s0370-1573(98)00080-5},
  number = {2–3},
  journal = {Physics Reports},
  publisher = {Elsevier BV},
  author = {Birkinshaw,  M},
  year = {1999},
  month = mar,
  pages = {97–195}
}

@article{Sunyaev_1970,
  title = {Small-scale fluctuations of relic radiation},
  volume = {7},
  ISSN = {1572-946X},
  url = {http://dx.doi.org/10.1007/BF00653471},
  DOI = {10.1007/bf00653471},
  number = {1},
  journal = {Astrophysics and Space Science},
  publisher = {Springer Science and Business Media LLC},
  author = {Sunyaev,  R. A. and Zeldovich,  Ya. B.},
  year = {1970},
  month = apr,
  pages = {3–19}
}

@ARTICLE{Giardiello_2024,
       author = {{Giardiello}, S. and {Duivenvoorden}, A.~J. and {Calabrese}, E. and {Galloni}, G. and {Hasselfield}, M. and {Hill}, J.~C. and {La Posta}, A. and {Louis}, T. and {Madhavacheril}, M. and {Pagano}, L.},
        title = "{Modeling beam chromaticity for high-resolution CMB analyses}",
      journal = {\prd},
         year = 2025,
        month = feb,
       volume = {111},
       number = {4},
          eid = {043502},
        pages = {043502},
          doi = {10.1103/PhysRevD.111.043502},
archivePrefix = {arXiv},
       eprint = {2411.10124},
 primaryClass = {astro-ph.CO},
    url = {https://arxiv.org/abs/2411.10124}
}

@article{Dunkley_2013,
  title = {The Atacama Cosmology Telescope: likelihood for small-scale CMB data},
  volume = {2013},
  ISSN = {1475-7516},
  url = {http://dx.doi.org/10.1088/1475-7516/2013/07/025},
  DOI = {10.1088/1475-7516/2013/07/025},
  number = {07},
  journal = {Journal of Cosmology and Astroparticle Physics},
  publisher = {IOP Publishing},
  author = {Dunkley,  J and Calabrese,  E and Sievers,  J and Addison,  G.E and Battaglia,  N and Battistelli,  E.S and Bond,  J.R and Das,  S and Devlin,  M.J and D\"{u}nner,  R and Fowler,  J.W and Gralla,  M and Hajian,  A and Halpern,  M and Hasselfield,  M and Hincks,  A.D and Hlozek,  R and Hughes,  J.P and Irwin,  K.D and Kosowsky,  A and Louis,  T and Marriage,  T.A and Marsden,  D and Menanteau,  F and Moodley,  K and Niemack,  M and Nolta,  M.R and Page,  L.A and Partridge,  B and Sehgal,  N and Spergel,  D.N and Staggs,  S.T and Switzer,  E.R and Trac,  H and Wollack,  E},
  year = {2013},
  month = jul,
  pages = {025–025}
}

@article{Battaglia_2013_kszpatchy,
	doi = {10.1088/0004-637x/776/2/83},
	url = {https://doi.org/10.1088\%2F0004-637x\%2F776\%2F2\%2F83},
	year = 2013,
	month = {oct},
	publisher = {American Astronomical Society},
	volume = {776},
	number = {2},
	pages = {83},  
	author = {N. Battaglia and A. Natarajan and H. Trac and R. Cen and A. Loeb},
	title = {Reionization on Large Scales. III. Predicions for Low-$\ell$ Cosmic Microwave Background Polarization and High-$\ell$ Kinetic Sunyaev–Zel'dovich Observables},
   journal = {The Astrophysical Journal}
}

@ARTICLE{battaglia_2013_model,
       author = {{Battaglia}, N. and {Trac}, H. and {Cen}, R. and {Loeb}, A.},
        title = "{Reionization on Large Scales. I. A Parametric Model Constructed from Radiation-hydrodynamic Simulations}",
      journal = {\apj},
         year = 2013,
        month = oct,
       volume = {776},
       number = {2},
          eid = {81},
        pages = {81},
          doi = {10.1088/0004-637X/776/2/81},
archivePrefix = {arXiv},
       eprint = {1211.2821},
 primaryClass = {astro-ph.CO},
       adsurl = {https://ui.adsabs.harvard.edu/abs/2013ApJ...776...81B}
}

@article{Zonca_2021,
  doi = {10.21105/joss.03783},
  url = {https://doi.org/10.21105/joss.03783},
  year = {2021},
  publisher = {The Open Journal},
  volume = {6},
  number = {67},
  pages = {3783},
  author = {Andrea Zonca and Ben Thorne and Nicoletta Krachmalnicoff and Julian Borrill},
  title = {{The Python Sky Model 3 software}},
  journal = {Journal of Open Source Software}
}

@article{Thorne_2017,
   title={{The Python Sky Model: software for simulating the Galactic microwave sky}},
   volume={469},
   ISSN={1365-2966},
   url={http://dx.doi.org/10.1093/mnras/stx949},
   DOI={10.1093/mnras/stx949},
   number={3},
   journal={Monthly Notices of the Royal Astronomical Society},
   publisher={Oxford University Press (OUP)},
   author={Thorne, B. and Dunkley, J. and Alonso, D. and Næss, S.},
   year={2017},
   month={May},
   pages={2821–2833}
}

@ARTICLE{Omori:2022uox,
       author = {{Omori}, Yuuki},
        title = "{AGORA: Multicomponent simulation for cross-survey science}",
      journal = {\mnras},
         year = 2024,
        month = jun,
       volume = {530},
       number = {4},
        pages = {5030-5068},
          doi = {10.1093/mnras/stae1031},
archivePrefix = {arXiv},
       eprint = {2212.07420},
 primaryClass = {astro-ph.CO},
    url  = {https://arxiv.org/abs/2212.07420}
}

@article{planck_poldust:2018,
collaboration = {Planck},
  title = {Planck2018 results: XI. Polarized dust foregrounds},
  volume = {641},
  ISSN = {1432-0746},
  url = {http://dx.doi.org/10.1051/0004-6361/201832618},
  DOI = {10.1051/0004-6361/201832618},
  journal = {Astronomy \& Astrophysics},
  publisher = {EDP Sciences},
  author = {Akrami,  Y. and Ashdown,  M. and Aumont,  J. and Baccigalupi,  C. and Ballardini,  M. and Banday,  A. J. and Barreiro,  R. B. and Bartolo,  N. and Basak,  S. and Benabed,  K. and Bernard,  J.-P. and Bersanelli,  M. and Bielewicz,  P. and Bond,  J. R. and Borrill,  J. and Bouchet,  F. R. and Boulanger,  F. and Bracco,  A. and Bucher,  M. and Burigana,  C. and Calabrese,  E. and Cardoso,  J.-F. and Carron,  J. and Chiang,  H. C. and Combet,  C. and Crill,  B. P. and de Bernardis,  P. and de Zotti,  G. and Delabrouille,  J. and Delouis,  J.-M. and Di Valentino,  E. and Dickinson,  C. and Diego,  J. M. and Ducout,  A. and Dupac,  X. and Efstathiou,  G. and Elsner,  F. and Enßlin,  T. A. and Falgarone,  E. and Fantaye,  Y. and Ferrière,  K. and Finelli,  F. and Forastieri,  F. and Frailis,  M. and Fraisse,  A. A. and Franceschi,  E. and Frolov,  A. and Galeotta,  S. and Galli,  S. and Ganga,  K. and Génova-Santos,  R. T. and Ghosh,  T. and González-Nuevo,  J. and Górski,  K. M. and Gruppuso,  A. and Gudmundsson,  J. E. and Guillet,  V. and Handley,  W. and Hansen,  F. K. and Herranz,  D. and Huang,  Z. and Jaffe,  A. H. and Jones,  W. C. and Keih\"{a}nen,  E. and Keskitalo,  R. and Kiiveri,  K. and Kim,  J. and Krachmalnicoff,  N. and Kunz,  M. and Kurki-Suonio,  H. and Lamarre,  J.-M. and Lasenby,  A. and Le Jeune,  M. and Levrier,  F. and Liguori,  M. and Lilje,  P. B. and Lindholm,  V. and López-Caniego,  M. and Lubin,  P. M. and Ma,  Y.-Z. and Macías-Pérez,  J. F. and Maggio,  G. and Maino,  D. and Mandolesi,  N. and Mangilli,  A. and Martin,  P. G. and Martínez-González,  E. and Matarrese,  S. and McEwen,  J. D. and Meinhold,  P. R. and Melchiorri,  A. and Migliaccio,  M. and Miville-Desch\^enes,  M.-A. and Molinari,  D. and Moneti,  A. and Montier,  L. and Morgante,  G. and Natoli,  P. and Pagano,  L. and Paoletti,  D. and Pettorino,  V. and Piacentini,  F. and Polenta,  G. and Puget,  J.-L. and Rachen,  J. P. and Reinecke,  M. and Remazeilles,  M. and Renzi,  A. and Rocha,  G. and Rosset,  C. and Roudier,  G. and Rubiño-Martín,  J. A. and Ruiz-Granados,  B. and Salvati,  L. and Sandri,  M. and Savelainen,  M. and Scott,  D. and Soler,  J. D. and Spencer,  L. D. and Tauber,  J. A. and Tavagnacco,  D. and Toffolatti,  L. and Tomasi,  M. and Trombetti,  T. and Valiviita,  J. and Vansyngel,  F. and Van Tent,  B. and Vielva,  P. and Villa,  F. and Vittorio,  N. and Wehus,  I. K. and Zacchei,  A. and Zonca,  A.},
  year = {2020},
  month = sep,
  pages = {A11}
}

@ARTICLE{Haslam,
       author = {{Haslam}, C.~G.~T. and {Salter}, C.~J. and {Stoffel}, H. and {Wilson}, W.~E.},
        title = "{A 408-MHZ All-Sky Continuum Survey. II. The Atlas of Contour Maps}",
      journal = {\aaps},
         year = 1982,
        month = jan,
       volume = {47},
        pages = {1},
       adsurl = {https://ui.adsabs.harvard.edu/abs/1982A&AS...47....1H}
}

@article{WMAP:2007,
    author = "Hinshaw, G. and others",
    collaboration = "WMAP",
    title = "{Three-year Wilkinson Microwave Anisotropy Probe (WMAP) observations: temperature analysis}",
    eprint = "astro-ph/0603451",
    archivePrefix = "arXiv",
    doi = "10.1086/513698",
    journal = "Astrophys. J. Suppl.",
    volume = "170",
    pages = "288",
    year = "2007",
    url = "https://arxiv.org/abs/astro-ph/0603451"
}

@article{Planck:2016_dust,
    author = "Aghanim, N. and others",
    collaboration = "Planck",
    title = "{Planck intermediate results. XLVIII. Disentangling Galactic dust emission and cosmic infrared background anisotropies}",
    eprint = "1605.09387",
    archivePrefix = "arXiv",
    primaryClass = "astro-ph.CO",
    doi = "10.1051/0004-6361/201629022",
    journal = "Astron. Astrophys.",
    volume = "596",
    pages = "A109",
    year = "2016",
    url = "https://arxiv.org/abs/1605.09387"
}

@ARTICLE{Bennett2013,
       author = {{Bennett}, C.~L. and others},
        title = "{Nine-year Wilkinson Microwave Anisotropy Probe (WMAP) Observations: Final Maps and Results}",
      journal = {\apjs},
         year = 2013,
        month = oct,
       volume = {208},
       number = {2},
          eid = {20},
        pages = {20},
          doi = {10.1088/0067-0049/208/2/20},
archivePrefix = {arXiv},
       eprint = {1212.5225},
 primaryClass = {astro-ph.CO},
       adsurl = {https://ui.adsabs.harvard.edu/abs/2013ApJS..208...20B}
}

@article{Remazeilles:2014mba,
    author = "Remazeilles, M. and Dickinson, C. and Banday, A. J. and Bigot-Sazy, M. -A. and Ghosh, T.",
    title = "{An improved source-subtracted and destriped 408 MHz all-sky map}",
    eprint = "1411.3628",
    archivePrefix = "arXiv",
    primaryClass = "astro-ph.IM",
    doi = "10.1093/mnras/stv1274",
    journal = "Mon. Not. Roy. Astron. Soc.",
    volume = "451",
    number = "4",
    pages = "4311--4327",
    year = "2015"
}

@article{Planck:2015mvg,
collaboration = {Planck},
    author = "Adam, R. and others",
    collaboration = "Planck",
    title = "{Planck 2015 results. X. Diffuse component separation: Foreground maps}",
    eprint = "1502.01588",
    archivePrefix = "arXiv",
    primaryClass = "astro-ph.CO",
    doi = "10.1051/0004-6361/201525967",
    journal = "Astron. Astrophys.",
    volume = "594",
    pages = "A10",
    year = "2016"
}

@article{Ali-Haimoud:2008kyz,
    author = "Ali-Haimoud, Y. and Hirata, C. M. and Dickinson, C.",
    title = "{A refined model for spinning dust radiation}",
    eprint = "0812.2904",
    archivePrefix = "arXiv",
    primaryClass = "astro-ph",
    doi = "10.1111/j.1365-2966.2009.14599.x",
    journal = "Mon. Not. Roy. Astron. Soc.",
    volume = "395",
    pages = "1055",
    year = "2009"
}

@ARTICLE{Giannantonio2016,
       author = {{Giannantonio}, T. and others},
        title = "{CMB lensing tomography with the DES Science Verification galaxies}",
      journal = {\mnras},
         year = 2016,
        month = mar,
       volume = {456},
       number = {3},
        pages = {3213-3244},
          doi = {10.1093/mnras/stv2678},
archivePrefix = {arXiv},
       eprint = {1507.05551},
 primaryClass = {astro-ph.CO},
       adsurl = {https://ui.adsabs.harvard.edu/abs/2016MNRAS.456.3213G}
}

@software{lenspix,
       author = {{Lewis}, Antony},
        title = "{LensPix: Fast MPI full sky transforms for HEALPix}",
 howpublished = {Astrophysics Source Code Library, record ascl:1102.025},
         year = 2011,
        month = feb,
          eid = {ascl:1102.025},
       adsurl = {https://ui.adsabs.harvard.edu/abs/2011ascl.soft02025L}
}

@ARTICLE{Mead:2020vgs,
       author = {{Mead}, A.~J. and {Brieden}, S. and {Tr{\"o}ster}, T. and {Heymans}, C.},
        title = "{HMCODE-2020: improved modelling of non-linear cosmological power spectra with baryonic feedback}",
      journal = {\mnras},
         year = 2021,
        month = mar,
       volume = {502},
       number = {1},
        pages = {1401-1422},
          doi = {10.1093/mnras/stab082},
archivePrefix = {arXiv},
       eprint = {2009.01858},
 primaryClass = {astro-ph.CO},
       adsurl = {https://ui.adsabs.harvard.edu/abs/2021MNRAS.502.1401M}}

@article{Mccarthy:2017yqf,
    author = "McCarthy, Ian G. and Bird, Simeon and Schaye, Joop and Harnois-Deraps, Joachim and Font, Andreea S. and Van Waerbeke, Ludovic",
    title = "{The BAHAMAS project: the CMB\textendash{}large-scale structure tension and the roles of massive neutrinos and galaxy formation}",
    eprint = "1712.02411",
    archivePrefix = "arXiv",
    primaryClass = "astro-ph.CO",
    doi = "10.1093/mnras/sty377",
    journal = "Mon. Not. Roy. Astron. Soc.",
    volume = "476",
    number = "3",
    pages = "2999--3030",
    year = "2018"
}

@ARTICLE{universemachine,
       author = {{Behroozi}, Peter and {Wechsler}, Risa H. and {Hearin}, Andrew P. and {Conroy}, Charlie},
        title = "{UNIVERSEMACHINE: The correlation between galaxy growth and dark matter halo assembly from z = 0-10}",
      journal = {\mnras},
         year = 2019,
        month = sep,
       volume = {488},
       number = {3},
        pages = {3143-3194},
          doi = {10.1093/mnras/stz1182},
archivePrefix = {arXiv},
       eprint = {1806.07893},
 primaryClass = {astro-ph.GA},
        url = {https://arxiv.org/abs/1806.07893}
}

@article{Miville-Deschenes:2008lza,
    author = "Miville-Deschenes, M. -A. and Ysard, N. and Lavabre, A. and Ponthieu, N. and Macias-Perez, J. F. and Aumont, J. and Bernard, J. P.",
    title = "{Separation of anomalous and synchrotron emissions using WMAP polarization data}",
    eprint = "0802.3345",
    archivePrefix = "arXiv",
    primaryClass = "astro-ph",
    doi = "10.1051/0004-6361:200809484",
    journal = "Astron. Astrophys.",
    volume = "490",
    pages = "1093",
    year = "2008",
    url = "https://arxiv.org/abs/0802.3345"
}

@article{Lenz:2019ugy,
    author = "Lenz, Daniel and Dor\'e, Olivier and Lagache, Guilaine",
    title = "{Large-scale Maps of the Cosmic Infrared Background from Planck}",
    eprint = "1905.00426",
    archivePrefix = "arXiv",
    primaryClass = "astro-ph.CO",
    doi = "10.3847/1538-4357/ab3c2b",
    journal = "Astrophys. J.",
    volume = "883",
    number = "1",
    pages = "75",
    year = "2019"
}

@article{Healpy,
  doi = {10.21105/joss.01298},
  url = {https://doi.org/10.21105/joss.01298},
  year = {2019},
  month = mar,
  publisher = {The Open Journal},
  volume = {4},
  number = {35},
  pages = {1298},
  author = {Andrea Zonca and Leo Singer and Daniel Lenz and Martin Reinecke and Cyrille Rosset and Eric Hivon and Krzysztof Gorski},
  title = {healpy: equal area pixelization and spherical harmonics transforms for data on the sphere in Python},
  journal = {Journal of Open Source Software}
}

@ARTICLE{Healpix,
   author = {{G{\'o}rski}, K.~M. and {Hivon}, E. and {Banday}, A.~J. and 
	{Wandelt}, B.~D. and {Hansen}, F.~K. and {Reinecke}, M. and 
	{Bartelmann}, M.},
    title = "{HEALPix: A Framework for High-Resolution Discretization and Fast Analysis of Data Distributed on the Sphere}",
  journal = {\apj},
   eprint = {astro-ph/0409513},
     year = 2005,
    month = apr,
   volume = 622,
    pages = {759-771},
      doi = {10.1086/427976},
   adsurl = {http://adsabs.harvard.edu/abs/2005ApJ...622..759G},
  url = {https://arxiv.org/abs/astro-ph/0409513}
}

@article{Atkins:2023yzu,
    author = "Atkins, Zachary and others",
    title = "{The Atacama Cosmology Telescope: map-based noise simulations for DR6}",
    eprint = "2303.04180",
    archivePrefix = "arXiv",
    primaryClass = "astro-ph.CO",
    doi = "10.1088/1475-7516/2023/11/073",
    journal = "JCAP",
    volume = "11",
    pages = "073",
    year = "2023"
}

@ARTICLE{beams_inprep,
       author = {{Duivenvoorden}, A. and others},
         year = "in prep.",
}

@ARTICLE{passbands_inprep,
       author = {{Alford}, T. and others},
         year = "in prep.",
}

@ARTICLE{dr6_maps,
       author = {{Naess}, Sigurd and {Guan}, Yilun and {Duivenvoorden}, Adriaan J. and {Hasselfield}, Matthew and {Wang}, Yuhan and {Abril-Cabezas}, Irene and {Addison}, Graeme E. and {Ade}, Peter A.~R. and {Aiola}, Simone and {Alford}, Tommy and {Alonso}, David and {Amiri}, Mandana and {An}, Rui and {Atkins}, Zachary and {Austermann}, Jason E. and {Barbavara}, Eleonora and {Battaglia}, Nicholas and {Battistelli}, Elia Stefano and {Beall}, James A. and {Bean}, Rachel and {Beheshti}, Ali and {Beringue}, Benjamin and {Bhandarkar}, Tanay and {Biermann}, Emily and {Bolliet}, Boris and {Bond}, J Richard and {Calabrese}, Erminia and {Capalbo}, Valentina and {Carrero}, Felipe and {Chen}, Stephen and {Chesmore}, Grace and {Cho}, Hsiao-mei and {Choi}, Steve K. and {Clark}, Susan E. and {Cordova Rosado}, Rodrigo and {Cothard}, Nicholas F. and {Coughlin}, Kevin and {Coulton}, William and {Crichton}, Devin and {Crowley}, Kevin T. and {Devlin}, Mark J. and {Dicker}, Simon and {Duell}, Cody J. and {Duff}, Shannon M. and {Dunkley}, Jo and {Dunner}, Rolando and {Embil Villagra}, Carmen and {Fankhanel}, Max and {Farren}, Gerrit S. and {Ferraro}, Simone and {Foster}, Allen and {Freundt}, Rodrigo and {Fuzia}, Brittany and {Gallardo}, Patricio A. and {Garrido}, Xavier and {Giardiello}, Serena and {Gill}, Ajay and {Givans}, Jahmour and {Gluscevic}, Vera and {Golec}, Joseph E. and {Gong}, Yulin and {Halpern}, Mark and {Harrison}, Ian and {Healy}, Erin and {Henderson}, Shawn and {Hensley}, Brandon and {Herv{\'\i}as-Caimapo}, Carlos and {Hill}, J. Colin and {Hilton}, Gene C. and {Hilton}, Matt and {Hincks}, Adam D. and {Hlo{\v{z}}ek}, Ren{\'e}e and {Ho}, Shuay-Pwu Patty and {Hood}, John and {Hornecker}, Erika and {Huber}, Zachary B. and {Hubmayr}, Johannes and {Huffenberger}, Kevin M. and {Hughes}, John P. and {Ikape}, Margaret and {Irwin}, Kent and {Isopi}, Giovanni and {Jense}, Hidde T. and {Joshi}, Neha and {Keller}, Ben and {Kim}, Joshua and {Knowles}, Kenda and {Koopman}, Brian J. and {Kosowsky}, Arthur and {Kramer}, Darby and {Kusiak}, Aleksandra and {La Posta}, Adrien and {Lagu{\"e}}, Alex and {Lakey}, Victoria and {Lee}, Eunseong and {Li}, Yaqiong and {Li}, Zack and {Limon}, Michele and {Lokken}, Martine and {Louis}, Thibaut and {Lungu}, Marius and {MacCrann}, Niall and {MacInnis}, Amanda and {Madhavacheril}, Mathew S. and {Maldonado}, Diego and {Maldonado}, Felipe and {Mallaby-Kay}, Maya and {Marques}, Gabriela A. and {van Marrewijk}, Joshiwa and {McCarthy}, Fiona and {McMahon}, Jeff and {Mehta}, Yogesh and {Menanteau}, Felipe and {Moodley}, Kavilan and {Morris}, Thomas W. and {Mroczkowski}, Tony and {Namikawa}, Toshiya and {Nati}, Federico and {Nerval}, Simran K. and {Newburgh}, Laura and {Nicola}, Andrina and {Niemack}, Michael D. and {Nolta}, Michael R. and {Orlowski-Scherer}, John and {Page}, Lyman A. and {Pandey}, Shivam and {Partridge}, Bruce and {Perez Sarmiento}, Karen and {Prince}, Heather and {Puddu}, Roberto and {Qu}, Frank J. and {Quiroga}, Rodrigo and {Ragavan}, Damien C. and {Ried Guachalla}, Bernardita and {Rogers}, Keir K. and {Rojas}, Felipe and {Sakuma}, Tai and {Schaan}, Emmanuel and {Schmitt}, Benjamin L. and {Sehgal}, Neelima and {Shaikh}, Shabbir and {Sherwin}, Blake D. and {Sierra}, Carlos and {Sievers}, Jon and {Sif{\'o}n}, Crist{\'o}bal and {Simon}, Sara and {Sonka}, Rita and {Spergel}, David N. and {Staggs}, Suzanne T. and {Storer}, Emilie and {Surrao}, Kristen and {Switzer}, Eric R. and {Tampier}, Niklas and {Thornton}, Robert and {Trac}, Hy and {Tucker}, Carole and {Ullom}, Joel and {Vale}, Leila R. and {Van Engelen}, Alexander and {Van Lanen}, Jeff and {Vargas}, Cristian and {Vavagiakis}, Eve M. and {Wagoner}, Kasey and {Wenzl}, Lukas and {Wollack}, Edward J. and {Zheng}, Kaiwen},
        title = "{The Atacama Cosmology Telescope: DR6 Maps}",
      journal = {arXiv e-prints},
         year = 2025,
        month = mar,
          eid = {arXiv:2503.14451},
        pages = {arXiv:2503.14451},
          doi = {10.48550/arXiv.2503.14451},
archivePrefix = {arXiv},
       eprint = {2503.14451},
 primaryClass = {astro-ph.CO},
       adsurl = {https://ui.adsabs.harvard.edu/abs/2025arXiv250314451N}
}

@ARTICLE{dr6_lcdm,
       author = {{Louis}, Thibaut and {La Posta}, Adrien and {Atkins}, Zachary and {Jense}, Hidde T. and {Abril-Cabezas}, Irene and {Addison}, Graeme E. and {Ade}, Peter A.~R. and {Aiola}, Simone and {Alford}, Tommy and {Alonso}, David and {Amiri}, Mandana and {An}, Rui and {Austermann}, Jason E. and {Barbavara}, Eleonora and {Battaglia}, Nicholas and {Battistelli}, Elia Stefano and {Beall}, James A. and {Bean}, Rachel and {Beheshti}, Ali and {Beringue}, Benjamin and {Bhandarkar}, Tanay and {Biermann}, Emily and {Bolliet}, Boris and {Bond}, J Richard and {Calabrese}, Erminia and {Capalbo}, Valentina and {Carrero}, Felipe and {Chen}, Stephen and {Chesmore}, Grace and {Cho}, Hsiao-mei and {Choi}, Steve K. and {Clark}, Susan E. and {Cothard}, Nicholas F. and {Coughlin}, Kevin and {Coulton}, William and {Crichton}, Devin and {Crowley}, Kevin T. and {Darwish}, Omar and {Devlin}, Mark J. and {Dicker}, Simon and {Duell}, Cody J. and {Duff}, Shannon M. and {Duivenvoorden}, Adriaan J. and {Dunkley}, Jo and {Dunner}, Rolando and {Embil Villagra}, Carmen and {Fankhanel}, Max and {Farren}, Gerrit S. and {Ferraro}, Simone and {Foster}, Allen and {Freundt}, Rodrigo and {Fuzia}, Brittany and {Gallardo}, Patricio A. and {Garrido}, Xavier and {Gerbino}, Martina and {Giardiello}, Serena and {Gill}, Ajay and {Givans}, Jahmour and {Gluscevic}, Vera and {Goldstein}, Samuel and {Golec}, Joseph E. and {Gong}, Yulin and {Guan}, Yilun and {Halpern}, Mark and {Harrison}, Ian and {Hasselfield}, Matthew and {Healy}, Erin and {Henderson}, Shawn and {Hensley}, Brandon and {Herv{\'\i}as-Caimapo}, Carlos and {Hill}, J. Colin and {Hilton}, Gene C. and {Hilton}, Matt and {Hincks}, Adam D. and {Hlo{\v{z}}ek}, Ren{\'e}e and {Ho}, Shuay-Pwu Patty and {Hood}, John and {Hornecker}, Erika and {Huber}, Zachary B. and {Hubmayr}, Johannes and {Huffenberger}, Kevin M. and {Hughes}, John P. and {Ikape}, Margaret and {Irwin}, Kent and {Isopi}, Giovanni and {Joshi}, Neha and {Keller}, Ben and {Kim}, Joshua and {Knowles}, Kenda and {Koopman}, Brian J. and {Kosowsky}, Arthur and {Kramer}, Darby and {Kusiak}, Aleksandra and {Lague}, Alex and {Lakey}, Victoria and {Lee}, Eunseong and {Li}, Yaqiong and {Li}, Zack and {Limon}, Michele and {Lokken}, Martine and {Lungu}, Marius and {MacCrann}, Niall and {MacInnis}, Amanda and {Madhavacheril}, Mathew S. and {Maldonado}, Diego and {Maldonado}, Felipe and {Mallaby-Kay}, Maya and {Marques}, Gabriela A. and {van Marrewijk}, Joshiwa and {McCarthy}, Fiona and {McMahon}, Jeff and {Mehta}, Yogesh and {Menanteau}, Felipe and {Moodley}, Kavilan and {Morris}, Thomas W. and {Mroczkowski}, Tony and {Naess}, Sigurd and {Namikawa}, Toshiya and {Nati}, Federico and {Nerval}, Simran K. and {Newburgh}, Laura and {Nicola}, Andrina and {Niemack}, Michael D. and {Nolta}, Michael R. and {Orlowski-Scherer}, John and {Pagano}, Luca and {Page}, Lyman A. and {Pandey}, Shivam and {Partridge}, Bruce and {Perez Sarmiento}, Karen and {Prince}, Heather and {Puddu}, Roberto and {Qu}, Frank J. and {Ragavan}, Damien C. and {Ried Guachalla}, Bernardita and {Rogers}, Keir K. and {Rojas}, Felipe and {Sakuma}, Tai and {Schaan}, Emmanuel and {Schmitt}, Benjamin L. and {Sehgal}, Neelima and {Shaikh}, Shabbir and {Sherwin}, Blake D. and {Sierra}, Carlos and {Sievers}, Jon and {Sif{\'o}n}, Crist{\'o}bal and {Simon}, Sara and {Sonka}, Rita and {Spergel}, David N. and {Staggs}, Suzanne T. and {Storer}, Emilie and {Surrao}, Kristen and {Switzer}, Eric R. and {Tampier}, Niklas and {Thornton}, Robert and {Trac}, Hy and {Tucker}, Carole and {Ullom}, Joel and {Vale}, Leila R. and {Van Engelen}, Alexander and {Van Lanen}, Jeff and {Vargas}, Cristian and {Vavagiakis}, Eve M. and {Wagoner}, Kasey and {Wang}, Yuhan and {Wenzl}, Lukas and {Wollack}, Edward J. and {Zheng}, Kaiwen},
        title = "{The Atacama Cosmology Telescope: DR6 Power Spectra, Likelihoods and $\Lambda$CDM Parameters}",
      journal = {arXiv e-prints},
         year = 2025,
        month = mar,
          eid = {arXiv:2503.14452},
        pages = {arXiv:2503.14452},
          doi = {10.48550/arXiv.2503.14452},
archivePrefix = {arXiv},
       eprint = {2503.14452},
 primaryClass = {astro-ph.CO},
       adsurl = {https://ui.adsabs.harvard.edu/abs/2025arXiv250314452L}
}

@ARTICLE{dr6_extended,
       author = {{Calabrese}, Erminia and {Hill}, J. Colin and {Jense}, Hidde T. and {La Posta}, Adrien and {Abril-Cabezas}, Irene and {Addison}, Graeme E. and {Ade}, Peter A.~R. and {Aiola}, Simone and {Alford}, Tommy and {Alonso}, David and {Amiri}, Mandana and {An}, Rui and {Atkins}, Zachary and {Austermann}, Jason E. and {Barbavara}, Eleonora and {Barbieri}, Nicola and {Battaglia}, Nicholas and {Battistelli}, Elia Stefano and {Beall}, James A. and {Bean}, Rachel and {Beheshti}, Ali and {Beringue}, Benjamin and {Bhandarkar}, Tanay and {Biermann}, Emily and {Bolliet}, Boris and {Bond}, J Richard and {Capalbo}, Valentina and {Carrero}, Felipe and {Chen}, Stephen and {Chesmore}, Grace and {Cho}, Hsiao-mei and {Choi}, Steve K. and {Clark}, Susan E. and {Cothard}, Nicholas F. and {Coughlin}, Kevin and {Coulton}, William and {Crichton}, Devin and {Crowley}, Kevin T. and {Darwish}, Omar and {Devlin}, Mark J. and {Dicker}, Simon and {Duell}, Cody J. and {Duff}, Shannon M. and {Duivenvoorden}, Adriaan J. and {Dunkley}, Jo and {Dunner}, Rolando and {Embil Villagra}, Carmen and {Fankhanel}, Max and {Farren}, Gerrit S. and {Ferraro}, Simone and {Foster}, Allen and {Freundt}, Rodrigo and {Fuzia}, Brittany and {Gallardo}, Patricio A. and {Garrido}, Xavier and {Gerbino}, Martina and {Giardiello}, Serena and {Gill}, Ajay and {Givans}, Jahmour and {Gluscevic}, Vera and {Goldstein}, Samuel and {Golec}, Joseph E. and {Gong}, Yulin and {Guan}, Yilun and {Halpern}, Mark and {Harrison}, Ian and {Hasselfield}, Matthew and {He}, Adam and {Healy}, Erin and {Henderson}, Shawn and {Hensley}, Brandon and {Herv{\'\i}as-Caimapo}, Carlos and {Hilton}, Gene C. and {Hilton}, Matt and {Hincks}, Adam D. and {Hlo{\v{z}}ek}, Ren{\'e}e and {Ho}, Shuay-Pwu Patty and {Hood}, John and {Hornecker}, Erika and {Huber}, Zachary B. and {Hubmayr}, Johannes and {Huffenberger}, Kevin M. and {Hughes}, John P. and {Ikape}, Margaret and {Irwin}, Kent and {Isopi}, Giovanni and {Joshi}, Neha and {Keller}, Ben and {Kim}, Joshua and {Knowles}, Kenda and {Koopman}, Brian J. and {Kosowsky}, Arthur and {Kramer}, Darby and {Kusiak}, Aleksandra and {Lague}, Alex and {Lakey}, Victoria and {Lattanzi}, Massimiliano and {Lee}, Eunseong and {Li}, Yaqiong and {Li}, Zack and {Limon}, Michele and {Lokken}, Martine and {Louis}, Thibaut and {Lungu}, Marius and {MacCrann}, Niall and {MacInnis}, Amanda and {Madhavacheril}, Mathew S. and {Maldonado}, Diego and {Maldonado}, Felipe and {Mallaby-Kay}, Maya and {Marques}, Gabriela A. and {van Marrewijk}, Joshiwa and {McCarthy}, Fiona and {McMahon}, Jeff and {Mehta}, Yogesh and {Menanteau}, Felipe and {Moodley}, Kavilan and {Morris}, Thomas W. and {Mroczkowski}, Tony and {Naess}, Sigurd and {Namikawa}, Toshiya and {Nati}, Federico and {Nerval}, Simran K. and {Newburgh}, Laura and {Nicola}, Andrina and {Niemack}, Michael D. and {Nolta}, Michael R. and {Orlowski-Scherer}, John and {Pagano}, Luca and {Page}, Lyman A. and {Pandey}, Shivam and {Partridge}, Bruce and {Perez Sarmiento}, Karen and {Prince}, Heather and {Puddu}, Roberto and {Qu}, Frank J. and {Ragavan}, Damien C. and {Ried Guachalla}, Bernardita and {Rogers}, Keir K. and {Rojas}, Felipe and {Sakuma}, Tai and {Schaan}, Emmanuel and {Schmitt}, Benjamin L. and {Sehgal}, Neelima and {Shaikh}, Shabbir and {Sherwin}, Blake D. and {Sierra}, Carlos and {Sievers}, Jon and {Sif{\'o}n}, Crist{\'o}bal and {Simon}, Sara and {Sonka}, Rita and {Spergel}, David N. and {Staggs}, Suzanne T. and {Storer}, Emilie and {Surrao}, Kristen and {Switzer}, Eric R. and {Tampier}, Niklas and {Thiele}, Leander and {Thornton}, Robert and {Trac}, Hy and {Tucker}, Carole and {Ullom}, Joel and {Vale}, Leila R. and {Van Engelen}, Alexander and {Van Lanen}, Jeff and {Vargas}, Cristian and {Vavagiakis}, Eve M. and {Wagoner}, Kasey and {Wang}, Yuhan and {Wenzl}, Lukas and {Wollack}, Edward J. and {Zheng}, Kaiwen},
        title = "{The Atacama Cosmology Telescope: DR6 Constraints on Extended Cosmological Models}",
      journal = {arXiv e-prints},
         year = 2025,
        month = mar,
          eid = {arXiv:2503.14454},
        pages = {arXiv:2503.14454},
          doi = {10.48550/arXiv.2503.14454},
archivePrefix = {arXiv},
       eprint = {2503.14454},
 primaryClass = {astro-ph.CO},
       adsurl = {https://ui.adsabs.harvard.edu/abs/2025arXiv250314454C}
}

@ARTICLE{wmap_spergel_2003,
       author = {{Spergel}, D.~N. and {Verde}, L. and {Peiris}, H.~V. and {Komatsu}, E. and
         {Nolta}, M.~R. and {Bennett}, C.~L. and {Halpern}, M. and
         {Hinshaw}, G. and {Jarosik}, N. and {Kogut}, A. and {Limon}, M. and
         {Meyer}, S.~S. and {Page}, L. and {Tucker}, G.~S. and {Weiland}, J.~L. and
         {Wollack}, E. and {Wright}, E.~L.},
        title = "{First-Year Wilkinson Microwave Anisotropy Probe (WMAP) Observations: Determination of Cosmological Parameters}",
      journal = {\apjs},
         year = 2003,
        month = sep,
       volume = {148},
       number = {1},
        pages = {175-194},
          doi = {10.1086/377226},
archivePrefix = {arXiv},
       eprint = {astro-ph/0302209},
 primaryClass = {astro-ph},
       adsurl = {https://ui.adsabs.harvard.edu/abs/2003ApJS..148..175S}
}

@ARTICLE{Balkenhol_SPT,
       author = {{Balkenhol}, L. and {Dutcher}, D. and {Spurio Mancini}, A. and {Doussot}, A. and {Benabed}, K. and {Galli}, S. and {Ade}, P.~A.~R. and {Anderson}, A.~J. and {Ansarinejad}, B. and {Archipley}, M. and {Bender}, A.~N. and {Benson}, B.~A. and {Bianchini}, F. and {Bleem}, L.~E. and {Bouchet}, F.~R. and {Bryant}, L. and {Camphuis}, E. and {Carlstrom}, J.~E. and {Cecil}, T.~W. and {Chang}, C.~L. and {Chaubal}, P. and {Chichura}, P.~M. and {Chou}, T. -L. and {Coerver}, A. and {Crawford}, T.~M. and {Cukierman}, A. and {Daley}, C. and {de Haan}, T. and {Dibert}, K.~R. and {Dobbs}, M.~A. and {Everett}, W. and {Feng}, C. and {Ferguson}, K.~R. and {Foster}, A. and {Gambrel}, A.~E. and {Gardner}, R.~W. and {Goeckner-Wald}, N. and {Gualtieri}, R. and {Guidi}, F. and {Guns}, S. and {Halverson}, N.~W. and {Hivon}, E. and {Holder}, G.~P. and {Holzapfel}, W.~L. and {Hood}, J.~C. and {Huang}, N. and {Knox}, L. and {Korman}, M. and {Kuo}, C. -L. and {Lee}, A.~T. and {Lowitz}, A.~E. and {Lu}, C. and {Millea}, M. and {Montgomery}, J. and {Nakato}, Y. and {Natoli}, T. and {Noble}, G.~I. and {Novosad}, V. and {Omori}, Y. and {Padin}, S. and {Pan}, Z. and {Paschos}, P. and {Prabhu}, K. and {Quan}, W. and {Rahimi}, M. and {Rahlin}, A. and {Reichardt}, C.~L. and {Rouble}, M. and {Ruhl}, J.~E. and {Schiappucci}, E. and {Smecher}, G. and {Sobrin}, J.~A. and {Stark}, A.~A. and {Stephen}, J. and {Suzuki}, A. and {Tandoi}, C. and {Thompson}, K.~L. and {Thorne}, B. and {Tucker}, C. and {Umilta}, C. and {Vieira}, J.~D. and {Wang}, G. and {Whitehorn}, N. and {Wu}, W.~L.~K. and {Yefremenko}, V. and {Young}, M.~R. and {Zebrowski}, J.~A. and {SPT-3G Collaboration}},
        title = "{Measurement of the CMB temperature power spectrum and constraints on cosmology from the SPT-3G 2018 TT, TE, and EE dataset}",
      journal = {\prd},
         year = 2023,
        month = jul,
       volume = {108},
       number = {2},
          eid = {023510},
        pages = {023510},
          doi = {10.1103/PhysRevD.108.023510},
archivePrefix = {arXiv},
       eprint = {2212.05642},
 primaryClass = {astro-ph.CO},
       adsurl = {https://ui.adsabs.harvard.edu/abs/2023PhRvD.108b3510B}
}

@ARTICLE{Reichardt_2012,
       author = {{Reichardt}, C.~L. and {Shaw}, L. and {Zahn}, O. and {Aird}, K.~A. and {Benson}, B.~A. and {Bleem}, L.~E. and {Carlstrom}, J.~E. and {Chang}, C.~L. and {Cho}, H.~M. and {Crawford}, T.~M. and {Crites}, A.~T. and {de Haan}, T. and {Dobbs}, M.~A. and {Dudley}, J. and {George}, E.~M. and {Halverson}, N.~W. and {Holder}, G.~P. and {Holzapfel}, W.~L. and {Hoover}, S. and {Hou}, Z. and {Hrubes}, J.~D. and {Joy}, M. and {Keisler}, R. and {Knox}, L. and {Lee}, A.~T. and {Leitch}, E.~M. and {Lueker}, M. and {Luong-Van}, D. and {McMahon}, J.~J. and {Mehl}, J. and {Meyer}, S.~S. and {Millea}, M. and {Mohr}, J.~J. and {Montroy}, T.~E. and {Natoli}, T. and {Padin}, S. and {Plagge}, T. and {Pryke}, C. and {Ruhl}, J.~E. and {Schaffer}, K.~K. and {Shirokoff}, E. and {Spieler}, H.~G. and {Staniszewski}, Z. and {Stark}, A.~A. and {Story}, K. and {van Engelen}, A. and {Vanderlinde}, K. and {Vieira}, J.~D. and {Williamson}, R.},
        title = "{A Measurement of Secondary Cosmic Microwave Background Anisotropies with Two Years of South Pole Telescope Observations}",
      journal = {\apj},
         year = 2012,
        month = aug,
       volume = {755},
       number = {1},
          eid = {70},
        pages = {70},
          doi = {10.1088/0004-637X/755/1/70},
archivePrefix = {arXiv},
       eprint = {1111.0932},
 primaryClass = {astro-ph.CO},
       adsurl = {https://ui.adsabs.harvard.edu/abs/2012ApJ...755...70R}
}

@ARTICLE{choi_atacama_2020,
       author = {{Choi}, Steve K. and {Hasselfield}, Matthew and {Ho}, Shuay-Pwu Patty and {Koopman}, Brian and {Lungu}, Marius and {Abitbol}, Maximilian H. and {Addison}, Graeme E. and {Ade}, Peter A.~R. and {Aiola}, Simone and {Alonso}, David and {Amiri}, Mandana and {Amodeo}, Stefania and {Angile}, Elio and {Austermann}, Jason E. and {Baildon}, Taylor and {Battaglia}, Nick and {Beall}, James A. and {Bean}, Rachel and {Becker}, Daniel T. and {Bond}, J. Richard and {Bruno}, Sarah Marie and {Calabrese}, Erminia and {Calafut}, Victoria and {Campusano}, Luis E. and {Carrero}, Felipe and {Chesmore}, Grace E. and {Cho}, Hsiao-mei and {Clark}, Susan E. and {Cothard}, Nicholas F. and {Crichton}, Devin and {Crowley}, Kevin T. and {Darwish}, Omar and {Datta}, Rahul and {Denison}, Edward V. and {Devlin}, Mark J. and {Duell}, Cody J. and {Duff}, Shannon M. and {Duivenvoorden}, Adriaan J. and {Dunkley}, Jo and {D{\"u}nner}, Rolando and {Essinger-Hileman}, Thomas and {Fankhanel}, Max and {Ferraro}, Simone and {Fox}, Anna E. and {Fuzia}, Brittany and {Gallardo}, Patricio A. and {Gluscevic}, Vera and {Golec}, Joseph E. and {Grace}, Emily and {Gralla}, Megan and {Guan}, Yilun and {Hall}, Kirsten and {Halpern}, Mark and {Han}, Dongwon and {Hargrave}, Peter and {Henderson}, Shawn and {Hensley}, Brandon and {Hill}, J. Colin and {Hilton}, Gene C. and {Hilton}, Matt and {Hincks}, Adam D. and {Hlo{\v{z}}ek}, Ren{\'e}e and {Hubmayr}, Johannes and {Huffenberger}, Kevin M. and {Hughes}, John P. and {Infante}, Leopoldo and {Irwin}, Kent and {Jackson}, Rebecca and {Klein}, Jeff and {Knowles}, Kenda and {Kosowsky}, Arthur and {Lakey}, Vincent and {Li}, Dale and {Li}, Yaqiong and {Li}, Zack and {Lokken}, Martine and {Louis}, Thibaut and {MacInnis}, Amanda and {Madhavacheril}, Mathew and {Maldonado}, Felipe and {Mallaby-Kay}, Maya and {Marsden}, Danica and {Maurin}, Lo{\"\i}c and {McMahon}, Jeff and {Menanteau}, Felipe and {Moodley}, Kavilan and {Morton}, Tim and {Naess}, Sigurd and {Namikawa}, Toshiya and {Nati}, Federico and {Newburgh}, Laura and {Nibarger}, John P. and {Nicola}, Andrina and {Niemack}, Michael D. and {Nolta}, Michael R. and {Orlowski-Sherer}, John and {Page}, Lyman A. and {Pappas}, Christine G. and {Partridge}, Bruce and {Phakathi}, Phumlani and {Prince}, Heather and {Puddu}, Roberto and {Qu}, Frank J. and {Rivera}, Jesus and {Robertson}, Naomi and {Rojas}, Felipe and {Salatino}, Maria and {Schaan}, Emmanuel and {Schillaci}, Alessandro and {Schmitt}, Benjamin L. and {Sehgal}, Neelima and {Sherwin}, Blake D. and {Sierra}, Carlos and {Sievers}, Jon and {Sifon}, Cristobal and {Sikhosana}, Precious and {Simon}, Sara and {Spergel}, David N. and {Staggs}, Suzanne T. and {Stevens}, Jason and {Storer}, Emilie and {Sunder}, Dhaneshwar D. and {Switzer}, Eric R. and {Thorne}, Ben and {Thornton}, Robert and {Trac}, Hy and {Treu}, Jesse and {Tucker}, Carole and {Vale}, Leila R. and {Van Engelen}, Alexander and {Van Lanen}, Jeff and {Vavagiakis}, Eve M. and {Wagoner}, Kasey and {Wang}, Yuhan and {Ward}, Jonathan T. and {Wollack}, Edward J. and {Xu}, Zhilei and {Zago}, Fernando and {Zhu}, Ningfeng},
        title = "{The Atacama Cosmology Telescope: a measurement of the Cosmic Microwave Background power spectra at 98 and 150 GHz}",
      journal = {\jcap},
         year = 2020,
        month = dec,
       volume = {2020},
       number = {12},
          eid = {045},
        pages = {045},
          doi = {10.1088/1475-7516/2020/12/045},
archivePrefix = {arXiv},
       eprint = {2007.07289},
 primaryClass = {astro-ph.CO},
        url = {https://arxiv.org/abs/2007.07289}
}

@ARTICLE{Hivon2002,
       author = {{Hivon}, Eric and {G{\'o}rski}, Krzysztof M. and {Netterfield}, C. Barth and {Crill}, Brendan P. and {Prunet}, Simon and {Hansen}, Frode},
        title = "{MASTER of the Cosmic Microwave Background Anisotropy Power Spectrum: A Fast Method for Statistical Analysis of Large and Complex Cosmic Microwave Background Data Sets}",
      journal = {\apj},
         year = 2002,
        month = mar,
       volume = {567},
       number = {1},
        pages = {2-17},
          doi = {10.1086/338126},
archivePrefix = {arXiv},
       eprint = {astro-ph/0105302},
 primaryClass = {astro-ph},
       adsurl = {https://ui.adsabs.harvard.edu/abs/2002ApJ...567....2H}
}

@article{Atkins:2024jlo,
  title = {The Atacama Cosmology Telescope: semi-analytic covariance matrices for the DR6 CMB power spectra},
  volume = {2025},
  ISSN = {1475-7516},
  url = {http://dx.doi.org/10.1088/1475-7516/2025/05/015},
  DOI = {10.1088/1475-7516/2025/05/015},
  number = {05},
  journal = {Journal of Cosmology and Astroparticle Physics},
  publisher = {IOP Publishing},
  author = {Atkins,  Zachary and Li,  Zack and Alonso,  David and Richard Bond,  J. and Calabrese,  Erminia and Duivenvoorden,  Adriaan J. and Dunkley,  Jo and Giardiello,  Serena and Hervías-Caimapo,  Carlos and Colin Hill,  J. and Jense,  Hidde T. and Kim,  Joshua and Louis,  Thibaut and Moodley,  Kavilan and Morris,  Thomas W. and Naess,  Sigurd and Niemack,  Michael D. and Page,  Lyman and Posta,  Adrien La and Sifón,  Cristóbal and Wollack,  Edward J.},
  year = {2025},
  month = may,
  pages = {015}
}

@article{Naess:2020wgi,
    author = "Naess, Sigurd and others",
    title = "{The Atacama Cosmology Telescope: arcminute-resolution maps of 18 000 square degrees of the microwave sky from ACT 2008\textendash{}2018 data combined with Planck}",
    eprint = "2007.07290",
    archivePrefix = "arXiv",
    primaryClass = "astro-ph.IM",
    doi = "10.1088/1475-7516/2020/12/046",
    journal = "JCAP",
    volume = "12",
    pages = "046",
    year = "2020"
}

@article{Kokron:2024,
  title = {Contributions of extragalactic CO emission lines to ground-based CMB observations},
  volume = {110},
  ISSN = {2470-0029},
  url = {http://dx.doi.org/10.1103/PhysRevD.110.103535},
  DOI = {10.1103/physrevd.110.103535},
  number = {10},
  journal = {Physical Review D},
  publisher = {American Physical Society (APS)},
  author = {Kokron,  Nickolas and Bernal,  José Luis and Dunkley,  Jo},
  year = {2024},
  month = nov 
}

@article{Delabrouille:2009,
       author = {{Delabrouille}, J. and {Cardoso}, J. -F. and {Le Jeune}, M. and {Betoule}, M. and {Fay}, G. and {Guilloux}, F.},
        title = "{A full sky, low foreground, high resolution CMB map from WMAP}",
      journal = {\aap},
         year = 2009,
        month = jan,
       volume = {493},
       number = {3},
        pages = {835-857},
          doi = {10.1051/0004-6361:200810514},
archivePrefix = {arXiv},
       eprint = {0807.0773},
 primaryClass = {astro-ph},
       adsurl = {https://ui.adsabs.harvard.edu/abs/2009A&A...493..835D}
}

@ARTICLE{Delabrouille:2011,
       author = {{Delabrouille}, J. and {Betoule}, M. and {Melin}, J. -B. and {Miville-Desch{\^e}nes}, M. -A. and {Gonzalez-Nuevo}, J. and {Le Jeune}, M. and {Castex}, G. and {de Zotti}, G. and {Basak}, S. and {Ashdown}, M. and {Aumont}, J. and {Baccigalupi}, C. and {Banday}, A.~J. and {Bernard}, J. -P. and {Bouchet}, F.~R. and {Clements}, D.~L. and {da Silva}, A. and {Dickinson}, C. and {Dodu}, F. and {Dolag}, K. and {Elsner}, F. and {Fauvet}, L. and {Fa{\"y}}, G. and {Giardino}, G. and {Leach}, S. and {Lesgourgues}, J. and {Liguori}, M. and {Mac{\'\i}as-P{\'e}rez}, J.~F. and {Massardi}, M. and {Matarrese}, S. and {Mazzotta}, P. and {Montier}, L. and {Mottet}, S. and {Paladini}, R. and {Partridge}, B. and {Piffaretti}, R. and {Prezeau}, G. and {Prunet}, S. and {Ricciardi}, S. and {Roman}, M. and {Schaefer}, B. and {Toffolatti}, L.},
        title = "{The pre-launch Planck Sky Model: a model of sky emission at submillimetre to centimetre wavelengths}",
      journal = {\aap},
         year = 2013,
        month = may,
       volume = {553},
          eid = {A96},
        pages = {A96},
          doi = {10.1051/0004-6361/201220019},
archivePrefix = {arXiv},
       eprint = {1207.3675},
 primaryClass = {astro-ph.CO}
}

@article{Planck:2018_pslkl,
collaboration = {Planck},
    author = "Aghanim, N. and others",
    collaboration = "Planck",
    title = "{Planck 2018 results. V. CMB power spectra and likelihoods}",
    eprint = "1907.12875",
    archivePrefix = "arXiv",
    primaryClass = "astro-ph.CO",
    doi = "10.1051/0004-6361/201936386",
    journal = "Astron. Astrophys.",
    volume = "641",
    pages = "A5",
    year = "2020"
}

@article{Planck:2013_pslkl,
collaboration = {Planck},
    author = "Ade, P. A. R. and others",
    collaboration = "Planck",
    title = "{Planck 2013 results. XV. CMB power spectra and likelihood}",
    eprint = "1303.5075",
    archivePrefix = "arXiv",
    primaryClass = "astro-ph.CO",
    reportNumber = "CERN-PH-TH-2013-128",
    doi = "10.1051/0004-6361/201321573",
    journal = "Astron. Astrophys.",
    volume = "571",
    pages = "A15",
    year = "2014"
}

@article{Planck:2015_pslkl,
collaboration = {Planck},
    author = "Aghanim, N. and others",
    collaboration = "Planck",
    title = "{Planck 2015 results. XI. CMB power spectra, likelihoods, and robustness of parameters}",
    eprint = "1507.02704",
    archivePrefix = "arXiv",
    primaryClass = "astro-ph.CO",
    doi = "10.1051/0004-6361/201526926",
    journal = "Astron. Astrophys.",
    volume = "594",
    pages = "A11",
    year = "2016"
}

@article{SPT:2017,
    author = "Henning, J. W. and others",
    collaboration = "SPT",
    title = "{Measurements of the Temperature and E-Mode Polarization of the CMB from 500 Square Degrees of SPTpol Data}",
    eprint = "1707.09353",
    archivePrefix = "arXiv",
    primaryClass = "astro-ph.CO",
    reportNumber = "FERMILAB-PUB-17-297-AE",
    doi = "10.3847/1538-4357/aa9ff4",
    journal = "Astrophys. J.",
    volume = "852",
    number = "2",
    pages = "97",
    year = "2018"
}

@ARTICLE{Websky,
       author = {{Stein}, George and {Alvarez}, Marcelo A. and {Bond}, J. Richard and {van Engelen}, Alexander and {Battaglia}, Nicholas},
        title = "{The Websky extragalactic CMB simulations}",
      journal = {\jcap},
         year = 2020,
        month = oct,
       volume = {2020},
       number = {10},
          eid = {012},
        pages = {012},
          doi = {10.1088/1475-7516/2020/10/012},
archivePrefix = {arXiv},
       eprint = {2001.08787},
 primaryClass = {astro-ph.CO},
       url = {https://arxiv.org/abs/2001.08787}
}

@article{Krachmalnicoff:2018imw,
    author = "Krachmalnicoff, N. and others",
    title = "{S--PASS view of polarized Galactic synchrotron at 2.3 GHz as a contaminant to CMB observations}",
    eprint = "1802.01145",
    archivePrefix = "arXiv",
    primaryClass = "astro-ph.GA",
    doi = "10.1051/0004-6361/201832768",
    journal = "Astron. Astrophys.",
    volume = "618",
    pages = "A166",
    year = "2018"
}

@article{Louis:2017,
    author = "Louis, Thibaut and others",
    title = "{The Atacama Cosmology Telescope: Two-Season ACTPol Spectra and Parameters}",
    eprint = "1610.02360",
    archivePrefix = "arXiv",
    primaryClass = "astro-ph.CO",
    doi = "10.1088/1475-7516/2017/06/031",
    journal = "JCAP",
    volume = "06",
    pages = "031",
    year = "2017"
}

@article{Reichardt_SPT:2020,
    author = "Reichardt, C. L. and others",
    title = "{An Improved Measurement of the Secondary Cosmic Microwave Background Anisotropies from the SPT-SZ + SPTpol Surveys}",
    eprint = "2002.06197",
    archivePrefix = "arXiv",
    primaryClass = "astro-ph.CO",
    reportNumber = "FERMILAB-PUB-20-399-AE",
    doi = "10.3847/1538-4357/abd407",
    journal = "Astrophys. J.",
    volume = "908",
    number = "2",
    pages = "199",
    year = "2021"
}

@article{George_SPT:2014,
    author = "George, E. M. and others",
    title = "{A measurement of secondary cosmic microwave background anisotropies from the 2500-square-degree SPT-SZ survey}",
    eprint = "1408.3161",
    archivePrefix = "arXiv",
    primaryClass = "astro-ph.CO",
    reportNumber = "FERMILAB-PUB-14-372-A",
    doi = "10.1088/0004-637X/799/2/177",
    journal = "Astrophys. J.",
    volume = "799",
    number = "2",
    pages = "177",
    year = "2015"
}

@Article{matplotlib,
  Author    = {Hunter, J. D.},
  Title     = {Matplotlib: A 2D graphics environment},
  Journal   = {Computing in Science \& Engineering},
  Volume    = {9},
  Number    = {3},
  Pages     = {90--95},
  publisher = {IEEE COMPUTER SOC},
  doi       = {10.1109/MCSE.2007.55},
  year      = 2007,
  url = {https://ieeexplore.ieee.org/document/4160265}
}

@article{getdist,
    author = "Lewis, Antony",
    title = "{GetDist: a Python package for analysing Monte Carlo samples}",
    eprint = "1910.13970",
    archivePrefix = "arXiv",
    primaryClass = "astro-ph.IM",
    month = "10",
    year = "2019",
    url = {https://arxiv.org/pdf/1910.13970.pdf}
}

@article{         numpy,
     title         = {Array programming with {NumPy}},
     author        = {Charles R. Harris and K. Jarrod Millman and St{\'{e}}fan J.
                     van der Walt and Ralf Gommers and Pauli Virtanen and David
                     Cournapeau and Eric Wieser and Julian Taylor and Sebastian
                     Berg and Nathaniel J. Smith and Robert Kern and Matti Picus
                     and Stephan Hoyer and Marten H. van Kerkwijk and Matthew
                     Brett and Allan Haldane and Jaime Fern{\'{a}}ndez del
                     R{\'{i}}o and Mark Wiebe and Pearu Peterson and Pierre
                     G{\'{e}}rard-Marchant and Kevin Sheppard and Tyler Reddy and
                     Warren Weckesser and Hameer Abbasi and Christoph Gohlke and
                     Travis E. Oliphant},
     year          = {2020},
     month         = sep,
     journal       = {Nature},
     volume        = {585},
     number        = {7825},
     pages         = {357--362},
     doi           = {10.1038/s41586-020-2649-2},
    url            = {https://arxiv.org/pdf/2006.10256.pdf},
     publisher     = {Springer Science and Business Media {LLC}}
}

@ARTICLE{scipy,
      author  = {Virtanen, Pauli and Gommers, Ralf and Oliphant, Travis E. and
                Haberland, Matt and Reddy, Tyler and Cournapeau, David and
                Burovski, Evgeni and Peterson, Pearu and Weckesser, Warren and
                Bright, Jonathan and {van der Walt}, St{\'e}fan J. and
                Brett, Matthew and Wilson, Joshua and Millman, K. Jarrod and
                Mayorov, Nikolay and Nelson, Andrew R. J. and Jones, Eric and
                Kern, Robert and Larson, Eric and Carey, C J and
                Polat, {\.I}lhan and Feng, Yu and Moore, Eric W. and
                {VanderPlas}, Jake and Laxalde, Denis and Perktold, Josef and
                Cimrman, Robert and Henriksen, Ian and Quintero, E. A. and
                Harris, Charles R. and Archibald, Anne M. and
                Ribeiro, Ant{\^o}nio H. and Pedregosa, Fabian and
                {van Mulbregt}, Paul and {SciPy 1.0 Contributors}},
      title   = {{{SciPy} 1.0: Fundamental Algorithms for Scientific
                Computing in Python}},
      journal = {Nature Methods},
      year    = {2020},
      volume  = {17},
      pages   = {261--272},
      adsurl  = {https://rdcu.be/b08Wh},
        url = {https://arxiv.org/pdf/1907.10121.pdf},
      doi     = {10.1038/s41592-019-0686-2},
}

@article{Rosenberg_npipe:2022,
    author = "Rosenberg, Erik and Gratton, Steven and Efstathiou, George",
    title = "{CMB power spectra and cosmological parameters from Planck PR4 with CamSpec}",
    eprint = "2205.10869",
    archivePrefix = "arXiv",
    primaryClass = "astro-ph.CO",
    doi = "10.1093/mnras/stac2744",
    journal = "Mon. Not. Roy. Astron. Soc.",
    volume = "517",
    number = "3",
    pages = "4620--4636",
    year = "2022"
}

@article{SimonsObservatory:2025wwn,
    author = "Abitbol, M. and others",
    collaboration = "Simons Observatory",
    title = "{The Simons Observatory: Science Goals and Forecasts for the Enhanced Large Aperture Telescope}",
    eprint = "2503.00636",
    archivePrefix = "arXiv",
    primaryClass = "astro-ph.IM",
    month = "3",
    year = "2025",
    url = "https://arxiv.org/abs/2503.00636"
}

@ARTICLE{Morris2025,
       author = {{Morris}, Thomas W. and {Battistelli}, Elia and {Bustos}, Ricardo and {Choi}, Steve K. and {Duivenvoorden}, Adriaan J. and {Dunkley}, Jo and {D{\"u}nner}, Rolando and {Halpern}, Mark and {Guan}, Yilun and {van Marrewijk}, Joshiwa and {Mroczkowski}, Tony and {Naess}, Sigurd and {Niemack}, Michael D. and {Page}, Lyman A. and {Partridge}, Bruce and {Puddu}, Roberto and {Salatino}, Maria and {Sif{\'o}n}, Crist{\'o}bal and {Wang}, Yuhan and {Wollack}, Edward J.},
        title = "{The Atacama Cosmology Telescope: Quantifying atmospheric emission above Cerro Toco}",
      journal = {\prd},
         year = 2025,
        month = apr,
       volume = {111},
       number = {8},
          eid = {082001},
        pages = {082001},
          doi = {10.1103/PhysRevD.111.082001},
archivePrefix = {arXiv},
       eprint = {2410.13064},
 primaryClass = {astro-ph.IM},
       url = {https://www.arxiv.org/abs/2410.13064}
}

@article{Efstathiou:2025ckq,
    author = "Efstathiou, George and McCarthy, Fiona",
    title = "{The Power Spectrum of the Thermal Sunyaev-Zeldovich Effect}",
    eprint = "2502.10232",
    archivePrefix = "arXiv",
    primaryClass = "astro-ph.CO",
    month = "2",
    year = "2025",
    url = "https://arxiv.org/abs/2502.10232"
}

@article{ACT:2020goa,
    author = "Han, Dongwon and others",
    title = "{The Atacama Cosmology Telescope: delensed power spectra and parameters}",
    eprint = "2007.14405",
    archivePrefix = "arXiv",
    primaryClass = "astro-ph.CO",
    doi = "10.1088/1475-7516/2021/01/031",
    journal = "JCAP",
    volume = "01",
    pages = "031",
    year = "2021",
    url = "https://arxiv.org/abs/2007.14405"
}

@article{CCAT-Prime:2021lly,
    author = "Aravena, Manuel and others",
    collaboration = "CCAT-Prime",
    title = "{CCAT-prime Collaboration: Science Goals and Forecasts with Prime-Cam on the Fred Young Submillimeter Telescope}",
    eprint = "2107.10364",
    archivePrefix = "arXiv",
    primaryClass = "astro-ph.CO",
    doi = "10.3847/1538-4365/ac9838",
    journal = "Astrophys. J. Suppl.",
    volume = "264",
    number = "1",
    pages = "7",
    year = "2023",
    url = "https://arxiv.org/abs/2107.10364"
}

@ARTICLE{CMB_S4_Science_Book,
       author = {{CMB-S4 Collaboration}},
        title = "{CMB-S4 Science Book, First Edition}",
      journal = {arXiv e-prints},
         year = 2016,
        month = oct,
          eid = {arXiv:1610.02743},
        pages = {arXiv:1610.02743},
          doi = {10.48550/arXiv.1610.02743},
archivePrefix = {arXiv},
       eprint = {1610.02743},
 primaryClass = {astro-ph.CO},
       adsurl = {https://ui.adsabs.harvard.edu/abs/2016arXiv161002743A}
}

@article{planck_collaboration_planck_2014,
    title = {Planck 2013 results. {XXX}. {Cosmic} infrared background measurements and implications for star formation},
    volume = {571},
    copyright = {© ESO, 2014},
    issn = {0004-6361, 1432-0746},
    url = {https://www.aanda.org/articles/aa/abs/2014/11/aa22093-13/aa22093-13.html},
    doi = {10.1051/0004-6361/201322093},
    language = {en},
    urldate = {2022-11-08},
    journal = {Astronomy \& Astrophysics},
    author = {Planck Collaboration, ea},
    month = nov,
    year = {2014},
    note = {Publisher: EDP Sciences},
    pages = {A30},
}

@ARTICLE{Cobaya,
       author = {{Torrado}, Jes{\'u}s and {Lewis}, Antony},
        title = "{Cobaya: code for Bayesian analysis of hierarchical physical models}",
      journal = {\jcap},
         year = 2021,
        month = may,
       volume = {2021},
       number = {5},
          eid = {057},
        pages = {057},
          doi = {10.1088/1475-7516/2021/05/057},
archivePrefix = {arXiv},
       eprint = {2005.05290},
 primaryClass = {astro-ph.IM},
        url = {https://arxiv.org/abs/2005.05290}
}

@article{Pan-ExperimentGalacticScienceGroup:2025vcd,
    author = "Borrill, Julian and others",
    collaboration = "Pan-Experiment Galactic Science Group",
    title = "{Full-sky Models of Galactic Microwave Emission and Polarization at Sub-arcminute Scales for the Python Sky Model}",
    eprint = "2502.20452",
    archivePrefix = "arXiv",
    primaryClass = "astro-ph.CO",
    month = "2",
    year = "2025",
    url = {https://arxiv.org/abs/2502.20452}
}

@ARTICLE{2022ApJ...929..166H,
       author = {{Hensley}, Brandon S. and {Clark}, Susan E. and {Fanfani}, Valentina and {Krachmalnicoff}, Nicoletta and {Fabbian}, Giulio and {Poletti}, Davide and {Puglisi}, Giuseppe and {Coppi}, Gabriele and {Nibauer}, Jacob and {Gerasimov}, Roman and {Galitzki}, Nicholas and {Choi}, Steve K. and {Ashton}, Peter C. and {Baccigalupi}, Carlo and {Baxter}, Eric and {Burkhart}, Blakesley and {Calabrese}, Erminia and {Chluba}, Jens and {Errard}, Josquin and {Frolov}, Andrei V. and {Herv{\'\i}as-Caimapo}, Carlos and {Huffenberger}, Kevin M. and {Johnson}, Bradley R. and {Jost}, Baptiste and {Keating}, Brian and {McCarrick}, Heather and {Nati}, Federico and {Sathyanarayana Rao}, Mayuri and {van Engelen}, Alexander and {Walker}, Samantha and {Wolz}, Kevin and {Xu}, Zhilei and {Zhu}, Ningfeng and {Zonca}, Andrea},
        title = "{The Simons Observatory: Galactic Science Goals and Forecasts}",
      journal = {\apj},
         year = 2022,
        month = apr,
       volume = {929},
       number = {2},
          eid = {166},
        pages = {166},
          doi = {10.3847/1538-4357/ac5e36},
archivePrefix = {arXiv},
       eprint = {2111.02425},
 primaryClass = {astro-ph.GA},
       url = {https://arxiv.org/abs/2111.02425}
}

@ARTICLE{Fixsen2009,
       author = {{Fixsen}, D.~J.},
        title = "{The Temperature of the Cosmic Microwave Background}",
      journal = {\apj},
         year = 2009,
        month = dec,
       volume = {707},
       number = {2},
        pages = {916-920},
          doi = {10.1088/0004-637X/707/2/916},
archivePrefix = {arXiv},
       eprint = {0911.1955},
 primaryClass = {astro-ph.CO},
    url = {https://arxiv.org/abs/0911.1955}
}

@ARTICLE{Morris2022,
       author = {{Morris}, Thomas W. and {Bustos}, Ricardo and {Calabrese}, Erminia and {Choi}, Steve K. and {Duivenvoorden}, Adriaan J. and {Dunkley}, Jo and {D{\"u}nner}, Rolando and {Gallardo}, Patricio A. and {Hasselfield}, Matthew and {Hincks}, Adam D. and {Mroczkowski}, Tony and {Naess}, Sigurd and {Niemack}, Michael D. and {Page}, Lyman and {Partridge}, Bruce and {Salatino}, Maria and {Staggs}, Suzanne and {Treu}, Jesse and {Wollack}, Edward J. and {Xu}, Zhilei},
        title = "{The Atacama Cosmology Telescope: Modeling bulk atmospheric motion}",
      journal = {\prd},
         year = 2022,
        month = feb,
       volume = {105},
       number = {4},
          eid = {042004},
        pages = {042004},
          doi = {10.1103/PhysRevD.105.042004},
archivePrefix = {arXiv},
       eprint = {2111.01319},
 primaryClass = {astro-ph.IM},
        url = {https://arxiv.org/abs/2111.01319}
}

@ARTICLE{CMBHD,
       author = {{The CMB-HD Collaboration} and {:} and {Aiola}, Simone and {Akrami}, Yashar and {Basu}, Kaustuv and {Boylan-Kolchin}, Michael and {Brinckmann}, Thejs and {Bryan}, Sean and {Casey}, Caitlin M. and {Chluba}, Jens and {Clesse}, Sebastien and {Cyr-Racine}, Francis-Yan and {Di Mascolo}, Luca and {Dicker}, Simon and {Essinger-Hileman}, Thomas and {Farren}, Gerrit S. and {Fedderke}, Michael A. and {Ferraro}, Simone and {Fuller}, George M. and {Galitzki}, Nicholas and {Gluscevic}, Vera and {Grin}, Daniel and {Han}, Dongwon and {Hasselfield}, Matthew and {Hlozek}, Renee and {Holder}, Gil and {Hotinli}, Selim C. and {Jain}, Bhuvnesh and {Johnson}, Bradley and {Johnson}, Matthew and {Klaassen}, Pamela and {MacInnis}, Amanda and {Madhavacheril}, Mathew and {Mandal}, Sayan and {Mauskopf}, Philip and {Meerburg}, Daan and {Meyers}, Joel and {Miranda}, Vivian and {Mroczkowski}, Tony and {Mukherjee}, Suvodip and {Munchmeyer}, Moritz and {Munoz}, Julian and {Naess}, Sigurd and {Nagai}, Daisuke and {Namikawa}, Toshiya and {Newburgh}, Laura and {Nguyen}, Ho Nam and {Niemack}, Michael and {Oppenheimer}, Benjamin D. and {Pierpaoli}, Elena and {Raghunathan}, Srinivasan and {Schaan}, Emmanuel and {Sehgal}, Neelima and {Sherwin}, Blake and {Simon}, Sara M. and {Slosar}, Anze and {Smith}, Kendrick and {Spergel}, David and {Switzer}, Eric R. and {Trivedi}, Pranjal and {Tsai}, Yu-Dai and {van Engelen}, Alexander and {Wandelt}, Benjamin D. and {Wollack}, Edward J. and {Wu}, Kimmy},
        title = "{Snowmass2021 CMB-HD White Paper}",
      journal = {arXiv e-prints},
         year = 2022,
        month = mar,
          eid = {arXiv:2203.05728},
        pages = {arXiv:2203.05728},
          doi = {10.48550/arXiv.2203.05728},
archivePrefix = {arXiv},
       eprint = {2203.05728},
 primaryClass = {astro-ph.CO},
       url = {https://arxiv.org/abs/2203.05728}
}

@ARTICLE{Park2016,
       author = {{Park}, Hyunbae and {Komatsu}, Eiichiro and {Shapiro}, Paul R. and {Koda}, Jun and {Mao}, Yi},
        title = "{The Impact of Nonlinear Structure Formation on the Power Spectrum of Transverse Momentum Fluctuations and the Kinetic Sunyaev-Zel'dovich Effect}",
      journal = {\apj},
         year = 2016,
        month = feb,
       volume = {818},
       number = {1},
          eid = {37},
        pages = {37},
          doi = {10.3847/0004-637X/818/1/37},
archivePrefix = {arXiv},
       eprint = {1506.05177},
 primaryClass = {astro-ph.CO},
       adsurl = {https://ui.adsabs.harvard.edu/abs/2016ApJ...818...37P}
}

@ARTICLE{Alvarez2016,
       author = {{Alvarez}, Marcelo A.},
        title = "{The Kinetic Sunyaev-Zel{\textquoteright}dovich Effect from Reionization: Simulated Full-sky Maps at Arcminute Resolution}",
      journal = {\apj},
         year = 2016,
        month = jun,
       volume = {824},
       number = {2},
          eid = {118},
        pages = {118},
          doi = {10.3847/0004-637X/824/2/118},
archivePrefix = {arXiv},
       eprint = {1511.02846},
 primaryClass = {astro-ph.CO},
       adsurl = {https://ui.adsabs.harvard.edu/abs/2016ApJ...824..118A}
}

@ARTICLE{2005ARA&A..43..727L,
       author = {{Lagache}, Guilaine and {Puget}, Jean-Loup and {Dole}, Herv{\'e}},
        title = "{Dusty Infrared Galaxies: Sources of the Cosmic Infrared Background}",
      journal = {\araa},
         year = 2005,
        month = sep,
       volume = {43},
       number = {1},
        pages = {727-768},
          doi = {10.1146/annurev.astro.43.072103.150606},
archivePrefix = {arXiv},
       eprint = {astro-ph/0507298},
 primaryClass = {astro-ph},
       adsurl = {https://ui.adsabs.harvard.edu/abs/2005ARA&A..43..727L}
}

@ARTICLE{2012AdAst2012E..52T,
       author = {{Tucci}, Marco and {Toffolatti}, Luigi},
        title = "{The Impact of Polarized Extragalactic Radio Sources on the Detection of CMB Anisotropies in Polarization}",
      journal = {Advances in Astronomy},
         year = 2012,
        month = dec,
       volume = {2012},
          eid = {624987},
        pages = {624987},
          doi = {10.1155/2012/624987},
archivePrefix = {arXiv},
       eprint = {1204.0427},
 primaryClass = {astro-ph.CO},
       adsurl = {https://ui.adsabs.harvard.edu/abs/2012AdAst2012E..52T}
}

@BOOK{2011piim.book.....D,
       author = {{Draine}, Bruce T.},
        title = "{Physics of the Interstellar and Intergalactic Medium}",
         year = 2011,
       adsurl = {https://ui.adsabs.harvard.edu/abs/2011piim.book.....D}
}

@ARTICLE{2014JCAP...08..010C,
       author = {{Calabrese}, Erminia and {Hlo{\v{z}}ek}, Ren{\'e}e and {Battaglia}, Nick and {Bond}, J. Richard and {de Bernardis}, Francesco and {Devlin}, Mark J. and {Hajian}, Amir and {Henderson}, Shawn and {Hil}, J. Colin and {Kosowsky}, Arthur and {Louis}, Thibaut and {McMahon}, Jeff and {Moodley}, Kavilan and {Newburgh}, Laura and {Niemack}, Michael D. and {Page}, Lyman A. and {Partridge}, Bruce and {Sehgal}, Neelima and {Sievers}, Jonathan L. and {Spergel}, David N. and {Staggs}, Suzanne T. and {Switzer}, Eric R. and {Trac}, Hy and {Wollack}, Edward J.},
        title = "{Precision epoch of reionization studies with next-generation CMB experiments}",
      journal = {\jcap},
         year = 2014,
        month = aug,
       volume = {2014},
       number = {8},
        pages = {010-010},
          doi = {10.1088/1475-7516/2014/08/010},
archivePrefix = {arXiv},
       eprint = {1406.4794},
 primaryClass = {astro-ph.CO},
       adsurl = {https://ui.adsabs.harvard.edu/abs/2014JCAP...08..010C}
}

@ARTICLE{Shaw2010,
       author = {{Shaw}, Laurie D. and {Nagai}, Daisuke and {Bhattacharya}, Suman and {Lau}, Erwin T.},
        title = "{Impact of Cluster Physics on the Sunyaev-Zel'dovich Power Spectrum}",
      journal = {\apj},
         year = 2010,
        month = dec,
       volume = {725},
       number = {2},
        pages = {1452-1465},
          doi = {10.1088/0004-637X/725/2/1452},
archivePrefix = {arXiv},
       eprint = {1006.1945},
 primaryClass = {astro-ph.CO},
       adsurl = {https://ui.adsabs.harvard.edu/abs/2010ApJ...725.1452S}
}

@ARTICLE{2019JCAP...02..056A,
       author = {{Ade}, Peter and {Aguirre}, James and {Ahmed}, Zeeshan and {Aiola}, Simone and {Ali}, Aamir and {Alonso}, David and {Alvarez}, Marcelo A. and {Arnold}, Kam and {Ashton}, Peter and {Austermann}, Jason and {Awan}, Humna and {Baccigalupi}, Carlo and {Baildon}, Taylor and {Barron}, Darcy and {Battaglia}, Nick and {Battye}, Richard and {Baxter}, Eric and {Bazarko}, Andrew and {Beall}, James A. and {Bean}, Rachel and {Beck}, Dominic and {Beckman}, Shawn and {Beringue}, Benjamin and {Bianchini}, Federico and {Boada}, Steven and {Boettger}, David and {Bond}, J. Richard and {Borrill}, Julian and {Brown}, Michael L. and {Bruno}, Sarah Marie and {Bryan}, Sean and {Calabrese}, Erminia and {Calafut}, Victoria and {Calisse}, Paolo and {Carron}, Julien and {Challinor}, Anthony and {Chesmore}, Grace and {Chinone}, Yuji and {Chluba}, Jens and {Cho}, Hsiao-Mei Sherry and {Choi}, Steve and {Coppi}, Gabriele and {Cothard}, Nicholas F. and {Coughlin}, Kevin and {Crichton}, Devin and {Crowley}, Kevin D. and {Crowley}, Kevin T. and {Cukierman}, Ari and {D'Ewart}, John M. and {D{\"u}nner}, Rolando and {de Haan}, Tijmen and {Devlin}, Mark and {Dicker}, Simon and {Didier}, Joy and {Dobbs}, Matt and {Dober}, Bradley and {Duell}, Cody J. and {Duff}, Shannon and {Duivenvoorden}, Adri and {Dunkley}, Jo and {Dusatko}, John and {Errard}, Josquin and {Fabbian}, Giulio and {Feeney}, Stephen and {Ferraro}, Simone and {Flux{\`a}}, Pedro and {Freese}, Katherine and {Frisch}, Josef C. and {Frolov}, Andrei and {Fuller}, George and {Fuzia}, Brittany and {Galitzki}, Nicholas and {Gallardo}, Patricio A. and {Tomas Galvez Ghersi}, Jose and {Gao}, Jiansong and {Gawiser}, Eric and {Gerbino}, Martina and {Gluscevic}, Vera and {Goeckner-Wald}, Neil and {Golec}, Joseph and {Gordon}, Sam and {Gralla}, Megan and {Green}, Daniel and {Grigorian}, Arpi and {Groh}, John and {Groppi}, Chris and {Guan}, Yilun and {Gudmundsson}, Jon E. and {Han}, Dongwon and {Hargrave}, Peter and {Hasegawa}, Masaya and {Hasselfield}, Matthew and {Hattori}, Makoto and {Haynes}, Victor and {Hazumi}, Masashi and {He}, Yizhou and {Healy}, Erin and {Henderson}, Shawn W. and {Hervias-Caimapo}, Carlos and {Hill}, Charles A. and {Hill}, J. Colin and {Hilton}, Gene and {Hilton}, Matt and {Hincks}, Adam D. and {Hinshaw}, Gary and {Hlo{\v{z}}ek}, Ren{\'e}e and {Ho}, Shirley and {Ho}, Shuay-Pwu Patty and {Howe}, Logan and {Huang}, Zhiqi and {Hubmayr}, Johannes and {Huffenberger}, Kevin and {Hughes}, John P. and {Ijjas}, Anna and {Ikape}, Margaret and {Irwin}, Kent and {Jaffe}, Andrew H. and {Jain}, Bhuvnesh and {Jeong}, Oliver and {Kaneko}, Daisuke and {Karpel}, Ethan D. and {Katayama}, Nobuhiko and {Keating}, Brian and {Kernasovskiy}, Sarah S. and {Keskitalo}, Reijo and {Kisner}, Theodore and {Kiuchi}, Kenji and {Klein}, Jeff and {Knowles}, Kenda and {Koopman}, Brian and {Kosowsky}, Arthur and {Krachmalnicoff}, Nicoletta and {Kuenstner}, Stephen E. and {Kuo}, Chao-Lin and {Kusaka}, Akito and {Lashner}, Jacob and {Lee}, Adrian and {Lee}, Eunseong and {Leon}, David and {Leung}, Jason S. -Y. and {Lewis}, Antony and {Li}, Yaqiong and {Li}, Zack and {Limon}, Michele and {Linder}, Eric and {Lopez-Caraballo}, Carlos and {Louis}, Thibaut and {Lowry}, Lindsay and {Lungu}, Marius and {Madhavacheril}, Mathew and {Mak}, Daisy and {Maldonado}, Felipe and {Mani}, Hamdi and {Mates}, Ben and {Matsuda}, Frederick and {Maurin}, Lo{\"\i}c and {Mauskopf}, Phil and {May}, Andrew and {McCallum}, Nialh and {McKenney}, Chris and {McMahon}, Jeff and {Meerburg}, P. Daniel and {Meyers}, Joel and {Miller}, Amber and {Mirmelstein}, Mark and {Moodley}, Kavilan and {Munchmeyer}, Moritz and {Munson}, Charles and {Naess}, Sigurd and {Nati}, Federico and {Navaroli}, Martin and {Newburgh}, Laura and {Nguyen}, Ho Nam and {Niemack}, Michael and {Nishino}, Haruki and {Orlowski-Scherer}, John and {Page}, Lyman and {Partridge}, Bruce and {Peloton}, Julien and {Perrotta}, Francesca and {Piccirillo}, Lucio and {Pisano}, Giampaolo and {Poletti}, Davide and {Puddu}, Roberto and {Puglisi}, Giuseppe and {Raum}, Chris and {Reichardt}, Christian L. and {Remazeilles}, Mathieu and {Rephaeli}, Yoel and {Riechers}, Dominik and {Rojas}, Felipe and {Roy}, Anirban and {Sadeh}, Sharon and {Sakurai}, Yuki and {Salatino}, Maria and {Sathyanarayana Rao}, Mayuri and {Schaan}, Emmanuel and {Schmittfull}, Marcel and {Sehgal}, Neelima and {Seibert}, Joseph},
        collaboration = "{Simons Observatory}",
        title = "{The Simons Observatory: science goals and forecasts}",
      journal = {\jcap},
         year = 2019,
        month = feb,
       volume = {2019},
       number = {2},
          eid = {056},
        pages = {056},
          doi = {10.1088/1475-7516/2019/02/056},
archivePrefix = {arXiv},
       eprint = {1808.07445},
 primaryClass = {astro-ph.CO},
       adsurl = {https://ui.adsabs.harvard.edu/abs/2019JCAP...02..056A}
}

@ARTICLE{Zahn2005,
       author = {{Zahn}, Oliver and {Zaldarriaga}, Matias and {Hernquist}, Lars and {McQuinn}, Matthew},
        title = "{The Influence of Nonuniform Reionization on the CMB}",
      journal = {\apj},
         year = 2005,
        month = sep,
       volume = {630},
       number = {2},
        pages = {657-666},
          doi = {10.1086/431947},
archivePrefix = {arXiv},
       eprint = {astro-ph/0503166},
 primaryClass = {astro-ph},
       adsurl = {https://ui.adsabs.harvard.edu/abs/2005ApJ...630..657Z}
}

@ARTICLE{McQuinn2005,
       author = {{McQuinn}, Matthew and {Furlanetto}, Steven R. and {Hernquist}, Lars and {Zahn}, Oliver and {Zaldarriaga}, Matias},
        title = "{The Kinetic Sunyaev-Zel'dovich Effect from Reionization}",
      journal = {\apj},
         year = 2005,
        month = sep,
       volume = {630},
       number = {2},
        pages = {643-656},
          doi = {10.1086/432049},
archivePrefix = {arXiv},
       eprint = {astro-ph/0504189},
 primaryClass = {astro-ph},
       adsurl = {https://ui.adsabs.harvard.edu/abs/2005ApJ...630..643M}
}

@ARTICLE{Iliev2007,
       author = {{Iliev}, Ilian T. and {Pen}, Ue-Li and {Bond}, J. Richard and {Mellema}, Garrelt and {Shapiro}, Paul R.},
        title = "{The Kinetic Sunyaev-Zel'dovich Effect from Radiative Transfer Simulations of Patchy Reionization}",
      journal = {\apj},
         year = 2007,
        month = may,
       volume = {660},
       number = {2},
        pages = {933-944},
          doi = {10.1086/513687},
archivePrefix = {arXiv},
       eprint = {astro-ph/0609592},
 primaryClass = {astro-ph},
       adsurl = {https://ui.adsabs.harvard.edu/abs/2007ApJ...660..933I}
}

@ARTICLE{Chen2023,
       author = {{Chen}, Nianyi and {Trac}, Hy and {Mukherjee}, Suvodip and {Cen}, Renyue},
        title = "{Patchy Kinetic Sunyaev-Zel'dovich Effect with Controlled Reionization History and Morphology}",
      journal = {\apj},
         year = 2023,
        month = feb,
       volume = {943},
       number = {2},
          eid = {138},
        pages = {138},
          doi = {10.3847/1538-4357/ac8481},
archivePrefix = {arXiv},
       eprint = {2203.04337},
 primaryClass = {astro-ph.CO},
       adsurl = {https://ui.adsabs.harvard.edu/abs/2023ApJ...943..138C}
}

@ARTICLE{Zahn2012,
       author = {{Zahn}, O. and {Reichardt}, C.~L. and {Shaw}, L. and {Lidz}, A. and {Aird}, K.~A. and {Benson}, B.~A. and {Bleem}, L.~E. and {Carlstrom}, J.~E. and {Chang}, C.~L. and {Cho}, H.~M. and {Crawford}, T.~M. and {Crites}, A.~T. and {de Haan}, T. and {Dobbs}, M.~A. and {Dor{\'e}}, O. and {Dudley}, J. and {George}, E.~M. and {Halverson}, N.~W. and {Holder}, G.~P. and {Holzapfel}, W.~L. and {Hoover}, S. and {Hou}, Z. and {Hrubes}, J.~D. and {Joy}, M. and {Keisler}, R. and {Knox}, L. and {Lee}, A.~T. and {Leitch}, E.~M. and {Lueker}, M. and {Luong-Van}, D. and {McMahon}, J.~J. and {Mehl}, J. and {Meyer}, S.~S. and {Millea}, M. and {Mohr}, J.~J. and {Montroy}, T.~E. and {Natoli}, T. and {Padin}, S. and {Plagge}, T. and {Pryke}, C. and {Ruhl}, J.~E. and {Schaffer}, K.~K. and {Shirokoff}, E. and {Spieler}, H.~G. and {Staniszewski}, Z. and {Stark}, A.~A. and {Story}, K. and {van Engelen}, A. and {Vanderlinde}, K. and {Vieira}, J.~D. and {Williamson}, R.},
        title = "{Cosmic Microwave Background Constraints on the Duration and Timing of Reionization from the South Pole Telescope}",
      journal = {\apj},
         year = 2012,
        month = sep,
       volume = {756},
       number = {1},
          eid = {65},
        pages = {65},
          doi = {10.1088/0004-637X/756/1/65},
archivePrefix = {arXiv},
       eprint = {1111.6386},
 primaryClass = {astro-ph.CO},
       adsurl = {https://ui.adsabs.harvard.edu/abs/2012ApJ...756...65Z}
}

@ARTICLE{2016A&A...596A.108P,
collaboration = "Planck",
       author = {{Planck Collaboration} and {Adam}, R. and {Aghanim}, N. and {Ashdown}, M. and {Aumont}, J. and {Baccigalupi}, C. and {Ballardini}, M. and {Banday}, A.~J. and {Barreiro}, R.~B. and {Bartolo}, N. and {Basak}, S. and {Battye}, R. and {Benabed}, K. and {Bernard}, J. -P. and {Bersanelli}, M. and {Bielewicz}, P. and {Bock}, J.~J. and {Bonaldi}, A. and {Bonavera}, L. and {Bond}, J.~R. and {Borrill}, J. and {Bouchet}, F.~R. and {Boulanger}, F. and {Bucher}, M. and {Burigana}, C. and {Calabrese}, E. and {Cardoso}, J. -F. and {Carron}, J. and {Chiang}, H.~C. and {Colombo}, L.~P.~L. and {Combet}, C. and {Comis}, B. and {Couchot}, F. and {Coulais}, A. and {Crill}, B.~P. and {Curto}, A. and {Cuttaia}, F. and {Davis}, R.~J. and {de Bernardis}, P. and {de Rosa}, A. and {de Zotti}, G. and {Delabrouille}, J. and {Di Valentino}, E. and {Dickinson}, C. and {Diego}, J.~M. and {Dor{\'e}}, O. and {Douspis}, M. and {Ducout}, A. and {Dupac}, X. and {Elsner}, F. and {En{\ss}lin}, T.~A. and {Eriksen}, H.~K. and {Falgarone}, E. and {Fantaye}, Y. and {Finelli}, F. and {Forastieri}, F. and {Frailis}, M. and {Fraisse}, A.~A. and {Franceschi}, E. and {Frolov}, A. and {Galeotta}, S. and {Galli}, S. and {Ganga}, K. and {G{\'e}nova-Santos}, R.~T. and {Gerbino}, M. and {Ghosh}, T. and {Gonz{\'a}lez-Nuevo}, J. and {G{\'o}rski}, K.~M. and {Gruppuso}, A. and {Gudmundsson}, J.~E. and {Hansen}, F.~K. and {Helou}, G. and {Henrot-Versill{\'e}}, S. and {Herranz}, D. and {Hivon}, E. and {Huang}, Z. and {Ili{\'c}}, S. and {Jaffe}, A.~H. and {Jones}, W.~C. and {Keih{\"a}nen}, E. and {Keskitalo}, R. and {Kisner}, T.~S. and {Knox}, L. and {Krachmalnicoff}, N. and {Kunz}, M. and {Kurki-Suonio}, H. and {Lagache}, G. and {L{\"a}hteenm{\"a}ki}, A. and {Lamarre}, J. -M. and {Langer}, M. and {Lasenby}, A. and {Lattanzi}, M. and {Lawrence}, C.~R. and {Le Jeune}, M. and {Levrier}, F. and {Lewis}, A. and {Liguori}, M. and {Lilje}, P.~B. and {L{\'o}pez-Caniego}, M. and {Ma}, Y. -Z. and {Mac{\'\i}as-P{\'e}rez}, J.~F. and {Maggio}, G. and {Mangilli}, A. and {Maris}, M. and {Martin}, P.~G. and {Mart{\'\i}nez-Gonz{\'a}lez}, E. and {Matarrese}, S. and {Mauri}, N. and {McEwen}, J.~D. and {Meinhold}, P.~R. and {Melchiorri}, A. and {Mennella}, A. and {Migliaccio}, M. and {Miville-Desch{\^e}nes}, M. -A. and {Molinari}, D. and {Moneti}, A. and {Montier}, L. and {Morgante}, G. and {Moss}, A. and {Naselsky}, P. and {Natoli}, P. and {Oxborrow}, C.~A. and {Pagano}, L. and {Paoletti}, D. and {Partridge}, B. and {Patanchon}, G. and {Patrizii}, L. and {Perdereau}, O. and {Perotto}, L. and {Pettorino}, V. and {Piacentini}, F. and {Plaszczynski}, S. and {Polastri}, L. and {Polenta}, G. and {Puget}, J. -L. and {Rachen}, J.~P. and {Racine}, B. and {Reinecke}, M. and {Remazeilles}, M. and {Renzi}, A. and {Rocha}, G. and {Rossetti}, M. and {Roudier}, G. and {Rubi{\~n}o-Mart{\'\i}n}, J.~A. and {Ruiz-Granados}, B. and {Salvati}, L. and {Sandri}, M. and {Savelainen}, M. and {Scott}, D. and {Sirri}, G. and {Sunyaev}, R. and {Suur-Uski}, A. -S. and {Tauber}, J.~A. and {Tenti}, M. and {Toffolatti}, L. and {Tomasi}, M. and {Tristram}, M. and {Trombetti}, T. and {Valiviita}, J. and {Van Tent}, F. and {Vielva}, P. and {Villa}, F. and {Vittorio}, N. and {Wandelt}, B.~D. and {Wehus}, I.~K. and {White}, M. and {Zacchei}, A. and {Zonca}, A.},
        title = "{Planck intermediate results. XLVII. Planck constraints on reionization history}",
      journal = {\aap},
         year = 2016,
        month = dec,
       volume = {596},
          eid = {A108},
        pages = {A108},
          doi = {10.1051/0004-6361/201628897},
archivePrefix = {arXiv},
       eprint = {1605.03507},
 primaryClass = {astro-ph.CO},
       adsurl = {https://ui.adsabs.harvard.edu/abs/2016A&A...596A.108P}
}

@ARTICLE{Shaw2012,
       author = {{Shaw}, Laurie D. and {Rudd}, Douglas H. and {Nagai}, Daisuke},
        title = "{Deconstructing the Kinetic SZ Power Spectrum}",
      journal = {\apj},
         year = 2012,
        month = sep,
       volume = {756},
       number = {1},
          eid = {15},
        pages = {15},
          doi = {10.1088/0004-637X/756/1/15},
archivePrefix = {arXiv},
       eprint = {1109.0553},
 primaryClass = {astro-ph.CO},
       adsurl = {https://ui.adsabs.harvard.edu/abs/2012ApJ...756...15S}
}

@article{Trac2022,
   title={AMBER: A Semi-numerical Abundance Matching Box for the Epoch of Reionization},
   volume={927},
   ISSN={1538-4357},
   url={http://dx.doi.org/10.3847/1538-4357/ac5116},
   DOI={10.3847/1538-4357/ac5116},
   number={2},
   journal={The Astrophysical Journal},
   publisher={American Astronomical Society},
   author={Trac, Hy and Chen, Nianyi and Holst, Ian and Alvarez, Marcelo A. and Cen, Renyue},
   year={2022},
   month=mar, pages={186} 
}

@article{kramer2025, 
year = {2025}, 
title = {{Cross-correlating the patchy screening and kinetic Sunyaev-Zel'dovich effects as a new probe of reionization}}, 
author = {Kramer, Darby and Engelen, Alexander van and Cain, Christopher and MacCrann, Niall and Trac, Hy and Grayson, Skylar and Scannapieco, Evan and Sherwin, Blake}, 
journal = {arXiv}, 
doi = {10.48550/arxiv.2501.07623}, 
eprint = {2501.07623}
}

@ARTICLE{trac_cen_2007,
       author = {{Trac}, Hy and {Cen}, Renyue},
        title = "{Radiative Transfer Simulations of Cosmic Reionization. I. Methodology and Initial Results}",
      journal = {\apj},
         year = 2007,
        month = dec,
       volume = {671},
       number = {1},
        pages = {1-13},
          doi = {10.1086/522566},
archivePrefix = {arXiv},
       eprint = {astro-ph/0612406},
 primaryClass = {astro-ph},
       adsurl = {https://ui.adsabs.harvard.edu/abs/2007ApJ...671....1T}
}

\appendix 
\section{More Details on Consistency with SPT and \emph{Planck}}

\begin{table*}[h]
\centering
\begin{tabular} { l  c c c c c c}
\noalign{\vskip 3pt}\hline\noalign{\vskip 1.5pt}\hline\noalign{\vskip 5pt}
 \multicolumn{1}{c}{\bf } &  \multicolumn{1}{c}{\bf ACT} &  \multicolumn{1}{c}{\bf SPT} &  \multicolumn{1}{c}{\bf SPT} &  \multicolumn{1}{c}{\bf ACT-SPT} &  \multicolumn{1}{c}{\bf ACT-\texttt{plik}$_{cut}$} &  \multicolumn{1}{c}{\bf ACT-SPT-\texttt{plik}$_{cut}$}\\
  \multicolumn{1}{c}{\bf } &  \multicolumn{1}{c}{(DR6 model)} &  \multicolumn{1}{c}{(R21 model)} &  \multicolumn{1}{c}{(DR6 model)} &  \multicolumn{1}{c}{(DR6 model)} &  \multicolumn{1}{c}{(DR6 model)} &  \multicolumn{1}{c}{(DR6 model)}\\
\noalign{\vskip 3pt}\cline{2-7}\noalign{\vskip 3pt}

Parameter &   &   &   &   &  &  \\
\hline
{\boldmath$a_\mathrm{tSZ} $} &  $3.35\pm 0.34 $ & $3.41\pm 0.53              $ & $3.49\pm 0.37              $ & $3.52^{+0.27}_{-0.20}      $ & $3.28\pm 0.34              $ & $3.46^{+0.29}_{-0.25}      $\\

{\boldmath$a_\mathrm{kSZ} $} &  $1.48^{+0.71}_{-1.1}       $ & $2.7\pm 1.0                $ & $< 1.90                    $ & $< 1.01                    $ & $2.06^{+0.88}_{-0.79}      $ & $1.34^{+0.57}_{-0.71}      $\\

{\boldmath$\alpha_\mathrm{tSZ}$} & $-0.53^{+0.22}_{-0.19}$ &      ---                     & $-0.40^{+0.36}_{-0.27}     $ & $-0.41^{+0.16}_{-0.13}     $ & $-0.65\pm 0.19             $ & $-0.52^{+0.17}_{-0.14}     $\\

{\boldmath$a_c            $} & $3.69\pm 0.47              $ & ---                          & $4.25\pm 0.35              $ & $4.01\pm 0.25              $ & $3.64\pm 0.44              $ & $3.93\pm 0.26              $\\

{\boldmath$a_p^{\rm ACT} $} & $7.65\pm 0.34              $ & --- &  --- & $7.02\pm 0.24              $ & $7.66\pm 0.33              $ & $7.06\pm 0.24              $\\

{\boldmath$a_p^\mathrm{SPT}$} & --- & $7.26^{+0.73}_{-0.57}      $ & $6.97\pm 0.21              $ & $7.36\pm 0.17              $ & --- & $7.38\pm 0.17              $\\

{\boldmath$\beta_c        $} & $1.87\pm 0.10              $ & $2.24^{+0.18}_{-0.16}      $    & $2.015\pm 0.049            $ & $2.039\pm 0.038            $ & $1.874\pm 0.095            $ & $2.045\pm 0.038            $\\

{\boldmath$\beta_p        $} & --- & $1.46^{+0.15}_{-0.13}$  & --- & --- & --- & --- \\

{\boldmath$\xi_{yc}       $} & $0.088^{+0.036}_{-0.075}   $ & $0.076^{+0.033}_{-0.043}   $ & $0.115^{+0.057}_{-0.045}   $ & $< 0.0400                  $ & $0.099\pm 0.053            $ & $< 0.0484                  $\\

{\boldmath$a_s^\mathrm{ACT}            $} & $2.86\pm 0.21              $ & --- & --- & $3.14\pm 0.15              $ & $2.89\pm 0.19              $ & $3.15\pm 0.15              $\\

{\boldmath$a_s^\mathrm{SPT}$} & --- & $1.03^{+0.16}_{-0.20}      $ & $1.45\pm 0.18              $ & $1.215\pm 0.091            $ & --- & $1.216\pm 0.088            $\\

{\boldmath$\beta_s        $} & $-2.783^{+0.085}_{-0.076}  $ & $-2.75^{+0.17}_{-0.15}     $ & $-2.40^{+0.10}_{-0.088}    $ & $-2.632\pm 0.054           $ & $-2.781^{+0.081}_{-0.072}  $ & $-2.640\pm 0.054           $\\

{\boldmath$a_c^\mathrm{SPT, 1-Halo}$} & --- & $2.34^{+0.82}_{-0.98}      $& --- & --- &--- &--- \\

{\boldmath$a_c^\mathrm{SPT, 2-Halo}$} & --- & $1.82^{+0.26}_{-0.34}      $& --- & --- &--- &--- \\
\hline

{\boldmath$\chi^2_\mathrm{ACT}$} & 1591 (1651) & --- & --- & 1601 (1651)  & 1595 (1651) & 1608 (1651) \\

{\boldmath$\chi^2_\mathrm{SPT}$} & --- & 89.7 (88) & 93.1 (88) & 112.0 (88) & --- & 114.5 (88) \\

{\boldmath$\chi^2_\mathrm{Planck}$} & --- & --- & --- & --- & 1485 (1470) & 1484 (1470) \\
\hline
\end{tabular}

    \caption{Marginalized constraints ($68\%$) on the foreground parameters from the ACT DR6 and SPT data and their combination with {\it Planck} at $\ell<1000$ in TT and $\ell<600$ in polarization. The goodness-of-fit of the best-fitting models (with maximum posterior probability) is reported for the different datasets, along with the number of data points in parentheses. These constraints are marginalized over the $6$ $\Lambda$CDM parameters (with a prior on $\tau$) as well as over nuisance parameters.}
    \label{tab:results_fits}
\end{table*}

Table~\ref{tab:results_fits} summarizes the best-fitting foreground parameters obtained from various combinations of ACT, SPT, and \texttt{plik}$_{cut}$ datasets. For comparison, we also include constraints derived by fitting the foreground model used in R21 directly to SPT data. The differences between these results and those originally reported in R21 primarily stem from our marginalization over the six $\Lambda$CDM cosmological parameters, including the application of a prior on $\tau$.

\section{Cosmological Parameters for $\alpha_{\rm tSZ}=0$}
\label{app:alphasz}

\begin{figure}[htb]
    \centering
    \includegraphics[width=1.0\linewidth]{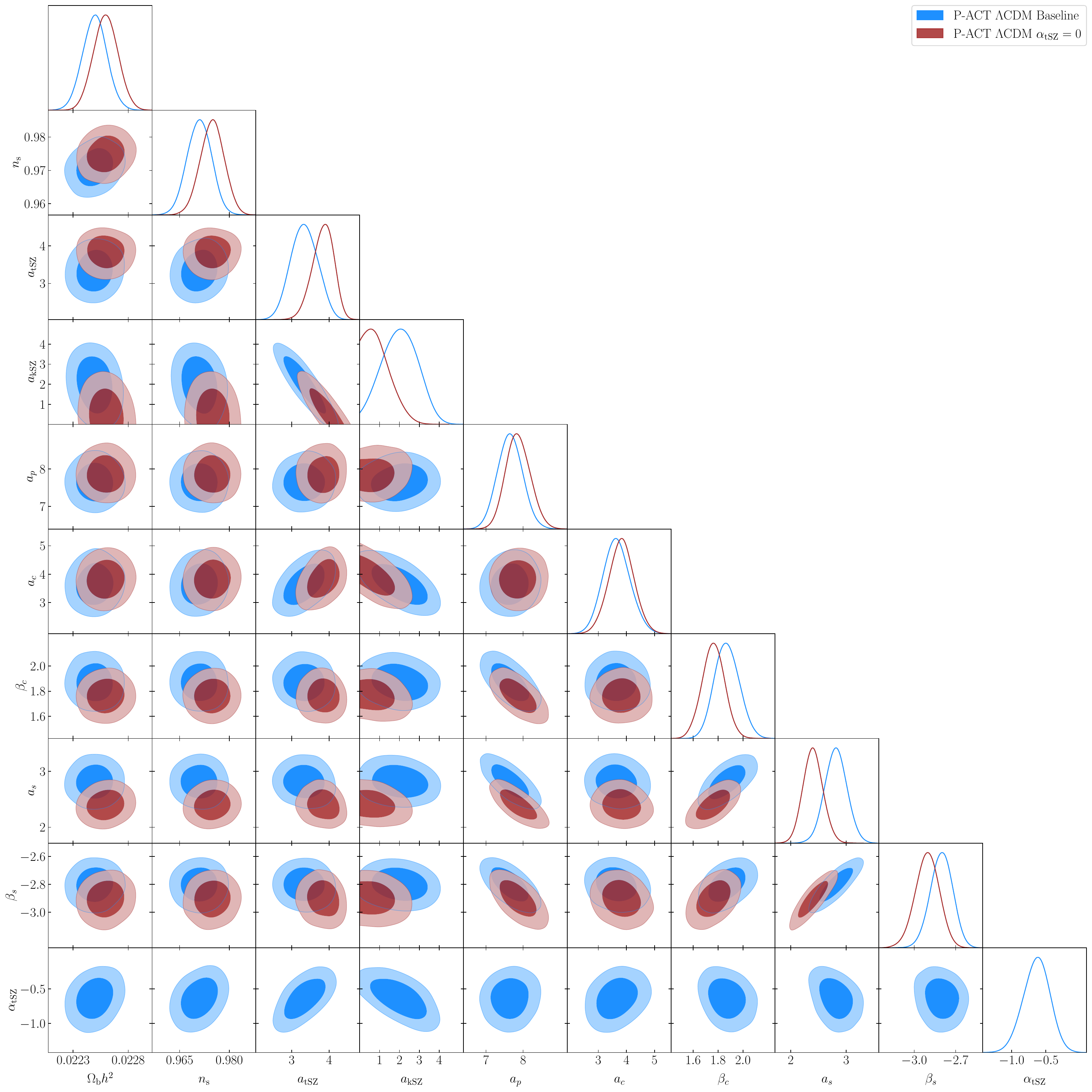}
    \caption{Extension of Figure~\ref{fig:pact_alphasz_2d}, showing 2D and 1D marginalized posterior distributions in the $\Lambda$CDM model using the P-ACT dataset combination for the baseline foreground model (blue), as compared to the foreground model variation with $\alpha_{\rm tSZ}=0$ (red). Results are shown here for several foreground parameters, as well as the cosmological parameters in Figure~\ref{fig:pact_alphasz_2d}.}
    \label{fig:pact_alphasz_2d_fg}
\end{figure}

In this appendix we show the full cosmological parameter constraints in the $\alpha_{\rm tSZ}=0$ foreground model variation in the $\Lambda$CDM and $\Lambda$CDM+$N_{\rm eff}$ cosmological models, for dataset combinations ACT and P-ACT. These are shown in Table~\ref{table:alphasz_full_constraints}. Figure~\ref{fig:pact_alphasz_2d_fg} shows 2D contours comparing the baseline and $\alpha_{\rm tSZ}=0$ model foreground parameters, as well as key cosmological parameters degenerate with $\alpha_{\rm tSZ}$.

\begin{figure*}[htb]
    \centering
    \includegraphics[width=0.45\textwidth]{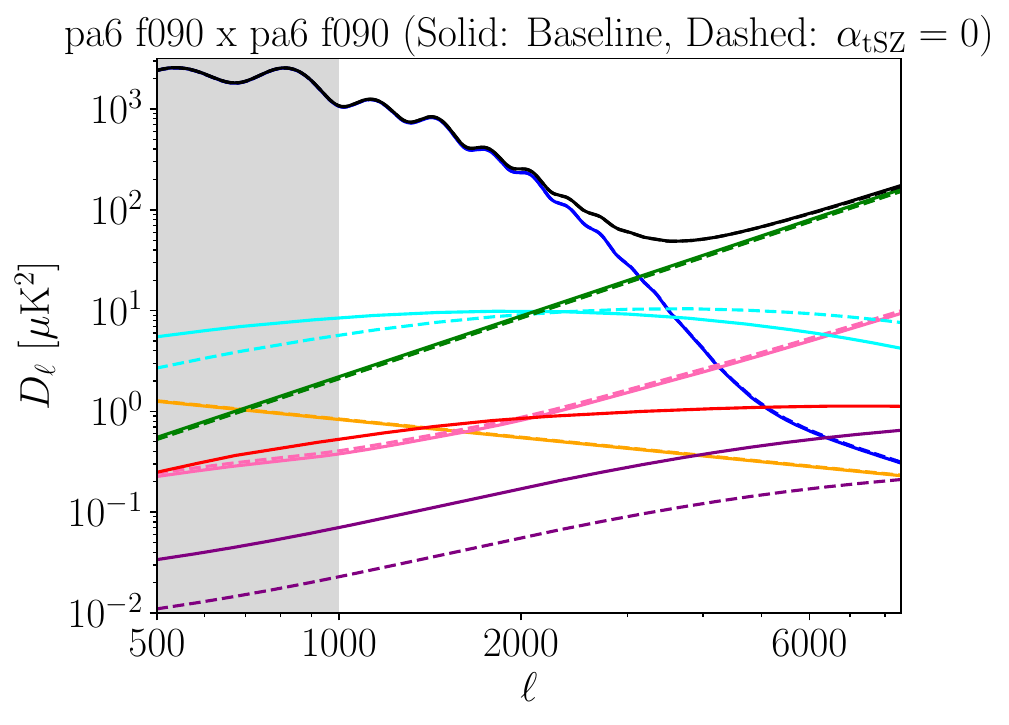}
    \includegraphics[width=0.45\textwidth]{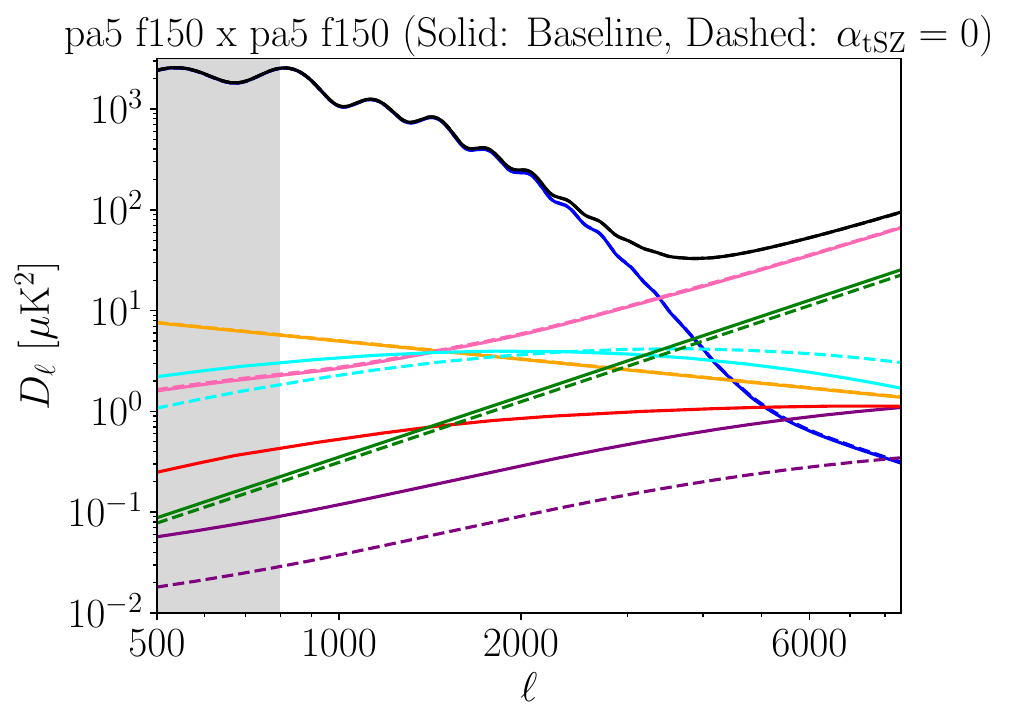}
 \includegraphics[width=0.6\textwidth]{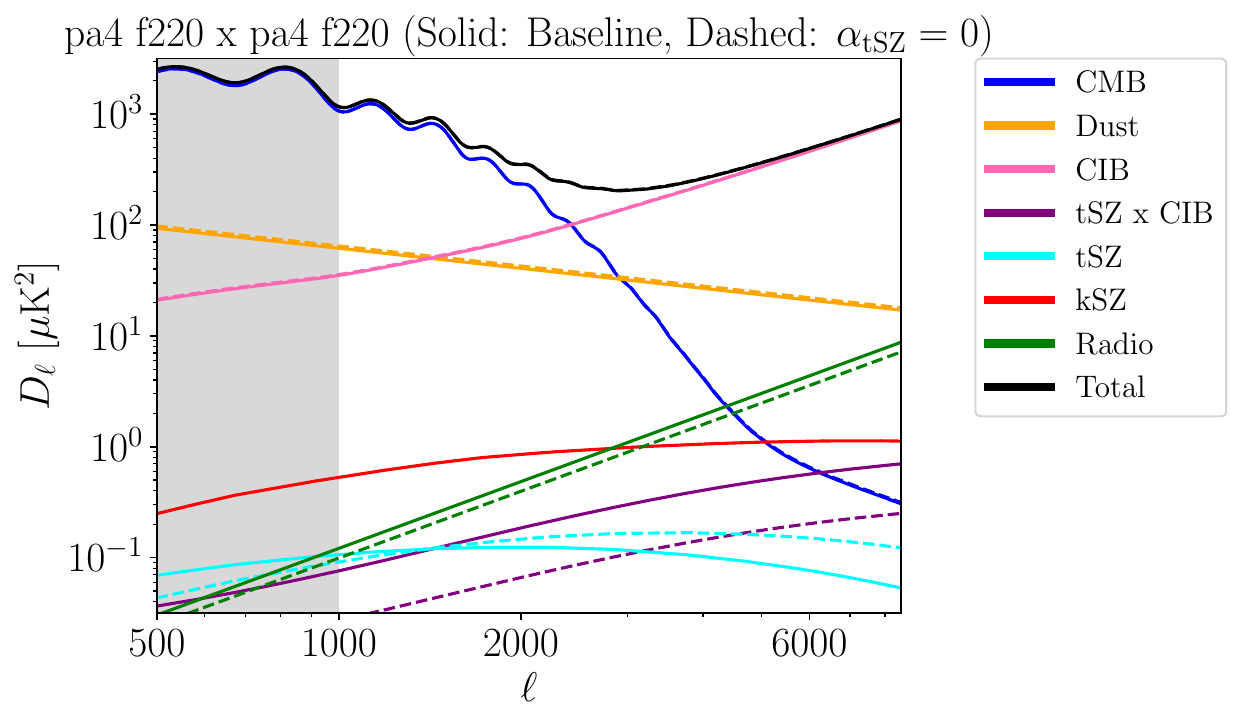}
    \caption{Comparison of the CMB and foregrounds (all in TT) evaluated at the MAP for the baseline model (solid lines) as compared with the $\alpha_{\rm tSZ}=0$ model (dashed lines) for a few array bands using only the DR6 data (the plots look very similar when using P-ACT instead of ACT alone). The CMB power spectrum is predicted to be slightly larger in the $\alpha_{\rm tSZ}=0$ case as compared with the baseline due to a decrease in predicted radio emission. The shaded gray band indicates multipole values that are not used in the likelihood analysis.} \label{fig:alphasz0_vs_baseline_output}
\end{figure*}

The parameter $\alpha_{\rm tSZ}$ primarily affects the high-$\ell$ tail of the multi-frequency spectra. Since the $\alpha_{\rm tSZ}=0$ model predicts slightly less radio emission in temperature than does the baseline model, the inferred CMB power spectrum in that case must be increased to compensate. This is shown in Figure~\ref{fig:alphasz0_vs_baseline_output}.

\begin{table*}[h]
\centering
    \renewcommand{\arraystretch}{1.1}
    
    \begin{tabular}{l c c c c }
    \hline
    \hline
     \multicolumn{1}{c}{Dataset}&  $\Omega_\mathrm{c}h^2$    &   $\Omega_\mathrm{b}h^2$  & $\ln 10^{10}A_s$ & $n_\mathrm{s}$  \\ 
     \hline
     
     \textbf{$\Lambda$CDM}  & &  \\ 
      \hline
    \hspace{10pt}ACT &$0.1242\pm 0.0020$ & $0.02265\pm 0.00016$ & $3.057\pm 0.013$ &$0.9695\pm 0.0073$  \\
    \hspace{10pt}P-ACT &$0.1190\pm 0.0012$ &$0.02260\pm 0.00011$ &$3.062^{+0.012}_{-0.014}$ &$0.9749\pm 0.0036$ \\
     \hline

    \textbf{$\Lambda$CDM+$N_{\rm eff}$}  & &   \\ 
      \hline
    \hspace{10pt}ACT &$0.1150\pm 0.0046$ &$0.02230\pm 0.00023$ &$3.032\pm 0.017$ &$0.926\pm 0.021$  \\
    \hspace{10pt}P-ACT &$0.1136\pm 0.0021$ &$0.02217\pm 0.00017$ &$3.037\pm 0.014$ &$0.9551\pm 0.0072$ \\
     \hline
    \hline
    
    \end{tabular}

    \begin{tabular}{l c c c}
    \hline
    \hline
     \multicolumn{1}{c}{Dataset} &  $H_0$ & $\tau$ &$N_{\rm eff}$ \\ 
     \hline
     
     \textbf{$\Lambda$CDM}  & &  \\ 
      \hline
    \hspace{10pt}ACT &$66.01\pm 0.75$ & $0.0572\pm 0.0057$ & 3.044 \\
    \hspace{10pt}P-ACT & $67.81\pm 0.51$ &$0.0628^{+0.0055}_{-0.0069}$ & 3.044\\
     \hline

    \textbf{$\Lambda$CDM+$N_{\rm eff}$}  & &   \\ 
      \hline
    \hspace{10pt}ACT  &$62.4\pm 1.8$ &$0.0524^{+0.0052}_{-0.0064}$  &$2.47\pm 0.26$ \\
    \hspace{10pt}P-ACT  &$64.9\pm 1.0$ &$0.0587^{+0.0053}_{-0.0062}$  &$2.64\pm 0.13$\\
     \hline
    \hline
    
    \end{tabular}

    \caption{Cosmological parameter constraints in the $\alpha_{\rm tSZ}=0$ foreground model variation in the $\Lambda$CDM and $\Lambda$CDM+$N_{\rm eff}$ cosmological models, for dataset combinations ACT and P-ACT.}\label{table:alphasz_full_constraints}
\end{table*}

\section{Inclusion of CO Templates}
\label{app:co}

\begin{figure*}[htb]
    \centering
    \includegraphics[width=1.0\textwidth]{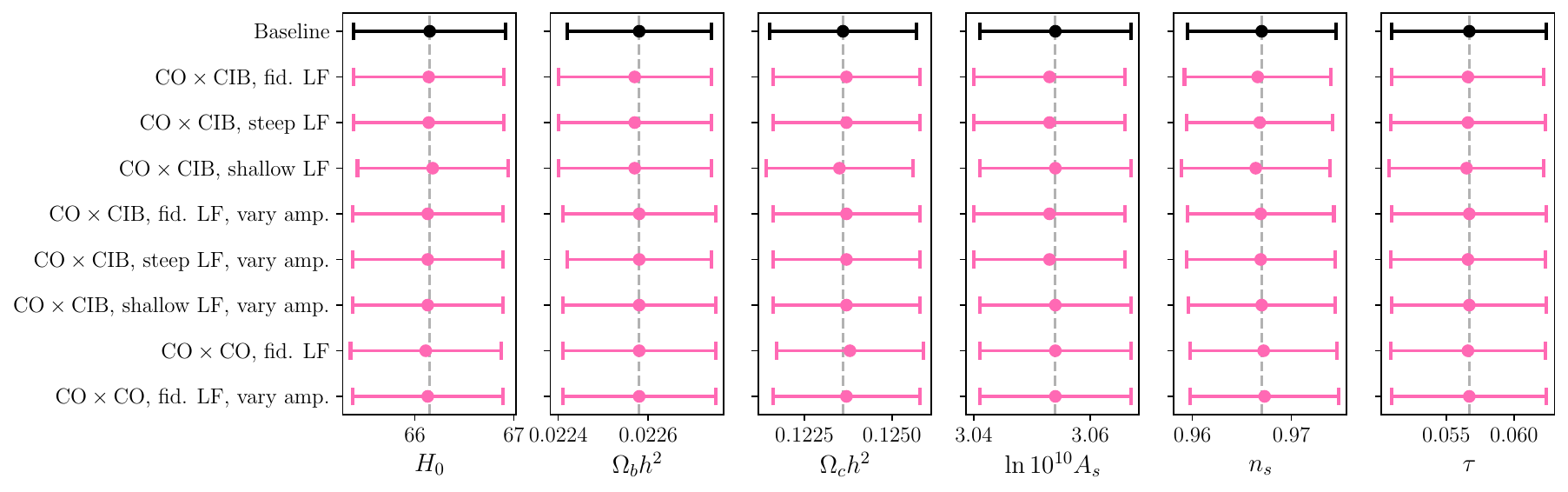}
    \includegraphics[width=1.0\textwidth]{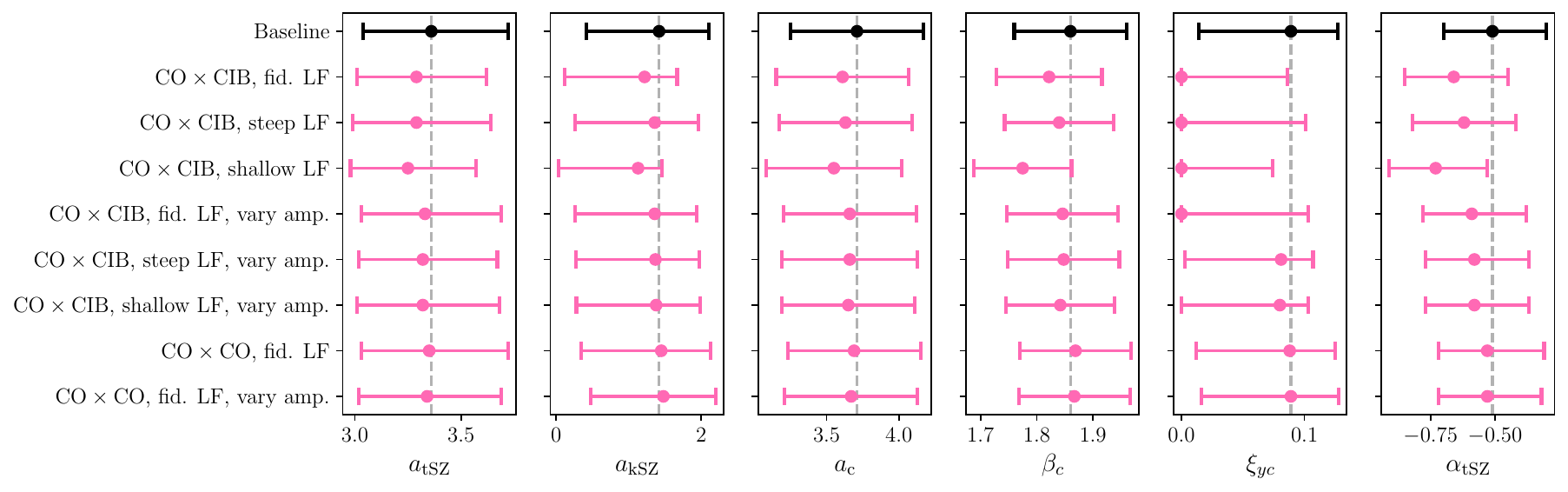}
    \includegraphics[width=0.75\textwidth]{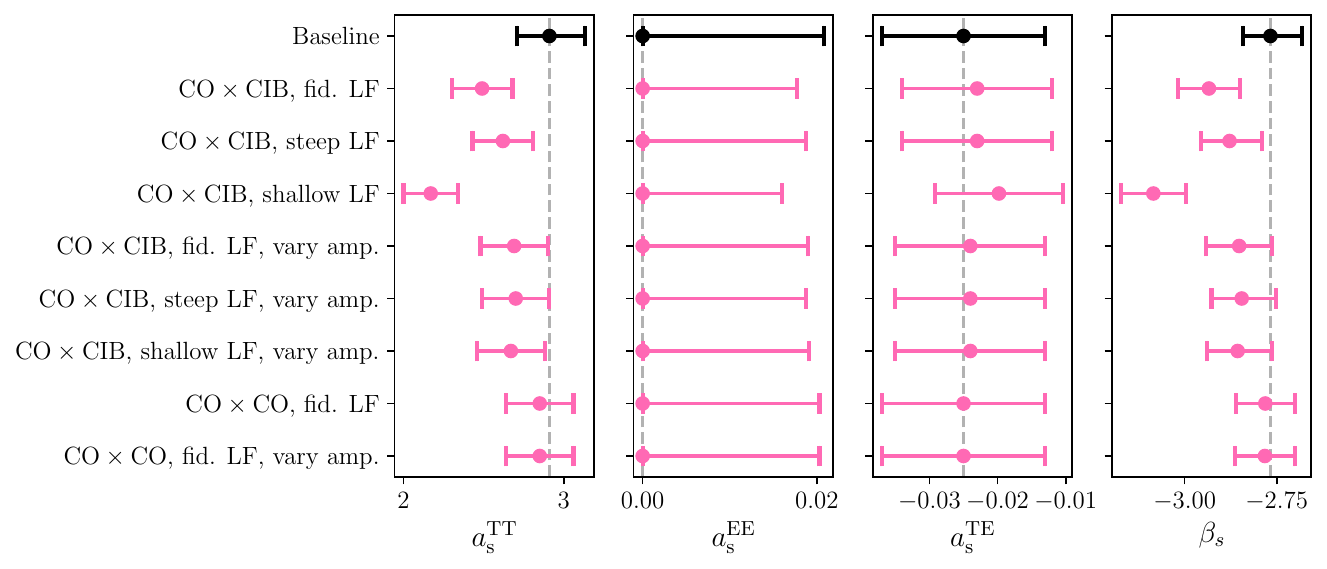}
    \caption{Cosmological and foreground parameter constraints (with 68\% CL error bars) for various foreground tests involving inclusion of CO $\times$ CO and CO $\times$ CIB templates. Constraints using the baseline foreground model are shown in black, while the constraints from tests are shown in pink. See the text for more detail on each of the tests. } 
    \label{fig:co_pars}
\end{figure*}

Rotational line emission from extragalactic CO has recently been investigated as another signal present in the microwave sky~\citep{Maniyar_2023, Kokron:2024, meta_CO_25}. Photons emitted via the CO($J \rightarrow J-1$) transitions can be redshifted into our observed passbands, with CO sources spanning a broad redshift range ($z\sim0-6$). Since the spectral energy distribution of each line is nearly a delta function, the CO emission amplitude is significantly suppressed by the passband width, making its auto-correlation particularly challenging to detect in broad-band CMB observations. However, the cross-correlation between CO and the CIB could be substantial, potentially comparable in magnitude to the kSZ signal at some frequencies. 

In~\cite{Kokron:2024}, the authors used the \verb|AGORA| simulations (described in \S\ref{sec.ng_sims}) to model the CO emission by linking galaxy infrared luminosities to CO luminosity functions, calibrated with observational data from submillimeter surveys.
In this appendix, we explore the robustness of parameter constraints with additions of CO and CO$\times$CIB templates, based on \cite{Kokron:2024}, which contain three model variations: fiducial, steep, and shallow luminosity functions (LFs). The CO$\times$CO and CO$\times$CIB power spectra are modeled as:
\begin{align}\label{eq:spectra_model_CO}
    D_{\ell,{\rm CO\times CO}}^{T_iT_j} &= a_{\rm CO\times CO}  D_{\ell,{\rm CO\times CO}}^{\rm LF}(\nu_i, \nu_j) \\
     D_{\ell,{\rm CO\times CIB}}^{T_iT_j} &= a_{\rm CO\times CIB}  D_{\ell,{\rm CO\times CIB}}^{\rm LF}(\nu_i, \nu_j) \,.
\end{align}

The templates $D_{\ell,{\rm CO\times CIB}}^{\rm LF}$ and $D_{\ell,{\rm CO\times CO}}^{\rm LF}$ are derived from~\citet{Kokron:2024} for the six different cross-spectra and the three luminosity functions. They are then rescaled by an overall amplitude, set to 0 in the baseline model. We study the impact of setting the amplitudes to $1$ or allowing them to vary. These amplitudes have no physical interpretation, and the templates strongly depend on the simulations used to construct them.

\begin{figure}[!ht]
    \centering
    \includegraphics[width=0.5\linewidth]{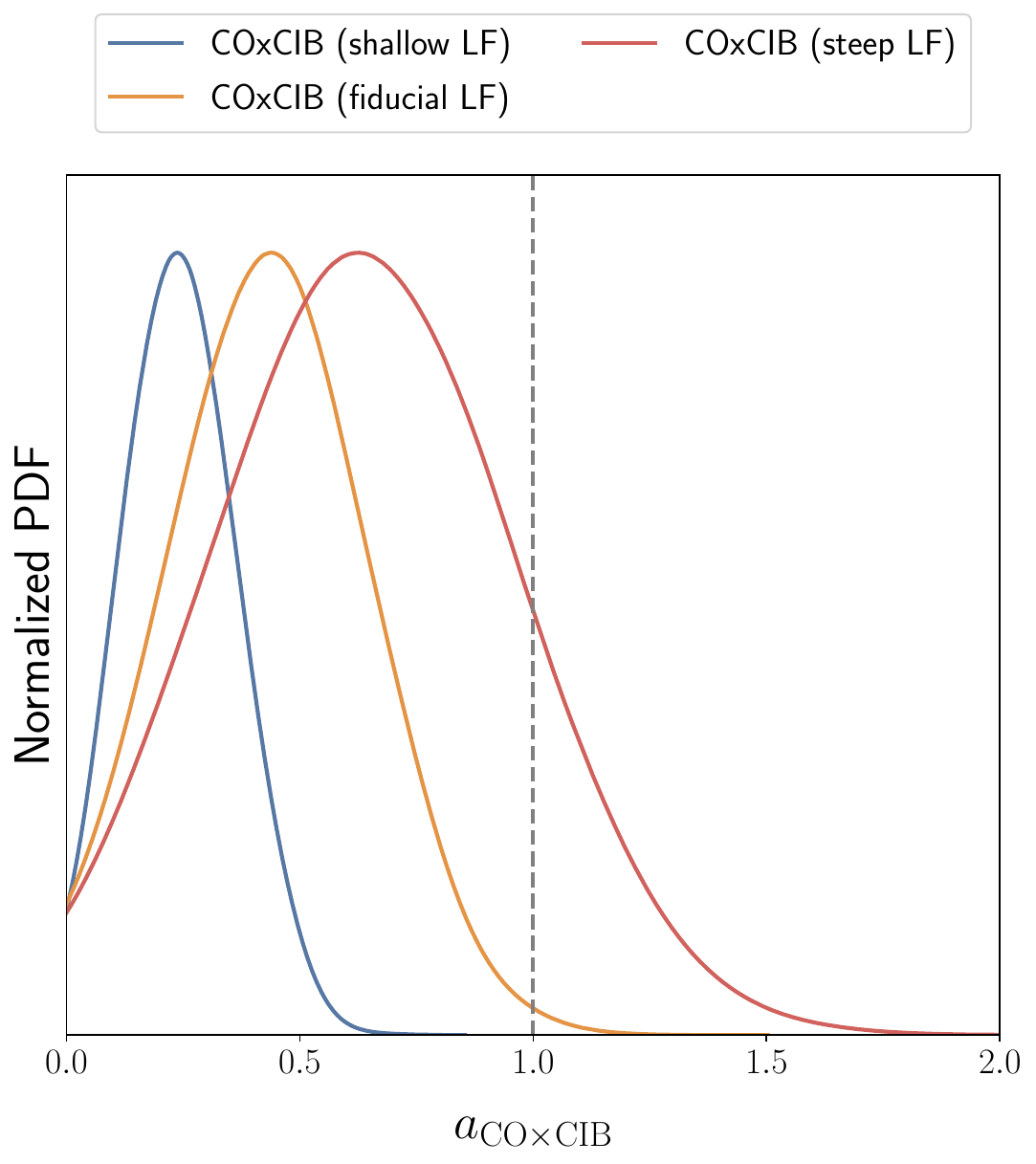}
    \caption{Posterior distribution for the overall amplitude of the CO $\times$ CIB templates for the three templates in~\citet{Kokron:2024}, corresponding to three assumptions on the CO luminosity function (LF). This amplitude is set to one (grey dashed line) in~\citet{Kokron:2024}.  }
    \label{fig:a_coxcib}
\end{figure}

Results are shown in Figure~\ref{fig:co_pars}. First, we note that the CO$\times$CO component, as expected, has no impact on cosmological parameters and very little influence on foreground parameters. For the CO$\times$CIB power spectra, we find that although it has negligible impact on the cosmological parameters, it can shift foreground parameters by up to $2.7\sigma$ ($a_s^\mathrm{TT}$ and $\beta_s$ for shallow LF). Qualitatively, we find that the shallower the luminosity function, the larger the effect, which is in agreement with the numerical findings and analytic scalings that are presented~\citet{Kokron:2024}. Marginalizing over the amplitude allows recovery of unbiased foreground parameters, while having a non-zero best-fitting value for $a_{\rm CO \times CIB}$. As shown in Figure~\ref{fig:a_coxcib}, the best-fitting amplitude (and the associated uncertainty) is inversely proportional to the steepness of the CO luminosity function and is consistently constrained to be different from zero at the $2.1\sigma$ level. The $\Delta \chi^2$ of each of these cases from the baseline are $-4.1$, $-3.8$, and $-4.0$ for the fiducial, steep, and shallow variants, respectively. These correspond to an approximately $2\sigma$ preference over the baseline model, which is not significant. Notably, this preference does not depend strongly on the template (steep, fiducial, and shallow) once marginalized over an overall amplitude. This may be a tentative hint that the CO$\times$CIB signal should be modeled more carefully in future analysis. A more comprehensive investigation of this effect and its modeling is left to future work.

\end{document}